%% file: main.tex
\PassOptionsToPackage{super,round,comma,sort&compress}{natbib}
\documentclass[12pt]{article}

\newif\ifarXiv
\newif\ifscience
\newif\ifSiwei
\newif\ifCombineRef
\newif\ifNature

\arXivtrue

\ifarXiv
\else
\Naturetrue
  \CombineReftrue
\fi

\usepackage{scicite}

\usepackage{times}

\usepackage{array}

\usepackage{booktabs}

\usepackage{arydshln}

\usepackage{placeins}

\usepackage[bottom]{footmisc}

\newenvironment{sciabstract}{%
\begin{quote} \bf}
{\end{quote}}

\usepackage{xspace}
\usepackage{amssymb,amsmath,epsfig}
\usepackage{bbding} 
\usepackage{algorithmicx,algorithm}
\usepackage[noend]{algpseudocode}

\algrenewcommand\algorithmicindent{0.8em}%

\algnewcommand\algorithmicforeach{\textbf{for each}}
\algdef{S}[FOR]{ForEach}[1]{\algorithmicforeach\ #1\ \algorithmicdo}
\newcommand{\setalglineno}[1]{\setcounter{ALG@line}{\numexpr#1-1}}

\usepackage[table]{xcolor} 
\usepackage{booktabs}
\usepackage{url}
\usepackage{float} 
\usepackage{longtable}
\usepackage{setspace}
\definecolor{myblue}{RGB}{245,245,245}
\usepackage[singlelinecheck=off]{caption}
\usepackage{lineno}

\ifarXiv
\else
\DeclareCaptionFont{linenumbers}{\internallinenumbers}
\fi

\usepackage[T1]{fontenc} 
\usepackage{array}

\usepackage{xparse}

\usepackage{tikz}
\usetikzlibrary{matrix, positioning, arrows.meta, calc, shapes, decorations,
  backgrounds, arrows, decorations.pathreplacing, automata}
\usepackage{pgfplots, siunitx}
\usepackage{hhline}
\usepackage{lmodern} 
\usepackage{tikz-qtree} 
\usepackage{forest}
\usetikzlibrary{patterns}

\usepackage{rotating}

\usepackage{array, longtable, makecell}
\usepackage{multirow}
\usepackage{textcomp}

\usepackage{mfirstuc}
\usepackage{titlecaps}

\usepackage{amsmath}

\usepackage{MnSymbol}

\usepackage{csvsimple}
\usepackage{pgf-pie}

\usepackage{pgfplots, pgfplotstable}

\usepackage{multibib} 
\newcites{siref}{References In Supplementary Materials}

\ifCombineRef
\newcommand{\mycite}[1]{\cite{{#1}}}
\newcommand{\mytwocite}[2]{\cite{{#1,#2}}}
\newcommand{\mythreecite}[3]{\cite{{#1,#2,#3}}}
\newcommand{\myfourcite}[4]{\cite{{#1,#2,#3,#4}}}
\newcommand{\myfivecite}[5]{\cite{{#1,#2,#3,#4,#5}}}
\fi

\makeatletter
\let\NAT@parse\undefined
\makeatother
\usepackage{hyperref}

\input{defs}

\title{\vspace{-1cm}
\myedit{Algorithm for Optimized mRNA Design\\Improves Stability and Immunogenicity}\\[-0.1cm]
\ifarXiv
\else 
  \ifscience
  \else
    \ifNature
    \else
    {\small \em draft of \today\vspace{-0.5cm}}
    \fi
  \fi
\fi
}

\author
{\normalsize He Zhang,$^{1,2\dag}$ Liang Zhang,$^{1,2\dag}$ Ang Lin,$^{3\dag}$ Congcong Xu,$^{3}$ Ziyu Li,$^{1}$\\
\normalsize Kaibo Liu,$^{1,2}$ Boxiang Liu,$^{1}$ Xiaopin Ma,$^{3}$ Fanfan Zhao,$^{3}$ Weiguo Yao,$^{3}$ \\ 
\normalsize Hangwen Li,$^{3}$ David H. Mathews,$^{4,5,6}$ Yujian Zhang,$^{3\ast}$ Liang Huang$^{1,2,\dag,\ast,\diamond}$ \\
\\[-0.2cm]
\normalsize{$^{1}$Baidu Research USA, Sunnyvale, CA 94089, USA}\\
\normalsize{$^{2}$School of EECS,}
\normalsize{Oregon State University, Corvallis, OR 97330, USA}\\
\normalsize{$^{3}$Stemirna Therapeutics Inc., Shanghai 201206, China}\\
\normalsize{$^{4}$Dept.~of Biochemistry \& Biophysics}, \normalsize{$^{5}$Center for RNA Biology, and}\\
\normalsize{$^{6}$Dept.~of Biostatistics \& Computational Biology,}\\
\normalsize{University of Rochester Medical Center, Rochester, NY 14642, USA}\\
\\[-0.2cm]
\normalsize{$^\dag$Equal contribution.} 
\normalsize{$^\ast$To whom correspondence should be addressed ($^\diamond$Lead contact).}\\
\normalsize{E-mail: liang.huang.sh@gmail.com; zhangyujian@stemirna.com.}
}

\ifarXiv
\newcommand{\myedit}[1]{{\color{black}{#1}}}
\newcommand{\frankedit}[1]{{\color{black}{#1}}}
  \else
    \ifNature 
    \newcommand{\myedit}[1]{{\color{black}{#1}}}
    \newcommand{\frankedit}[1]{{\color{black}{#1}}}
    \else
      \ifscience
        \newcommand{\myedit}[1]{{\color{black}{#1}}}
        \newcommand{\frankedit}[1]{{\color{black}{#1}}}
  \else
    \newcommand{\myedit}[1]{{\color{cyan}{#1}}}
    \newcommand{\frankedit}[1]{{\color{black}{#1}}}
  \fi
\fi
\fi

\date{}

\begin{document} 



\maketitle 


\baselineskip18pt

\vspace{-0.8cm}

\baselineskip22pt

\ifarXiv
\else
\begin{linenumbers}
\fi

\begin{sciabstract}
Messenger RNA (mRNA) vaccines are being used for COVID-19\myedit{~\cite{baden+:2021, polack+:2020, gebre+:2021}}, 
but still suffer from the \myedit{critical issue of mRNA instability and
degradation}, 
which is a major obstacle in the storage, distribution, and efficacy of the vaccine\myedit{~\cite{crommelin+:2021}}.
Previous work showed that 
\myedit{optimizing  secondary structure  stability lengthens mRNA half-life,
which,
together with optimal codons, increases 
protein expression}\myedit{~\cite{mauger+:2019}}.
Therefore, 
\myedit{a principled mRNA design algorithm must optimize 
{\em both} structural stability and codon usage to improve mRNA efficiency.}
However, due to 
synonymous codons, 
the mRNA design space is prohibitively large, e.g., there are
$\sim\!10^{632}$ 
mRNAs for the \sarscovtwo \spike protein,
\myedit{which poses insurmountable  challenges to previous methods}.
Here we 
provide a surprisingly simple solution to this hard problem
by reducing it to a classical problem in 
 {computational linguistics},
 \myedit{where finding the optimal mRNA is akin to
 finding the most likely sentence among similar sounding alternatives}\myedit{~\cite{hall:2005}}.
\myedit{Our algorithm, named {\em LinearDesign}, 
takes only 11 minutes for 
the \spike protein,
and can  
jointly optimize stability 
and codon usage}.
\myedit{Experimentally, without chemical modification, 
our designs} 
substantially improve \myedit{mRNA half-life} 
and protein expression \invitro, 
and {dramatically} increase antibody response by up to 23$\times$ \invivo,
\myedit{compared to the codon-optimized benchmark}.
\myedit{Our work enables the %
exploration of 
highly stable and efficient
designs that are previously unreachable} 
and is a timely 
tool not only for vaccines but also for mRNA medicine
encoding all therapeutic proteins
(e.g., monoclonal antibodies and anti-cancer drugs\myedit{~\cite{schlake+:2019,reinhard2020rna}}).
\end{sciabstract}


\label{sec:intro}
\input{intro}
\section*{Formulations and Algorithms} 
\label{sec:alg}
\input{formulation}

\input{alg}
\section*{\Insilico Results and Analysis} 
\label{sec:insilico}

\input{result_runtime}
\section*{\Invitro and \Invivo Experimental Results} 
\label{subsubsec:wetlab}
\input{result_wetlab}

\section*{Discussion}
\label{sec:discussion}
\input{discussion}


\ifarXiv
\renewcommand{\refname}{References}
\else
\renewcommand{\refname}{References {\normalfont\selectfont\normalsize 
\big(refs.~\cite{baden+:2021}--\!\cite{kariko2008incorporation} cited in the main text\big)}}
\fi
\ifscience
\bibliographystyle{Science} 
\else
\bibliographystyle{pnas-new} 
\fi
\bibliography{main}



\section*{Availability}
Web server: \url{http://rna.baidu.com}.
\ifNature
\myedit{The code and data has been submitted as part of this submission, 
and will be released at a permanent location upon acceptance.}
\else
Please contact {\tt rna@baidu.com} if you need more advanced features.
\fi

\ifNature
\section*{Conflict of Interest}
{
H.Z., L.Z., Z.L., K.L., B.L., and L.H.~(employees of \myedit{Baidu USA LLC}) have filed a provisional patent for the LinearDesign algorithm.
Sanofi has entered a non-exclusive licensing agreement with \myedit{Baidu} to use LinearDesign 
to design mRNAs for vaccines and therapeutics,
which involves milestone payments when LinearDesign-generated sequences enter clinical trials.}
\fi

\section*{Author Contributions}
L.H.~conceived and directed the project. 
L.H.~designed the basic algorithm for the Nussinov model and wrote a Python prototype,
and H.Z.~and L.Z.~extended this algorithm to the Turner model,
and implemented it in C++, which Z.L.~optimized.
L.H., H.Z., and L.Z.~designed the CAI integration algorithm
which L.Z.~and H.Z.~implemented.
L.Z.~implemented the beam search and $k$-best modules,
and handled design constraints.
K.L.~made the webserver. B.L.~implemented a baseline.
Y.Z.~supervised the \invitro and \invivo experiments.
A.L., X.M., F.Z.~performed the protein expression and immunogenicity assays,
and C.X.~performed chemical stability and structure compactness assays.
D.H.M.~discussed the approach and provided guidance for \insilico analysis and writing.
L.H., H.Z., L.Z., D.H.M., A.L., C.X., and Y.Z.~wrote the manuscript.
The work of H.Z., L.Z., K.L., and L.H.~were done at Baidu Research USA.

\section*{Acknowledgments}
We thank Rhiju Das (Stanford) for introducing the
mRNA design problem to us,
Robin Li (Baidu) for connecting Baidu Research with Stemirna,
Julia Li (Baidu Research) for coordinating resources for this project,
Goro Terai and Kiyoshi Asai (Univ.~of Tokyo) for sending us the CDSfold code,
Sharon Aviran (UC Davis) for \myedit{spotting a typo} in 
the hyperparameter~$\lambda$ in our earlier version,
\myedit{Alicia Sol\'orzano (Pfizer) for the question on LinearDesign's independence of the choice of UTRs},
\myedit{Jinzhong Lin (Fudan) for early discussions,}
and Sizhen Li (Oregon State Univ.) for proofreading and help on \LaTeX.
\myedit{We thank Sanofi and many other vaccine companies worldwide for licensing and early adoption of LinearDesign}. 
D.H.M.~is supported by National Institutes
of Health grant R01GM076485.

\ifNature
\section*{List of Supplementary Materials}
\begin{itemize}
\setlength\itemsep{-.5em}
\item Methods
  \vspace{-.3cm}
  \begin{itemize}
    \item Details of the LinearDesign Algorithm
    \item Details of \Invitro and \Invivo Experiments
  \end{itemize}
\item Supplementary Figures and Tables
\vspace{-.3cm}
  \begin{itemize}
    \item Figures~\ref{fig:word_lattice} to~\ref{fig:seqA_H_structure} 
      \begin{itemize}
          \item Fig.~\ref{fig:word_lattice}: word lattice parsing in natural languague processing
          \item Fig.~\ref{fig:CFG_DFA}: our algorithm solves mRNA design problem via lattice parsing
          \item Fig.~\ref{fig:algorithm}--~\ref{fig:linear}: pseudocode of \lineardesign algorithm
          \item Fig.~\ref{fig:si_alter_aa} and~\ref{fig:si_alter_aa_enzyme}: more examples of the DFA representations for other genetic codes, modified nucleotides and coding constraints
          \item Fig.~\ref{fig:LD_CDSfold_items_generated_table}--~\ref{fig:human_yeast_codon_oval_full}: more \insilico results and analysis
          \item Fig.~\ref{fig:seqA_H_structure}: secondary structures of sequences {\sc a}--{\sc h} 
      \end{itemize} 
    \item Tables~\ref{tab:seqs_info} to~\ref{tab:si_utrs} 
      \begin{itemize}
          \item Tab.~\ref{tab:seqs_info}: details of sequences {\sc a}--{\sc h} and other benchmark sequences
          \item Tab.~\ref{tab:si_utrs}: the numbers of base pairs formed between UTRs and the coding region of different sequences
      \end{itemize} 
  \end{itemize}
\item References starting from~\cite{Rivas:2013} are only cited in Supplementary Materials 
\item Sequences {\sc a}--{\sc g} used in the wet lab experiments (in the FASTA file {\tt {seq\_A\_to\_G.fasta}}; without stop codon) 
\item \lineardesign code (in the file {\tt {\lineardesign-codebase.zip}})
\end{itemize}
\fi

\ifCombineRef
\else
\bibliographystylesiref{pnas-new} 
\newcommand{\mycite}[1]{\citesiref{{si:#1}}}
\newcommand{\mytwocite}[2]{\citesiref{{si:#1,si:#2}}}
\newcommand{\mythreecite}[3]{\citesiref{{si:#1,si:#2,si:#3}}}
\newcommand{\myfourcite}[4]{\citesiref{{si:#1,si:#2,si:#3,si:#4}}}
\newcommand{\myfivecite}[5]{\citesiref{{si:#1,si:#2,si:#3,si:#4,si:#5}}}
\let\oldthebibliography=\thebibliography
\renewenvironment{thebibliography}[1]{%
   \oldthebibliography{#1}%
   \setcounter{enumiv}{0}%
}
\fi

\newpage
\input{methods}

\ifarXiv
\else
\end{linenumbers}
\fi
%
%

%









%
\input{si}

\clearpage

\ifCombineRef
\else
\bibliographysiref{si}
\fi

\clearpage

\end{document}

%% file: defs.tex
\newcommand{\nts}{\ensuremath{\text{\it nt}}\xspace}

\newcommand{\ks}{\ensuremath{k}}

\newcommand{\kbest}{\ks-best\xspace}
\newcommand{\best}{\ensuremath{\textit{best}}\xspace}
\newcommand{\back}{\ensuremath{\textit{back}}\xspace}
\newcommand{\backpointer}{\ensuremath{\textit{backpointer}}\xspace}

\newcommand{\notes}[1]{}

\newcommand{\vecp}{\ensuremath{\mathbf{p}}}

\newcommand{\ith}[1]{\ensuremath{i^{{th}}}}

\newcommand{\goesto}{\ensuremath{\rightarrow}\xspace}
\newcommand{\goestow}[1]{\ensuremath{\stackrel{#1}\rightarrow}\xspace}

\newcount\permx
\newcount\permy
\def\permdot#1#2{
\permx=#1 \advance\permx by-1
\permy=#2 \advance\permy by-1
\psframe[fillcolor=black, fillstyle=solid]
(\permx,\permy)(#1, #2)
}

\newcommand{\argmin}{\operatornamewithlimits{\mathrm{argmin}}}

\newcommand{\cands}{\ensuremath{\textit{cands}}\xspace}
\newcommand{\nodes}{\ensuremath{\textit{nodes}}\xspace}

\newcommand{\score}{\ensuremath{\textit{score}}\xspace}

\newcommand{\vecr}{\ensuremath{\mathbf{r}}\xspace}
\newcommand{\vecs}{\ensuremath{\mathbf{s}}\xspace}

\newcommand{\smallnt}[1]{\ensuremath{_{\mbox{\tiny PP}}}\xspace}

\iffalse

\else

\fi

\newcommand{\seq}{\ensuremath{\textit{seq}}\xspace}
\newcommand{\struct}{\ensuremath{\textit{struct}}\xspace}

\newcommand{\defeq}{\ensuremath{\stackrel{\Delta}{=}}\xspace}

\renewcommand{\to}{\ensuremath{\!\rightarrow\!}\xspace}

\newcommand{\smallurl}[1]{{\scriptsize \url{#1}}}

\newcommand{\md}{\ensuremath{\text{\tt .}}}

\newcommand{\bml}{\ensuremath{\text{\tt{\textbf (}}}}
\newcommand{\bmr}{\ensuremath{\text{\tt{\textbf )}}}}

\newcommand{\linearfold}{{LinearFold}\xspace}

\newcommand{\viennarnafold}{{Vienna RNAfold}\xspace}
\newcommand{\rnafold}{{RNAfold}\xspace}

\newcommand{\lineardesign}{{LinearDesign}\xspace}

\newcommand{\nucA}{\ensuremath{\text{\sc a}}}
\newcommand{\nucU}{\ensuremath{\text{\sc u}}}
\newcommand{\nucC}{\ensuremath{\text{\sc c}}}
\newcommand{\nucG}{\ensuremath{\text{\sc g}}}

\newcommand{\panel}[1]{\large \sf {#1}}

\newcommand{\freeenergy}{\ensuremath{\Delta G^\circ}\xspace}

\newcommand{\RNA}{\ensuremath{\mathit{RNA}}\xspace}
\newcommand{\mRNA}{\ensuremath{\mathit{mRNA}}\xspace}

\newcommand{\protein}{\ensuremath{\mathit{protein}}\xspace}
\newcommand{\MFE}{\ensuremath{\text{MFE}}\xspace}
\newcommand{\MFEplusCAI}{\ensuremath{\text{MFE+CAI}}\xspace}
\newcommand{\mfe}{\ensuremath{\textsc{mfe}}\xspace}
\newcommand{\CAI}{\ensuremath{\text{CAI}}\xspace}

\newcommand{\spike}{\ensuremath{\text{Spike}}\xspace}

\newcommand{\structures}{\ensuremath{\mathit{structures}}\xspace}

\newcommand{\outedges}{\ensuremath{\mathit{out\scalebox{0.5}{\_}edges}}\xspace}
\newcommand{\inedges}{\ensuremath{\mathit{in\scalebox{0.5}{\_}edges}}\xspace}

\definecolor{intnull}{RGB}{213,229,255}
\definecolor{inteins}{RGB}{128,179,255}
\definecolor{intvier}{RGB}{42,127,255}
\definecolor{intdrei}{RGB}{0,85,212}
\definecolor{intvier}{RGB}{0,51,128}
\definecolor{intfunf}{RGB}{0,34,85}

\newcommand{\invitro}{{\em in vitro}\xspace}

\newcommand{\invivo}{{\em in vivo}\xspace}
\newcommand{\insilico}{{\em in silico}\xspace}

\newcommand{\Invitro}{{\em In vitro}\xspace}
\newcommand{\Invivo}{{\em In vivo}\xspace}
\newcommand{\Insilico}{{\em In silico}\xspace}
\newcommand{\Insolution}{{In-solution}\xspace}

\newcommand{\sarscovtwo}{{SARS-CoV-2}\xspace}

\newcommand\dfastate[1]{\scalebox{0.9}{\ensuremath{#1}}\xspace}

\newcommand{\laedge}[3]{\ensuremath{\text{\footnotesize \ensuremath{#1}}\stackrel{#2}{\goesto}\text{\footnotesize \ensuremath{#3}}}\xspace} 

\newcommand{\laedgew}[3]{\ensuremath{\text{\footnotesize \ensuremath{#1}}\stackrel{#2}{\longrightarrow}\text{\footnotesize \ensuremath{#3}}}\xspace} 

\newcommand{\stateindex}[1]{\ensuremath{\mathrm{index}}(#1)\xspace}

\newcommand{\textviolet}{\textcolor{violet}}
\newcommand{\textred}{\textcolor{red}}
\newcommand{\textGray}{\textcolor{gray}}
\newcommand{\textblue}{\textcolor{blue}}

\newcommand{\longleadsto}{\resizebox{1.5em}{1ex}{$\leadsto$}} 

\newcommand{\labeledspan}[3]{\ensuremath{{\scriptstyle {#2}} \stackrel{\stackrel{{#1}}{\triangle}}{\rightarrow} {\scriptstyle {#3}}}\xspace}

\newcommand{\labeledtwospans}[5]{\ensuremath{{\scriptstyle {#2}} \stackrel{\stackrel{{#1}}{\triangle}}{\rightarrow} {\scriptstyle {#3}} \stackrel{\stackrel{{#4}}{\triangle}}{\rightarrow} {\scriptstyle {#5}}}\xspace}

\newcommand{\labeledpath}[3]{\ensuremath{\text{\small \ensuremath{#2}} \stackrel{\stackrel{{#1}}{\triangle}}{\leadsto} \text{\small \ensuremath{#3}}}\xspace} 

\newcommand{\labeledtwopaths}[5]{\ensuremath{{\scriptstyle {#2}} \stackrel{\stackrel{{#1}}{\triangle}}{\leadsto} {\scriptstyle {#3}} \stackrel{\stackrel{{#4}}{\triangle}}{\leadsto} {\scriptstyle {#5}} }\xspace} 

\newcommand{\codon}{\ensuremath{\text{codon}}\xspace}
\newcommand{\mfiveC}{\ensuremath{\text{m}^5\nucC}\xspace}
\newcommand{\msixA}{\ensuremath{\text{m}^6\nucA}\xspace}

\newcommand{\optimumgene}{OptimumGene\textsuperscript{\texttrademark}\xspace}

\newcommand{\celsius}{\textdegree{}C\xspace}
\newcommand{\mincost}{\ensuremath{\text{mincost}}\xspace}

\newsavebox\CBox
\def\textBF#1{\sbox\CBox{#1}\resizebox{\wd\CBox}{\ht\CBox}{\textbf{#1}}}

\newcommand{\MFECAIlambda}{\ensuremath{\MFE\CAI_{\lambda}}\xspace}

\newcommand{\bluediamond}{\small\scalebox{1.5}[1]{\color{blue}$\blacklozenge$}}
\newcommand{\blueDiamond}{\small\ensuremath{\color{blue}\Diamond\hspace{-0.475em}{\cdot}}\hspace{0.2em}} 

%% file: intro.tex


Messenger RNA (mRNA) vaccines \cite{wolff+:1990, pardi+:2018}
emerged as a promising tool against  COVID-19
thanks to 
their scalable production, safety, and efficacy\myedit{~\cite{baden+:2021, polack+:2020, gebre+:2021}}. 
However, among other limitations, they 
still suffer 
from the critical issue of chemical instability and degradation of the 
largely single-stranded and fragile
mRNA molecule
both in solution and \invivo.
This instability has become a major obstacle in the storage and distribution of the vaccine,
requiring the use of cold-chain technologies that {hinders its use 
in developing countries~\cite{crommelin+:2021}}.
More importantly, the \invivo instability of mRNAs leads to insufficient protein expression~\cite{mauger+:2019}, 
and in turn, 
compromised immunogenicity.
While 
chemical stability is hard to model, previous work established its correlation with 
\myedit{secondary structures,
as quantified by the well-studied thermodynamic folding stability.}
\myedit{This structural stability,} along 
with optimal codon usage, 
leads to greater 
protein expression \cite{mauger+:2019}.
Therefore, \myedit{a principled mRNA design algorithm needs to optimize 
{\em both} 
structural stability and codon usage
to enhance mRNA translation efficiency}.


\input{fig_framework}

However, this mRNA design problem is extremely challenging due to the exponentially large
search space.\footnote{\myedit{Our work only designs the coding region, 
and is independent of the untranslated regions (UTRs).}} 
Each amino acid is encoded by a triplet codon, i.e., three adjacent nucleotides,
but due to redundancies in the genetic code ($4^3\!=\!64$ codons for 20 amino acids), 
most amino acids have 
multiple codons.
This combinatorial explosion 
results in a prohibitively large number of candidates. 
For example, the \spike protein of SARS-CoV-2
with 1,273 amino acids
can be encoded by $\sim\!2.4 \times 10^{632}$ 
mRNA sequences (Fig.~\ref{fig:overview}A). 
\myedit{This poses an insurmountable computational challenge to 
jointly optimize structural stability and codon usage, 
and rules out  enumeration which takes
$\sim\!10^{616}$ {\em billion years} for the Spike protein (Fig.~\ref{fig:overview}B)}.
On the other hand, 
the conventional approach to mRNA design,
{\em codon optimization} \cite{gustafsson+:2004}, 
\myedit{only} optimizes codon usage but barely improves  stability, 
leaving out
the huge space of \myedit{highly stable mRNAs}
(Fig.~\ref{fig:overview}D).
Optimizing the other widely considered factor, GC-content, 
has a similar effect as it 
correlates  with codon usage in vertebrates\cite{nabiyouni+:2013}. 
As a result, 
the vast majority of \myedit{highly stable} designs 
 remains unexplored.



Here we provide a surprisingly simple algorithm, {\em LinearDesign},
to solve this challenging problem 
\myedit{by reducing it to the classical problem of ``lattice parsing''~\cite{hall:2005} in 
computational linguistics (Fig.~\ref{fig:overview}C).
We show that finding the optimal mRNA among vast space of candidates is analogous to finding the 
most grammatical sentence among numerous similar-sounding alternatives.
More specifically,} we formulate the mRNA design space using a deterministic finite-state automaton (DFA),
similar to a ``word lattice'' \cite{hall:2005},
which compactly encodes exponentially many  mRNA candidates (Figs.~\ref{fig:overview}--\ref{fig:main_CFG_DFA}).
\myedit{We then use lattice parsing
to find the most stable mRNA in the DFA without enumeration
(Fig.~\ref{fig:main_CFG_DFA}B), 
which can also jointly optimize stability and codon optimality 
by representing the latter in a {\em weighted} DFA (Fig.~\ref{fig:main_CFG_DFA}D).}
\myedit{This unexpected connection to natural language enables an efficient 
algorithm 
that scales {\em quadratically} with the mRNA sequence length in practice,} 
taking only 11 minutes to design the most stable mRNA encoding 
the \sarscovtwo \spike protein. 
\myedit{This optimal design has a mostly double-stranded secondary structure vastly different 
from the largely single-stranded wildtype, thus being much more stable than the latter} (Fig.~\ref{fig:overview}B). 
\myedit{In this sense, our work turns the enormous search space into a {\em blessing} (freedom of design) rather than an obstacle.}
We also develop an even faster (linear-time) variant 
(Fig.~\ref{fig:insilico}A) 
that provides suboptimal designs for vaccine development.
%
\myedit{More importantly}, 
experimental results confirmed that,
compared to the codon-optimized benchmark, 
our designs 
substantially improve chemical stability \invitro, protein expression {\em in cell}, 
and immunogenicity \invivo  (Figs.~\ref{fig:overview}\&~\ref{fig:wetlab}).
In particular, our designed molecules
 increase  antibody response by up to 23$\times$ in mice. 
Our work provides a timely and promising tool 
not only for mRNA vaccines, but also for  mRNA medicine 
which has shown great potential to revolutionize healthcare 
\cite{sahin+:2014},
as LinearDesign can  optimize mRNAs encoding
all therapeutic proteins including monoclonal antibodies\myedit{~\cite{schlake+:2019}} and anti-cancer drugs~\frankedit{\cite{reinhard2020rna}}.

\input{fig_dfa_cfg} 

%% file: fig_framework.tex

\begin{figure}[!t]
\centering
\vspace{-0.5cm}
\begin{tabular}{cc}
\raisebox{0cm}{\hspace{-.3cm}\includegraphics[width=.95\textwidth]{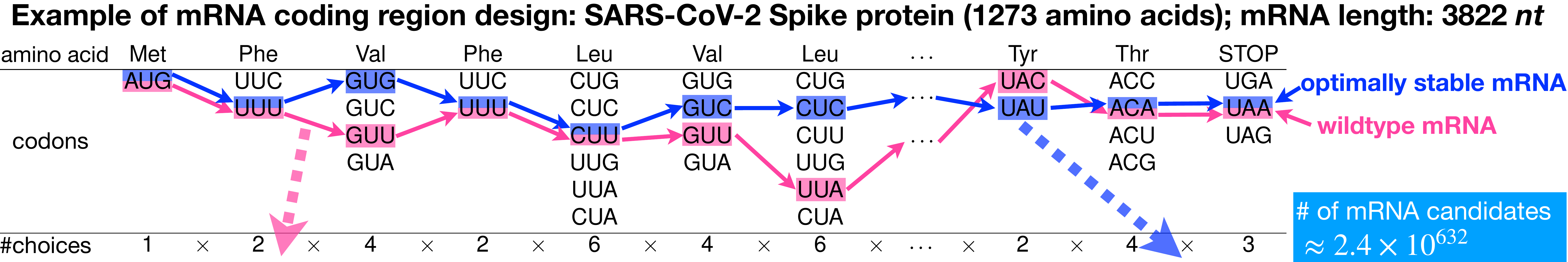}} &\\[-.1cm]
\raisebox{0cm}{\hspace{-.4cm}\includegraphics[width=.95\textwidth]{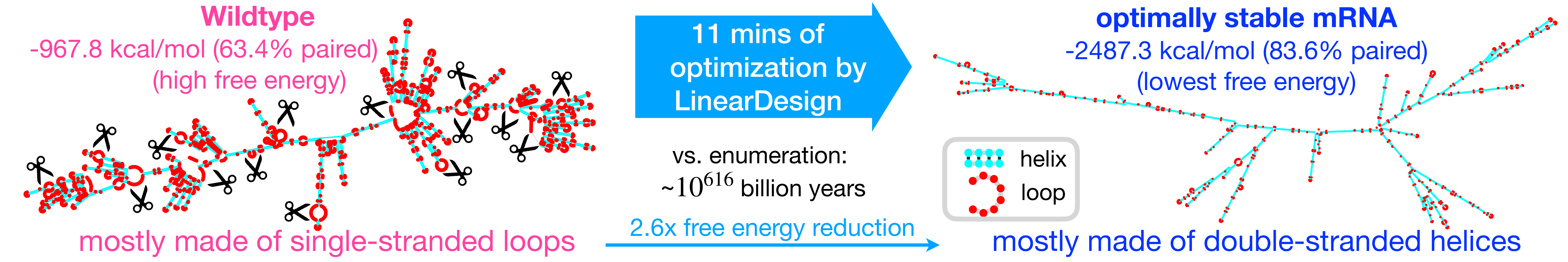}} &\\[-.0cm]
\raisebox{0cm}{\hspace{-.25cm}\includegraphics[width=.99\textwidth]{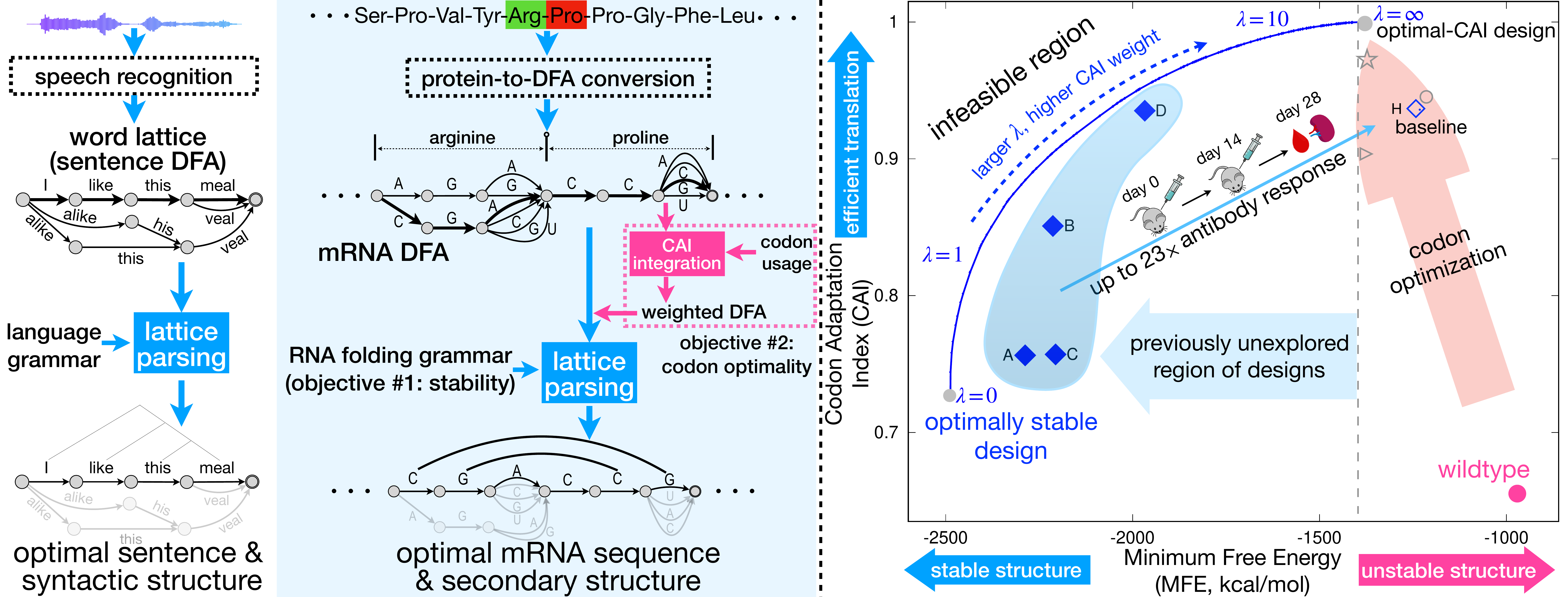}} &\raisebox{0cm}{}\\[-11.3cm]
\raisebox{0cm}{\hspace{-15.85cm}\panel{A}} &\raisebox{0cm}{} \\[2.cm]
\raisebox{0cm}{\hspace{-15.85cm}\panel{B}} &\raisebox{0cm}{} \\[2.cm]
\raisebox{0cm}{\hspace{-15.85cm}\panel{C}} & \raisebox{0cm}{\hspace{-15.5cm}\panel{D}}\\[5.4cm]

\end{tabular}
\caption{Overview of mRNA coding region design for two well-established objectives, 
{\em stability} 
and {\em codon optimality}, using SARS-CoV-2 Spike protein as an example. 
{\bf A}: Due to combinatorial explosion, there are $\sim\! 10^{632}$ mRNAs for the \spike protein.
The pink and blue paths represent the wildtype 
and the optimally stable (i.e., lowest energy) design found by our work, respectively.
{\bf B}: \myedit{There are vastly different secondary structures between these two sequences,
with the former being mostly single-stranded (thus prone to degradation \ScissorRight)
and the latter 
being mostly double-stranded. 
Our algorithm LinearDesign takes just 11 minutes for this optimization while enumeration needs $\sim\!\!10^{616}$ billion years.}
{\bf C}:~We borrow deterministic finite-state automaton (DFA) and lattice parsing from language (left) for mRNA design (right).
An mRNA DFA, inspired by ``word lattice'', 
compactly represents all mRNA candidates. 
Lattice parsing folds all sequences in this DFA 
with an RNA folding grammar
  to find the optimally stable mRNA
(Fig.~\ref{fig:main_CFG_DFA}B),
and can also incorporate codon optimality using a weighted DFA (Fig.~\ref{fig:main_CFG_DFA}D).
%
{\bf D}: 
Visualization of the mRNA design space for the \spike protein, with stability 
on the $x$-axis and 
codon optimality
on the $y$-axis. 
The conventional mRNA design method,  
used by the COVID-19 vaccines of 
BioNTech-Pfizer~({\color{gray}$\circ$}), Moderna~($\color{gray}{\medstar}$), and CureVac~($\color{gray}{\triangleright}$),
is {\em codon optimization}~\protect\cite{kon+:2021},  
which improves codon usage (the~\includegraphics[width=.01\textheight]{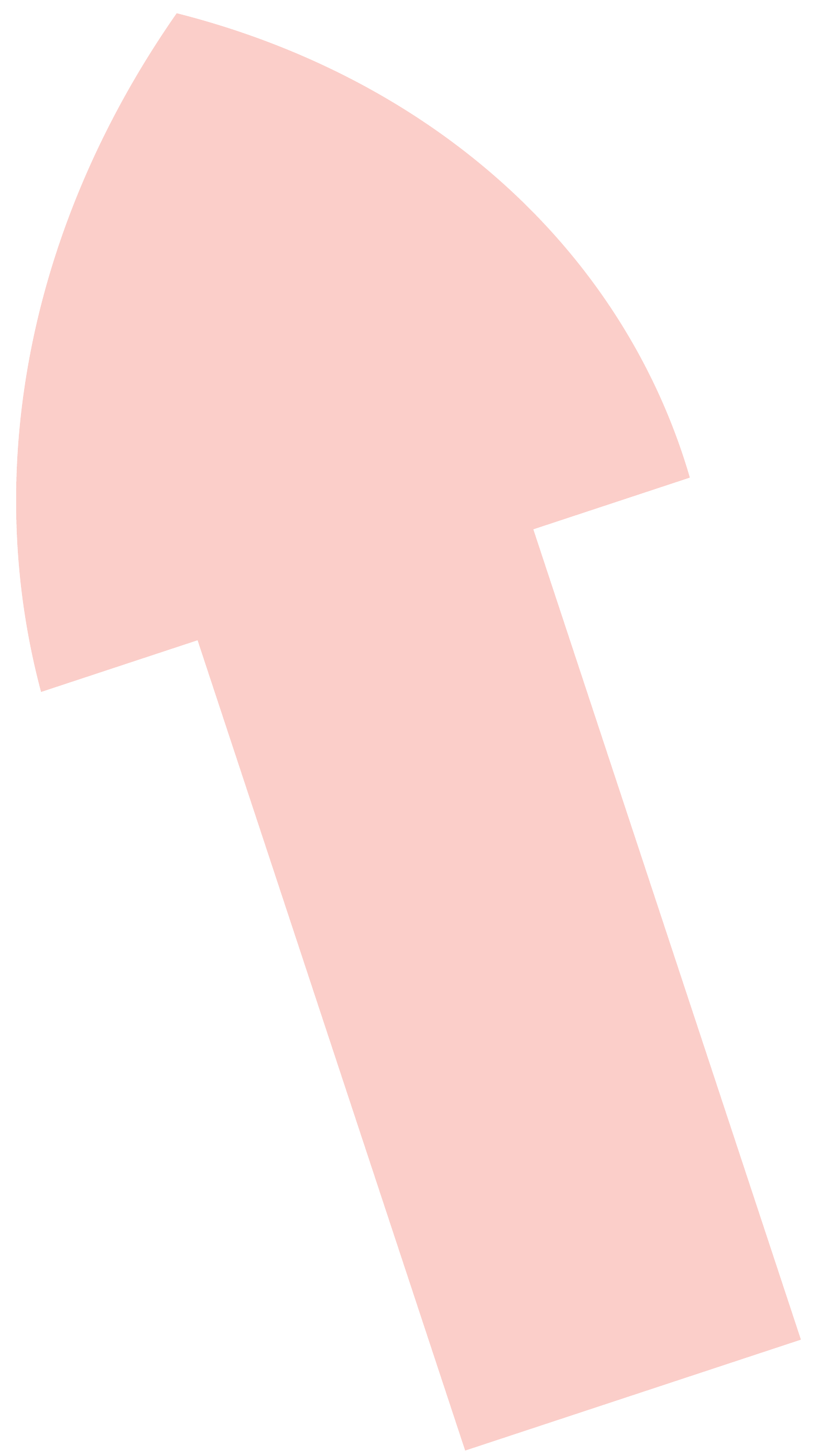} arrow), 
but leaves out the high-stability region (left of 
the dashed line). 
\lineardesign, by contrast, 
jointly optimizes stability and codon optimality (the blue curve, with $\lambda$ being the weight of the latter).
By considering other factors, 
we select a few of our designs (four shown here: {\small \bluediamond \sc a--d})
for 
experiments (Fig.~\ref{fig:wetlab}),
which show up to $23\times$  antibody responses  
over the codon-optimized baseline ({\small \sc h\blueDiamond}).
\label{fig:overview}
}
\end{figure}

%% file: fig_dfa_cfg.tex

\usetikzlibrary{positioning, calc, automata, arrows, shapes, backgrounds, quotes, decorations.text}
\tikzset{every picture/.style={/utils/exec={\sffamily}}}

\colorlet{DarkGray}{gray!30}
\colorlet{DarkerGray}{black!60}
\colorlet{ColorGray}{blue!45}

\colorlet{StartNode}{purple!60}
\colorlet{NodeGray}{gray!20}
\colorlet{EndNode}{purple!60}
\colorlet{MidNode}{purple!30}

\begin{figure}[!t]
\vspace{-0.2cm}
\centering
\begin{tabular}{cccc}


\hspace{-.3cm}\raisebox{.8cm}{\panel{\myedit{A}} \fontsize{9}{0}\selectfont \textbf{\myedit{\ \ \ codon DFAs}}}
& 
\hspace{-.3cm}\raisebox{.3cm}{\resizebox{.27\textwidth}{!}{
\begin{tikzpicture}[->,>=stealth',shorten >=1pt,auto,semithick]
\node[state, initial, initial text=, inner sep=-5pt, minimum size=.2cm, fill=StartNode] (n00) {\fontsize{12.5}{0}\selectfont };
\node[state, right=.8cm of n00, inner sep=-5pt, minimum size=.2cm, fill=NodeGray] (n10) {\fontsize{12.5}{0}\selectfont };
\node[state, right=.8cm of n10, inner sep=-5pt, minimum size=.2cm, fill=NodeGray] (n20) {\fontsize{12.5}{0}\selectfont };
\node[state, accepting, right=.8cm of n20, inner sep=-5pt, minimum size=.2cm, fill=EndNode] (n30) {\fontsize{12.5}{0}\selectfont };
\node[below=-1.1cm of n00, inner sep=-10pt] (n0a) {}; 
\node[right=.4cm of n0a] (n0b) {$D(\text{isoleucine})$}; 
\draw (n00) edge[above] node[red]{A} (n10)
(n10) edge[above] node[gray]{U} (n20)
(n20) edge[above, bend left=60] node[red]{A} (n30)
(n20) edge[above] node[blue]{C} (n30)
(n20) edge[below, bend right=60] node[gray, above]{U} (n30);
\end{tikzpicture}}}
&
\hspace{-.3cm}\raisebox{-.4cm}{\resizebox{.27\textwidth}{!}{
\begin{tikzpicture}[->,>=stealth',shorten >=1pt,auto,semithick]
\node[state, initial, initial text=, inner sep=-5pt, minimum size=.2cm, fill=StartNode] (n00) {\fontsize{12.5}{0}\selectfont };
\node[state, right=.8cm of n00, inner sep=-5pt, minimum size=.2cm, fill=NodeGray] (n10) {\fontsize{12.5}{0}\selectfont };
\node[state, below=0.35cm of n10, inner sep=-5pt, minimum size=.2cm, fill=NodeGray] (n11) {\fontsize{16.5}{0}\selectfont };
\node[state, right=.8cm of n10, inner sep=-5pt, minimum size=.2cm, fill=NodeGray] (n20) {\fontsize{12.5}{0}\selectfont };
\node[state, below=0.35cm of n20, inner sep=-5pt, minimum size=.2cm, fill=NodeGray] (n21) {\fontsize{12.5}{0}\selectfont };
\node[state, accepting, right=.8cm of n20, inner sep=-5pt, minimum size=.2cm, fill=EndNode] (n30) {\fontsize{12.5}{0}\selectfont };
\node[below=-1.1cm of n00, inner sep=-10pt] (n0a) {}; 
\node[right=.6cm of n0a] (n0b) {$D(\text{leucine})$}; 
\draw (n00) edge[above] node[gray]{C} (n10)
(n00) edge[above] node[gray, below]{U} (n11)
(n10) edge[above] node[gray]{U} (n20)
(n11) edge[above] node[gray]{U} (n21)
(n20) edge[above, bend left=80,inner sep=2pt] node[red]{A} (n30)
(n20) edge[above, bend left=30,inner sep=2pt] node[blue]{C} (n30)
(n20) edge[below, bend right=15,inner sep=2pt] node[violet, above]{G} (n30)
(n20) edge[below, bend right=58,inner sep=2pt] node[gray, above, yshift=-.2pt]{U} (n30)
(n21) edge[below, bend right=30] node[red, below, yshift=.4pt]{A} (n30)
(n21) edge[below, bend right=80, right=0.3] node[violet, below]{G} (n30);
\end{tikzpicture}}}
&
\hspace{-.3cm}\raisebox{.0cm}{\resizebox{.27\textwidth}{!}{
\begin{tikzpicture}[->,>=stealth',shorten >=1pt,auto,semithick]
\node[state, initial, initial text=, inner sep=-5pt, minimum size=.2cm, fill=StartNode] (n00) {\fontsize{12.5}{0}\selectfont };
\node[state, right=.8cm of n00, inner sep=-5pt, minimum size=.2cm, fill=NodeGray] (n10) {\fontsize{12.5}{0}\selectfont };
\node[state, right=.8cm of n10, inner sep=-5pt, minimum size=.2cm, fill=NodeGray] (n20) {\fontsize{12.5}{0}\selectfont };
\node[state, below=0.35cm of n20, inner sep=-5pt, minimum size=.2cm, fill=NodeGray] (n21) {\fontsize{12.5}{0}\selectfont };
\node[state, accepting, right=.8cm of n20, inner sep=-5pt, minimum size=.2cm, fill=EndNode] (n30) {\fontsize{12.5}{0}\selectfont };
\node[below=-1.1cm of n00, inner sep=-10pt] (n0a) {}; 
\node[right=.6cm of n0a] (n0b) {$D(\text{stop})$}; 
\draw (n00) edge[above] node[gray]{U} (n10)
(n10) edge[above] node[red]{A} (n20)
(n10) edge[above] node[violet, below]{G} (n21)
(n20) edge[above, bend left=60,inner sep=2pt] node[red]{A} (n30)
(n20) edge[below,inner sep=2pt] node[violet, above]{G} (n30)
(n21) edge[below] node[red, below]{A} (n30);
\end{tikzpicture}}}\\
\end{tabular}
\\[-.2cm]

\begin{tabular}{cc}
\hspace{-7.9cm}\raisebox{-.2cm}{\panel{\myedit{B}} \fontsize{9}{0}\selectfont \textbf{\ \ \ \myedit{mRNA DFA and lattice parsing}}} & \hspace{-2.6cm}\raisebox{-.2cm}{\panel{\myedit{C}} \fontsize{9}{0}\selectfont \textbf{\ \ \ \myedit{optimal sequence \& structure}}} \\[-0.3cm]
\hspace{-.6cm}\raisebox{-2cm}{
\resizebox{.82\textwidth}{!}{
\begin{tikzpicture}[->,>=stealth',shorten >=1pt,auto,semithick,
no arrow/.style={-,every loop/.append style={-}}]

\fill [blue!34] (0,.) -- (2.2,4.) to[out=60,in=-180] (3.5,4.7) -- (13.7,4.7) to[out=0,in=120] (15.,4.) -- (17.2,.) -- cycle;
\fill [blue!26] (2.8,.) -- (4.6,2.5) to[out=60,in=-180] (6.0,3.4) -- (12.4,3.4) to[out=0,in=120] (13.8,2.5) -- (15.6,.) -- cycle;
\fill [blue!18] (4.3,.) -- (5.5,1.8) to[out=60,in=-180] (6.7,2.55) -- (11.8,2.55) to[out=0,in=120] (13.,1.8) -- (14.2,.) -- cycle;
\fill [blue!10] (5.7,.) -- (6.45,1.2) to[out=60,in=-180] (7.55,2) -- (10.95,2) to[out=0,in=120] (12.05,1.2) -- (12.6,.) -- cycle;
\fill [red!25] (7.2,.) -- (7.5,.8) to[out=60,in=-180] (8.1,1.3) -- (10.6,1.3) to[out=0,in=120] (11.2,.8) -- (11.5,.) -- cycle;
\fill [red!25] (1.4,.) -- (1.6,.8) to[out=60,in=-180] (1.9,1.3) -- (2.3,1.3) to[out=0,in=120] (2.6,.8) -- (2.8,.) -- cycle;

\node[state, initial, initial text=, inner sep=-5pt, minimum size=.2cm, fill=StartNode] (n00) {\fontsize{12.5}{0}\selectfont };
\node[below=1.5cm of n00] (n0b) {$\mid$}; 
\node[state, right=.9cm of n00, inner sep=-5pt, minimum size=.2cm, fill=NodeGray] (new10) {\fontsize{12.5}{0}\selectfont };
\node[state, right=.9cm of new10, inner sep=-5pt, minimum size=.2cm, fill=NodeGray] (new20) {\fontsize{12.5}{0}\selectfont };
\node[state, right=.9cm of new20, inner sep=-5pt, minimum size=.2cm, fill=MidNode] (new30) {\fontsize{12.5}{0}\selectfont };
\node[below=1.5cm of new30] (new3b) {$\mid$}; 
\node[below=1.9cm of new30] (new3c) {$\circ$}; 

\node[state, right=.9cm of new30, inner sep=-5pt, minimum size=.2cm, fill=NodeGray] (n10) {\fontsize{12.5}{0}\selectfont };
\node[state, right=.9cm of n10, inner sep=-5pt, minimum size=.2cm, fill=NodeGray] (n20) {\fontsize{12.5}{0}\selectfont };
\node[state, right=.9cm of n20, inner sep=-5pt, minimum size=.2cm, fill=MidNode] (n30) {\fontsize{12.5}{0}\selectfont };
\node[below=1.5cm of n30] (n3b) {$\mid$}; 
\node[below=1.9cm of n30] (n3c) {$\circ$}; 

\node[state, right=.9cm of n30, inner sep=-5pt, minimum size=.2cm, fill=NodeGray] (n40) {\fontsize{12.5}{0}\selectfont };
\node[state, right=.9cm of n40, inner sep=-5pt, minimum size=.2cm, fill=NodeGray] (n50) {\fontsize{12.5}{0}\selectfont };
\node[state, right=.9cm of n50, inner sep=-5pt, minimum size=.2cm, fill=MidNode] (n60) {\fontsize{12.5}{0}\selectfont };
\node[below=1.5cm of n60] (n6b) {$\mid$}; 
\node[below=1.9cm of n60] (n6c) {$\circ$}; 
\node[state, right=.9cm of n60, inner sep=-5pt, minimum size=.2cm, fill=NodeGray] (n70) {\fontsize{12.5}{0}\selectfont };
\node[state, below=.7cm of n70, inner sep=-5pt, minimum size=.2cm, fill=NodeGray, opacity=1] (n71) {\fontsize{12.5}{0}\selectfont };
\node[state, right=.9cm of n70, inner sep=-5pt, minimum size=.2cm, fill=NodeGray] (n80) {\fontsize{12.5}{0}\selectfont };
\node[state, below=.7cm of n80, inner sep=-5pt, minimum size=.2cm, fill=NodeGray, opacity=1] (n81) {\fontsize{12.5}{0}\selectfont };
\node[state, accepting, right=.9cm of n80, inner sep=-5pt, minimum size=.2cm, fill=EndNode] (n90) {\fontsize{12.5}{0}\selectfont };
\node[below=1.5cm of n90] (n9b) {$\mid$}; 
\node[below=1.9cm of n90] (n9c) {$\circ$}; 

\draw (n00) edge[above] node[red]{A} (new10)
(n00) edge[above,line width=1.5pt, blue] node[black, yshift=15pt, brown]{\fontsize{8.5}{0}\selectfont \textbf{(}} (new10)
(new10) edge[above] node[gray]{U} (new20)
(new10) edge[above,line width=1.5pt, blue] node[black, yshift=15pt]{\fontsize{8.5}{0}\selectfont $\bullet$} (new20)
(new20) edge[above] node[violet]{G} (new30)
(new20) edge[above,line width=1.5pt, blue] node[black, yshift=15pt]{\fontsize{8.5}{0}\selectfont \textbf{(}} (new30)

(new30) edge[above] node[red]{A} (n10)
(new30) edge[above,line width=1.5pt, blue] node[black, yshift=15pt, brown]{\fontsize{8.5}{0}\selectfont \textbf{(}} (n10)
(n10) edge[above] node[blue]{C} (n20)
(n10) edge[above,line width=1.5pt, blue] node[black, yshift=15pt]{\fontsize{8.5}{0}\selectfont \textbf{(}} (n20)
(n20) edge[above] node[blue]{C} (n30)
(n20) edge[above,line width=1.5pt, blue] node[black, yshift=15pt]{\fontsize{8.5}{0}\selectfont $\bullet$} (n30)
(n20) edge[above, bend right=35, opacity=1] node[red, yshift=-3pt]{\fontsize{8.5}{0}\selectfont A} (n30)
(n20) edge[above, bend right=80, opacity=1] node[violet, yshift=-3pt, opacity=1]{\fontsize{8.5}{0}\selectfont G} (n30);
\draw[] (n20) to[out=-100, in=-180] (7.85,-.98) to[out=0,in=-80] node[gray, xshift=-8.5pt, yshift=-8.2pt]{\fontsize{8.5}{0}\selectfont U} (n30);
\draw (n30) edge[above] node[gray]{U} (n40)
(n30) edge[above,line width=1.5pt, blue] node[black, yshift=15pt]{\fontsize{8.5}{0}\selectfont $\bullet$} (n40)
(n40) edge[above] node[violet]{G} (n50)
(n40) edge[above,line width=1.5pt, blue] node[black, yshift=15pt]{\fontsize{8.5}{0}\selectfont $\bullet$} (n50)
(n50) edge[above] node[violet]{G} (n60)
(n50) edge[above,line width=1.5pt, blue] node[black, yshift=15pt]{\fontsize{8.5}{0}\selectfont \textbf{)}} (n60)
(n60) edge[above] node[gray]{U} (n70)
(n60) edge[above,line width=1.5pt, blue] node[black, yshift=15pt, brown]{\fontsize{8.5}{0}\selectfont \textbf{)}} (n70)
(n60) edge[above, opacity=1] node[red]{\fontsize{8.5}{0}\selectfont A} (n71)
(n70) edge[above] node[blue]{C} (n80)
(n70) edge[above,line width=1.5pt, blue] node[black, yshift=15pt]{\fontsize{8.5}{0}\selectfont \textbf{)}} (n80)
(n71) edge[above, opacity=1] node[violet]{\fontsize{8.5}{0}\selectfont G} (n81)
(n80) edge[above] node[gray]{U} (n90)
(n80) edge[above,line width=1.5pt, blue] node[black, yshift=15pt, brown]{\fontsize{8.5}{0}\selectfont \textbf{)}} (n90)
(n80) edge[above, bend right=30] node[red, yshift=-4pt]{\fontsize{8.5}{0}\selectfont A} (n90)
(n80) edge[above, bend right=65] node[blue, yshift=-4pt]{\fontsize{8.5}{0}\selectfont C} (n90);
\draw[] (n80) to[out=-90, in=-180] (16.35,-.88) to[out=0,in=-90] node[violet, xshift=-7pt, yshift=-8.2pt]{\fontsize{8.5}{0}\selectfont G} (n90);
\draw (n81) edge[above, bend right=70, opacity=1] node[blue, opacity=1, yshift=-2pt]{\fontsize{8.5}{0}\selectfont C} (n90);
\draw[] (n81) to[out=-45, in=-160] (16.8,-1.56) to[out=20,in=-50] node[gray, xshift=-8pt, yshift=-20pt]{\fontsize{8.5}{0}\selectfont U} (n90);

\node[right=.3cm of n00](n01) {};
\node[above=.8cm of n01](n01up) {};
\node[right=.3cm of n80](n89) {};
\node[above=.8cm of n89](n89up) {};

\node[right=.3cm of new20](new23) {};
\node[above=.8cm of new23](new23up) {};
\node[right=.3cm of n70](n78) {};
\node[above=.8cm of n78](n78up) {};

\node[right=.3cm of new30](new34) {};
\node[above=.8cm of new34](new34up) {};
\node[right=.3cm of n60](n67) {};
\node[above=.8cm of n67](n67up) {};

\node[right=.3cm of n10](n45) {};
\node[above=.8cm of n45](n45up) {};
\node[right=.3cm of n50](n56) {};
\node[above=.8cm of n56](n56up) {};

\draw (n01up) edge[above, bend left=46, no arrow, line width=2.5pt, brown] node{} (n89up);
\draw (new23up) edge[above, bend left=40, no arrow, line width=2.5pt] node{} (n78up);
\draw (new34up) edge[above, bend left=31, no arrow, line width=2.5pt, brown] node{} (n67up);
\draw (n45up) edge[above, bend left=25, no arrow, line width=2.5pt] node{} (n56up);

\draw (n0b) edge[below,<->,dashed] node{$D(\text{methionine})$} (new3b);
\draw (new3b) edge[below,<->,dashed] node{$D(\text{threonine})$} (n3b);
\draw (n3b) edge[below,<->,dashed] node{$D(\text{tryptophan})$} (n6b);
\draw (n6b) edge[below,<->,dashed] node{$D(\text{serine})$} (n9b);

\node[left=.8cm of n00](m0) {};
\node[above=.4cm of m0](m1) {};
\node[left=.3cm of m1](m11) {};
\node[right=.6cm of m1](m111) {};
\node[below=1.5cm of m0](m2) {};
\node[left=.3cm of m2](m22) {};
\node[right=.6cm of m2](m222) {};
\node[above=4.5cm of m0](m3) {};
\node[left=.3cm of m3](m33) {};
\node[right=.6cm of m3](m333) {};

\draw (m1) edge[above,<->] node[rotate=90]{\fontsize{9.5}{0}\selectfont mRNA DFA} (m2);
\draw (m1) edge[above,<->] node[rotate=90]{\fontsize{9.5}{0}\selectfont Lattice Parsing} (m3);
\draw (m11) edge[no arrow, dashed] (m111);
\draw (m22) edge[no arrow, dashed] (m222);
\draw (m33) edge[no arrow, dashed] (m333);
\end{tikzpicture}}}
&

\hspace{-.6cm}\raisebox{-1.1cm}{
\resizebox{.23\textwidth}{!}{
\begin{tikzpicture}[->,>=stealth',shorten >=1pt,auto,semithick,
no arrow/.style={-,every loop/.append style={-}}]

\fill [blue!34] (-.20,-1.3) -- (-.20,1.65) to[out=60,in=-180] (-.1,1.7) -- (1.62,1.7) to[out=0,in=120] (1.72,1.65) -- (1.72,-1.3) -- cycle;
\fill [blue!26] (-.1,-1.2) -- (-.1,1.05) to[out=60,in=-180] (.0,1.1) -- (1.52,1.1) to[out=0,in=120] (1.62,1.05) -- (1.62,-1.2) -- cycle;
\fill [blue!18] (.0,-1.1) -- (.0,.63)  to[out=60,in=-180] (.1,.68) -- (1.42,.68) to[out=0,in=120] (1.52,.63) -- (1.52,-1.1) -- cycle;
\fill [blue!10] (.1,-1) -- (.1,.19)  to[out=60,in=-180] (.2,.24) -- (1.32,.24) to[out=0,in=120] (1.42,.19) -- (1.42,-1) -- cycle;
\fill [red!25] (.2,-.9) -- (.2,-.4)  to[out=60,in=-180] (.3,-.3) -- (1.22,-.3) to[out=0,in=120] (1.32,-.4) -- (1.32,-.9) -- cycle;
\fill [red!25] (-.15,1.1) -- (-.15,1.35)  to[out=60,in=-180] (-.13,1.4) -- (.1,1.4) to[out=0,in=120] (.13,1.35) -- (.13,1.1) -- cycle;

\node(n00) {};
\node[right=.1cm of n00](n0) {};
\node[above=1.5cm of n0](n5prime) {\fontsize{7}{0}\selectfont 5'};
\node[above=1.3cm of n0, inner sep=-.5pt](n1) {\fontsize{8}{0}\selectfont \textred A};
\node[above=1.cm of n00, inner sep=-.5pt](n2) {\fontsize{8}{0}\selectfont \textGray U};
\node[above=.7cm of n0, inner sep=-.5pt](n3) {\fontsize{8}{0}\selectfont \textviolet G};
\node[above=.2cm of n0, inner sep=-.5pt](n4) {\fontsize{8}{0}\selectfont \textred A};
\node[above=-.3cm of n0, inner sep=-.5pt](n5) {\fontsize{8}{0}\selectfont \textblue C};
\node[right=.001cm of n00](n66) {};
\node[above=-.8cm of n66, inner sep=-.5pt](n6) {\fontsize{8}{0}\selectfont \textblue C};
\node[right=.46cm of n00](n77) {};
\node[above=-1cm of n77, inner sep=-.5pt](n7) {\fontsize{8}{0}\selectfont \textGray U};
\node[right=.73cm of n6, inner sep=-.5pt](n8) {\fontsize{8}{0}\selectfont \textviolet G};
\node[right=.55cm of n5, inner sep=-.5pt](n9) {\fontsize{8}{0}\selectfont \textviolet G};
\node[right=.55cm of n4, inner sep=-.5pt](n10) {\fontsize{8}{0}\selectfont \textGray U};
\node[right=.55cm of n3, inner sep=-.5pt](n11) {\fontsize{8}{0}\selectfont \textred C};
\node[right=.55cm of n1, inner sep=-.5pt](n12) {\fontsize{8}{0}\selectfont \textGray U};
\node[right=.31cm of n5prime](n3prime) {\fontsize{7}{0}\selectfont 3'};

\draw (n1) edge[no arrow, line width=.6pt, blue] node{} (n2);
\draw (n2) edge[no arrow, line width=.6pt, blue] node{} (n3);
\draw (n3) edge[no arrow, line width=.6pt, blue] node{} (n4);
\draw (n4) edge[no arrow, line width=.6pt, blue] node{} (n5);
\draw (n5) edge[no arrow, line width=.6pt, blue] node{} (n6);
\draw (n6) edge[no arrow, line width=.6pt, blue] node{} (n7);
\draw (n7) edge[no arrow, line width=.6pt, blue] node{} (n8);
\draw (n8) edge[no arrow, line width=.6pt, blue] node{} (n9);
\draw (n9) edge[no arrow, line width=.6pt, blue] node{} (n10);
\draw (n10) edge[no arrow, line width=.6pt, blue] node{} (n11);
\draw (n11) edge[no arrow, line width=.6pt, blue] node{} (n12);

\draw (n1) edge[no arrow, line width=1.4pt, brown] node{} (n12);
\draw (n3) edge[no arrow, line width=1.4pt] node{} (n11);
\draw (n4) edge[no arrow, line width=1.4pt, brown] node{} (n10);
\draw (n5) edge[no arrow, line width=1.4pt] node{} (n9);

\node[above=1.4cm of n0] (n5prime1) {};
\node[left=.7cm of n5prime1] (n5prime1bm) {};
\node[left=.5cm of n5prime1, inner sep=-.5pt] (n5prime1cm) {};
\node[above=.3cm of n0] (n34) {};
\node[left=.7cm of n34] (n34bm) {};
\node[left=.5cm of n34, inner sep=-.5pt] (n34cm) {};

\node[below=.3cm of n0] (n67) {};
\node[right=-.15cm of n67] (n67right) {};
\node[below=.45cm of n67right, inner sep=-5pt] (n67down) {};
\node[left=.2cm of n67down, inner sep=-.5pt] (n67bm) {};
\node[below=.3cm of n67right] (n67mid) {};
\node[left=.13cm of n67mid, inner sep=-.5pt] (n67cm) {};
\node[below=1.85cm of n34cm, inner sep=-.5pt] (n67dm) {};

\node[above=.0cm of n9] (n910) {};
\node[right=.7cm of n910] (n910bm) {};
\node[right=.5cm of n910, inner sep=-.5pt] (n910cm) {};

\node[right=.44cm of n5prime1] (n3prime1) {};
\node[right=.7cm of n3prime1] (n3prime1bm) {};
\node[right=.5cm of n3prime1, inner sep=-.5pt] (n3prime1cm) {}; 

\draw (n5prime1) edge[no arrow, line width=.2pt] node{} (n5prime1bm);
\draw (n34) edge[no arrow, line width=.2pt] node{} (n34bm);
\draw[no arrow, line width=.2pt] (n67right) -- (.17,-1.3) -- (-.45,-1.3);
\draw (n910) edge[no arrow, line width=.2pt] node{} (n910bm);
\draw (n3prime1) edge[no arrow, line width=.2pt] node{} (n3prime1bm);

\draw (n5prime1cm) edge[above, <->, dashed, bend right=20, line width=.2pt] node[below, rotate=-90, yshift=2pt]{\fontsize{6}{0}\selectfont methionine} (n34cm);
\draw (n34cm) edge[above, <->, dashed, bend right=0, line width=.2pt] node[below, rotate=-90, yshift=2pt]{\fontsize{6}{0}\selectfont threonine} (n67dm);
\draw (n67cm) edge[below, <->, dashed, bend right=0, line width=.2pt, bend right=55] node[rotate=40, yshift=2pt]{\fontsize{6}{0}\selectfont tryptophan} (n910cm);
\draw (n910cm) edge[below, <->, dashed, bend right=0, line width=.2pt] node[rotate=90, yshift=2pt]{\fontsize{6}{0}\selectfont serine} (n3prime1cm);
\end{tikzpicture}}}

\end{tabular}\\[.2cm]

\pgfplotsset{
    select row/.style={
        x filter/.code={\ifnum\coordindex=#1\else\def\pgfmathresult{}\fi}
    }
}

\pgfplotstableread[header=false]{
0.7 \textcolor{red}A\textcolor{blue}C\textcolor{gray}U
0.3 \textcolor{red}A\textcolor{blue}C\textcolor{purple}G
0.8 \textcolor{red}A\textcolor{blue}C\textcolor{red}A
1.0 \textcolor{red}A\textcolor{blue}C\textcolor{blue}C
}\datatablethreonine

\pgfplotstableread[header=false]{
0.6 \textcolor{red}A\textcolor{purple}G\textcolor{gray}U
1.0 \textcolor{red}A\textcolor{purple}G\textcolor{blue}C
0.3 \textcolor{gray}U\textcolor{blue}C\textcolor{purple}G
0.9 \textcolor{gray}U\textcolor{blue}C\textcolor{blue}C
0.6 \textcolor{gray}U\textcolor{blue}C\textcolor{red}A
0.8 \textcolor{gray}U\textcolor{blue}C\textcolor{gray}U
}\datatableserine

\begin{tabular}{cccc}
\hspace{-.5cm}\raisebox{0cm}{{\panel{\myedit{D}}} {\fontsize{9}{0}\selectfont \textbf{\ \myedit{codons, frequencies, relative adaptivenesses, and weighted mRNA DFA with CAI integration}}}} &  \\[-.05cm]


\hspace{3.1cm}\raisebox{0.1cm}{\resizebox{.22\textwidth}{!}{
\begin{tikzpicture}[scale=.5]

  \begin{axis}[
        xmin=0,xmax=1.2,
        xbar, bar shift=0pt,
        ytick={0,...,3},
        yticklabels from table={\datatablethreonine}{1},
        xtick = {-1},
        bar width=6mm, 
        width=5.5cm, height=5.3cm, 
  ]

\pgfplotsinvokeforeach{0}{
\addplot+[color=gray] table [select row=#1, y expr=#1] {\datatablethreonine};
}
\pgfplotsinvokeforeach{1}{
\addplot+[color=purple] table [select row=#1, y expr=#1] {\datatablethreonine};
}
\pgfplotsinvokeforeach{2}{
\addplot+[color=red] table [select row=#1, y expr=#1] {\datatablethreonine};
}
\pgfplotsinvokeforeach{3}{
\addplot+[blue, pattern=north east lines, pattern color=blue] table [select row=#1, y expr=#1] {\datatablethreonine};
}

\end{axis}




\draw (5,2) edge[above] node[][black, pos=1.2]{\tiny $-\log w(c)$} (6,2);
\draw (6,2) edge[above, ->,>=stealth',shorten >=1pt,auto,semithick] node[][black, pos=1.2]{} (6,0.5);

%
\end{tikzpicture}
}}

\hspace{-1.75cm}\raisebox{.87cm}{\resizebox{.05\textwidth}{!}{
\begin{tabular}{|c|}
\textcolor{black}{$w(c)$} \\[0.5em]
\textcolor{blue}{1}    \\[0.8em]

\textcolor{red}{0.8}  \\[0.8em]

\textcolor{purple}{0.3}  \\[0.8em]

\textcolor{gray}{0.7} \\[0.8em]

\end{tabular}

}}

\hspace{4.35cm}\raisebox{0.02cm}{\resizebox{.22\textwidth}{!}{
\begin{tikzpicture}[scale=.5][->,>=stealth',shorten >=1pt,auto,semithick]

  \begin{axis}[
        xmin=0,xmax=1.2,
        xbar, bar shift=0pt,
        xmin=0,
        ytick={0,...,5},
        yticklabels from table={\datatableserine}{1},
	xtick = {-1},
        bar width=4.5mm, 
        width=5.5cm, height=5.5cm, 
  ]

\pgfplotsinvokeforeach{0}{
\addplot+[color=gray] table [select row=#1, y expr=#1] {\datatableserine};
}

\pgfplotsinvokeforeach{1}{
\addplot+[blue, pattern=north east lines, pattern color=blue] table [select row=#1, y expr=#1] {\datatableserine};
}

\pgfplotsinvokeforeach{2}{
\addplot+[color=purple] table [select row=#1, y expr=#1] {\datatableserine};
}

\pgfplotsinvokeforeach{3}{
\addplot+[color=blue] table [select row=#1, y expr=#1] {\datatableserine};
}

\pgfplotsinvokeforeach{4}{
\addplot+[color=red] table [select row=#1, y expr=#1] {\datatableserine};
}

\pgfplotsinvokeforeach{5}{
\addplot+[color=gray] table [select row=#1, y expr=#1] {\datatableserine};
}

\end{axis}



\draw (5,2) edge[above] node[][black, pos=1.2]{\tiny $-\log w(c)$} (6,2);
\draw (6,2) edge[above, ->,>=stealth',shorten >=1pt,auto,semithick] node[][black, pos=1.2]{} (6,0.5);

\end{tikzpicture}

}}

\hspace{-1.75cm}\raisebox{.9cm}{\resizebox{.05\textwidth}{!}{
\begin{tabular}{|c|}
\smallskip
\smallskip
\textcolor{black}{$w(c)$} \\[0.05em]
\textcolor{gray}{0.8}    \\[0.05em]
\textcolor{red}{0.6}  \\[0.05em]
\textcolor{blue}{0.9}  \\[0.05em]
\textcolor{purple}{0.3} \\[0.05em]
\textcolor{blue}{1}    \\[0.05em]
\textcolor{gray}{0.6}    \\[0.05em]

\end{tabular}

}}\\[-0.45cm]

\hspace{-.5cm}\raisebox{3.5cm}{\resizebox{1\textwidth}{!}{
\begin{tikzpicture}[->,>=stealth',shorten >=1pt,auto,semithick]
\node[state, initial, initial text=,  inner sep=-5pt, minimum size=.2cm, fill=StartNode] (n0) {};
\node[below=1.2cm of n0] (n0bm) {$\mid$}; 

\node[state, right=1cm of n0,  inner sep=-5pt, minimum size=.2cm, fill=NodeGray] (n1) {};
\node[state, right=1cm of n1,  inner sep=-5pt, minimum size=.2cm, fill=NodeGray] (n2) {};

\node[state, right=1cm of n2,  inner sep=-5pt, minimum size=.2cm, fill=MidNode] (n00) {};
\node[below=1.2cm of n00] (n3bm) {$\mid$}; 
\node[below=1.6cm of n00] (n3cm) {$\circ$}; 

\node[state, right=1cm of n00,  inner sep=-5pt, minimum size=.2cm, fill=NodeGray] (n10) {};
\node[state, right=1cm of n10,  inner sep=-5pt, minimum size=.2cm, fill=NodeGray] (n20) {};
\node[state, right=1cm of n20,  inner sep=-5pt, minimum size=.2cm, fill=MidNode] (n30) {};

\node[below=1.2cm of n30] (n3b) {$\mid$}; 
\node[below=1.6cm of n30] (n3c) {$\circ$}; 

\node[state, right=1cm of n30,  inner sep=-5pt, minimum size=.2cm, fill=NodeGray] (n40) {};
\node[state, right=1cm of n40,  inner sep=-5pt, minimum size=.2cm, fill=NodeGray] (n50) {};
\node[state, right=1cm of n50,  inner sep=-5pt, minimum size=.2cm, fill=MidNode] (n60) {};

\node[below=1.2cm of n60] (n6b) {$\mid$}; 
\node[below=1.6cm of n60] (n36) {$\circ$}; 

\node[state, right=1cm of n60,  inner sep=-5pt, minimum size=.2cm, fill=NodeGray] (n70) {};
\node[state, right=1cm of n70,  inner sep=-5pt, minimum size=.2cm, fill=NodeGray] (n80) {};
\node[state, below=.8cm of n70,  inner sep=-5pt, minimum size=.2cm, fill=NodeGray] (n71) {};
\node[state, below=.8cm of n80,  inner sep=-5pt, minimum size=.2cm, fill=NodeGray] (n81) {};
\node[state, accepting, right=.9cm of n80,  inner sep=-5pt, minimum size=.2cm, fill=EndNode] (n90) {};

\node[below=1.2cm of n90] (n9b) {$\mid$}; 

\draw (n0bm) edge[below,<->,dashed] node{weighted $D(\text{methionine})$} (n3bm);
\draw (n3bm) edge[below,<->,dashed] node{weighted $D(\text{threonine})$} (n3b);
\draw (n3b) edge[below,<->,dashed] node{weighted $D(\text{tryptophan})$} (n6b);
\draw (n6b) edge[below,<->,dashed] node{weighted $D(\text{serine})$} (n9b);

\draw (n0) edge[above] node[gray]{\fontsize{10}{0}\selectfont A:0} (n1)
(n1) edge[above] node[gray]{\fontsize{10}{0}\selectfont U:0} (n2)
(n2) edge[above] node[violet]{\fontsize{10}{0}\selectfont G:0} (n00);

\draw (n00) edge[above] node[red]{\fontsize{10}{0}\selectfont A:0} (n10)
(n10) edge[above] node[blue]{\fontsize{10}{0}\selectfont C:0} (n20)
(n20) edge[above, bend left=60] node[blue]{\fontsize{9}{0}\selectfont C:0} (n30)
(n20) edge[above, bend left=20] node[yshift=-3pt][red]{\fontsize{9}{0}\selectfont A:0.3} (n30)
(n20) edge[below] node[yshift=2pt][violet]{\fontsize{9}{0}\selectfont G:1.1} (n30)
(n20) edge[below, bend right=45] node[gray]{\fontsize{9}{0}\selectfont U:0.4} (n30);

\draw (n30) edge[above] node[gray]{\fontsize{10}{0}\selectfont U:0} (n40)
(n40) edge[above] node[violet]{\fontsize{10}{0}\selectfont G:0} (n50)
(n50) edge[above] node[violet]{\fontsize{10}{0}\selectfont G:0} (n60);

\draw
(n60) edge[above] node[gray]{\fontsize{10}{0}\selectfont U:0} (n70)
(n60) edge[above] node[red]{\fontsize{10}{0}\selectfont \ \  A:0} (n71)

(n70) edge[above] node[blue]{\fontsize{10}{0}\selectfont C:0} (n80)
(n71) edge[above] node[violet]{\fontsize{10}{0}\selectfont G:0} (n81)

(n80) edge[above, bend left=60] node[gray]{\fontsize{9}{0}\selectfont U:0.3} (n90)
(n80) edge[above, bend left=20] node[yshift=-3pt][red]{\fontsize{9}{0}\selectfont A:0.5} (n90)
(n80) edge[below] node[yshift=3pt][blue]{\fontsize{9}{0}\selectfont C:0.1} (n90)
(n80) edge[below, bend right=41] node[yshift=2pt][violet]{\fontsize{9}{0}\selectfont G:1.4} (n90)

(n81) edge[above, bend right=40] node[sloped, anchor=center, above, xshift=-3pt ,yshift=-2pt][blue]{\fontsize{9}{0}\selectfont C:0} (n90)
(n81) edge[below, bend right=80] node[sloped, anchor=center, above, xshift=-3pt, yshift=-2pt][gray]{\fontsize{9}{0}\selectfont U:0.5} (n90);
\end{tikzpicture}}}\\[-3.5cm]
\\[-0.8cm]
\end{tabular}

\caption{
\myedit{Lattice parsing solves the mRNA design problem, either optimizing stability alone (objective 1; see {\bf A--C})
or jointly optimizing both stability and codon optimality (objectives 1 \& 2; see {\bf D}).
{\bf A}: Codon DFAs.
{\bf B}: An mRNA DFA (bottom)
and lattice parsing on that DFA (top).
In the 
 DFA, the optimal mRNA sequence under the simplified energy model 
(Fig.~\ref{fig:CFG_DFA}E)
is shown in the thick blue path,
together with its optimal structure shown in the dot-bracket format
(``$\bullet$'': unpaired; ``\bml'' and ``\bmr'': base pairs).
In lattice parsing,
the brown and black arcs also depict base pairs (two GC pairs and two AU pairs),
while the round trapezoid shadings 
depict the decomposition of the optimal structure.
Among all mRNA sequences encoded in the DFA, 
lattice parsing finds 
the optimal sequence with its optimal structure,
achieving the lowest free energy under this energy model 
(where GC and AU pairs have -3 and -2~kcal/mol, respectively). 
{\bf C}: 
Another illustration of the optimal sequence and secondary structure in {\bf B}.
Next, we show in {\bf D} 
how to do the joint optimization by integrating codon optimality in weighted DFAs.
{\bf D}: Top: bar charts showing the codon frequencies of threonine and serine. 
The relative adaptiveness $w(c)$ of a codon $c$ 
is the ratio between the frequencies of $c$ and its most frequent codon (shown in stripes).
Bottom: a weighted mRNA DFA 
encodes each candidate's CAI 
in the total weight of its corresponding path,
by taking $-\log w(c)$ as edge weights to represent the cost  
of choosing codon~$c$ 
(see Methods~\ref{methods:obj}).
This weighted DFA can be plugged back into lattice parsing for joint optimization between stability and codon optimality.} 
\label{fig:main_CFG_DFA} 
}
\end{figure}

%% file: formulation.tex


Previous work \cite{mauger+:2019}
established two main objectives for mRNA design, stability and codon optimality,
which synergize to increase protein expression.
To optimize for stability, 
given a protein sequence, 
we aim to find  
the 
mRNA sequence 
that has the {\em lowest
minimum folding free energy change} (\MFE)
among all possible mRNA sequences 
encoding
that protein (Fig.~\ref{fig:CFG_DFA}A),
which is a {\em minimization within a minimization}.
Naively, 
for each candidate mRNA sequence, we find its MFE structure among 
all of its possible secondary structures, 
and then choose the sequence with the lowest MFE. 
But that would take billions of years, 
so we need 
an efficient algorithm that solves the problem without enumeration. %

{Next, 
we also aim to jointly optimize mRNA stability 
and codon optimality (Fig.~\ref{fig:CFG_DFA}B).
The latter is often measured by the Codon Adaptation Index (\CAI) \cite{sharp+li:1987} defined as the geometric mean of the relative adaptiveness of each codon in the mRNA.} 
Because CAI is between 0 and 1 but MFE is generally proportional to the  sequence length, 
we multiply the logarithm of CAI by the number of codons in the mRNA, 
and use a hyperparameter $\lambda$ to balance \MFE and \CAI ($\lambda = 0$ being \MFE-only).
See Methods~\ref{methods:obj} for details.

We next describe our solution to these two optimization problems with two ideas borrowed from natural language:
DFA (lattice) representation and lattice parsing.

%% file: alg.tex
\paragraph{{Design Space Representation: DFA (Lattice)}}

Inspired by the ``word lattice'' representation of ambiguities in computational linguistics,
we represent the choice of codons for each amino acid using a similar lattice,
or more formally, a DFA, which is basically a directed graph with labeled edges (Figs.~\ref{fig:main_CFG_DFA}A \&~\ref{fig:CFG_DFA}C; see Methods~\ref{methods:DFA} for formal definitions). 
After building a codon DFA for each amino acid, we then concatenate them
into a single {\em mRNA DFA}, 
where each path from the start state 
to the final state 
represents a possible mRNA sequence
that encodes that protein (see Figs.~\ref{fig:main_CFG_DFA}B \&~\ref{fig:CFG_DFA}D).

\paragraph{{Objective 1 (Stability): Lattice Parsing}} 

RNA folding is well-known to be equivalent to 
natural language parsing, 
with a stochastic context-free grammar (SCFG) representing the folding energy model~\cite{Durbin+:1998} (see Fig.~\ref{fig:CFG_DFA}E--F).
But for mRNA design, the big question is,
how to fold all the mRNA sequences in the DFA  together?
{We borrow the idea of ``lattice parsing''~\cite{bar-hillel+:1961,hall:2005}, which 
generalizes 
single-sequence parsing
to handle
all sentences in the lattice simultaneously to find the most likely one (Figs.~\ref{fig:overview}C \&~\ref{fig:word_lattice}).
Similarly, we use lattice parsing to
fold all sequences in the mRNA DFA simultaneously to find the most stable one (Figs.~\ref{fig:main_CFG_DFA}B \& \ref{fig:CFG_DFA}G--H).
Note that single-sequence folding is a special case with a single-chain DFA.
This process can also be interpreted as SCFG-DFA intersection (Fig.~\ref{fig:CFG_DFA}A)
where the SCFG scores for stability and the DFA demarcates the set of candidates.
This algorithm runs in the same worst-case $O(n^3)$ time as single-sequence folding (but with a larger constant) 
where $n$ is the mRNA length (Methods~\ref{methods:CFG-DFA-DP}),
but for practical applications it only scales $O(n^2)$ (Fig.~\ref{fig:insilico}A).}

\paragraph{{Adding Objective 2 (Codon Optimality): Lattice Parsing with Weighted DFAs}} 

{We now extend DFAs to {\em weighted DFAs} (WDFAs) to integrate codon optimality on edge weights.
Since our joint optimization formulation factors \CAI onto 
the relative adaptiveness $w(c)$ of 
each individual codon $c$, 
we set edge weights in each codon DFA
so that a codon~$c$ has path cost $-\!\log w(c)$.
Then in a weighted mRNA DFA,
the cost of each start-end path
is the sum of $-\!\log w(c)$
for each codon $c$ in the corresponding mRNA,
which is proportional to its~$-\!\log\CAI$
(Fig.~\ref{fig:main_CFG_DFA}D).
Now lattice parsing takes a stochastic grammar (for stability)
and a weighted DFA (for codon optimality) and solves the joint optimization,
which can be viewed as the weighted intersection between SCFG and WDFA (Fig.~\ref{fig:CFG_DFA}B; Methods~\ref{methods:cai}).}
{By contrast, two previous efforts at stabilizing mRNAs~\cite{cohen+skiena:2003,Terai+:2016} using ad-hoc algorithms can not jointly optimize codon usage; see Methods~\ref{methods:prevwork} for details.}


{
\paragraph{Expressiveness of DFAs}
Our DFA framework
can also represent 
alternative genetic codes,
modified nucleotides, and coding constraints; see Figs.~\ref{fig:si_alter_aa}--\ref{fig:si_alter_aa_enzyme} and Methods~\ref{methods:generality_dfa}.}

\paragraph{Linear-time Approximation}

The exact design algorithm might still be slow for long sequences.
On the other hand, suboptimal designs may also be worth exploring for wet lab experiments (see 
below), 
due to the many other factors involved in mRNA design besides stability and codon usage.
So we developed an approximate search version that runs in linear time 
using beam search, keeping only the top $b$ most promising items per step ($b$ is the beam size), inspired by our previous work LinearFold~\cite{huang+:2019}.

%% file: result_runtime.tex



\input{fig_insilico} 

Fig.~\ref{fig:insilico}A benchmarked the runtime of \lineardesign on UniProt proteins \cite{UniProt:2005}. 
\lineardesign was shown in a combination of two optimization objectives, 
\MFE-only (objective 1) vs.~\MFEplusCAI (objectives 1~\& 2),
and two search modes, exact search 
vs.~beam search (beam size $b=500$).
Empirically, \lineardesign
scales 
quadratically with mRNA sequence length $n$ for practical applications ($n < 10,000$~\nts)
thanks to the DFA representation and lattice parsing (see Fig.~\ref{fig:LD_CDSfold_items_generated_table} for analysis). 
Next, our CAI-integrated exact search (\myedit{CAI weight $\lambda=4$}) had the same empirical complexity, and was only \myedit{$\sim\!15\%$} slower than the MFE-only version thanks to the convenience of adding CAI in our DFA representation.
Last, our beam search version ($b=500$) further speeds up our design and scales {\em linearly} with sequence length, taking only 2.7 minutes (vs.~10.7 minutes for exact search) 
on the \sarscovtwo \spike protein (for MFE-only), with an approximation error 
(i.e., 
energy gap \%, defined as $1- \mfe_\text{approx\_design} / \mfe_\text{exact\_design}$)
of just 1.2\%.
In fact, as sequences get longer, this percentage stabilizes, 
suggesting that beam search quality does not degrade with sequence length;
see Fig.~\ref{fig:breakdown_gap} for details. 

\myedit{As shown in Figs.~\ref{fig:insilico}B--C, 
for a GC-favoring codon preference (such as human),
the conventional  codon optimization method
does improve stability, but only slightly,
since its optimization directions (the pink arrows) are largely orthogonal to the stability optimization directions (the blue arrows).
By contrast, our \lineardesign can directly optimize stability and
find the optimally stable mRNAs ($\lambda=0$)
on both the Spike protein
and the enhanced gene fluorescent protein (eGFP), 
which have the lowest MFEs
that are $1.8\times$ lower than the optimal-CAI's ($\lambda=\infty$).
Also, our optimally stable designs have mostly double-stranded secondary structures (Fig.~\ref{fig:insilico}D),
which are predicted to be much less prone to degradation \cite{mauger+:2019}.
\lineardesign also finds the sequence with the lowest MFE for a given CAI and vice versa, thus forming the feasibility limit of mRNA design (the blue curves in Fig.~\ref{fig:insilico}B--C, with $\lambda$ ranging from 0 to $\infty$).}
\myedit{Furthermore, when the codon bias prefers AU-rich codons (such as yeast), 
codon optimization actually {\em worsens} stability 
(Fig.~\ref{fig:human_yeast_codon_oval_full}).}

%% file: fig_insilico.tex

\begin{figure}[!t]
\begin{tabular}{ccc}


 \raisebox{4.5cm}{\hspace{0.cm}\includegraphics[width=.44\linewidth]{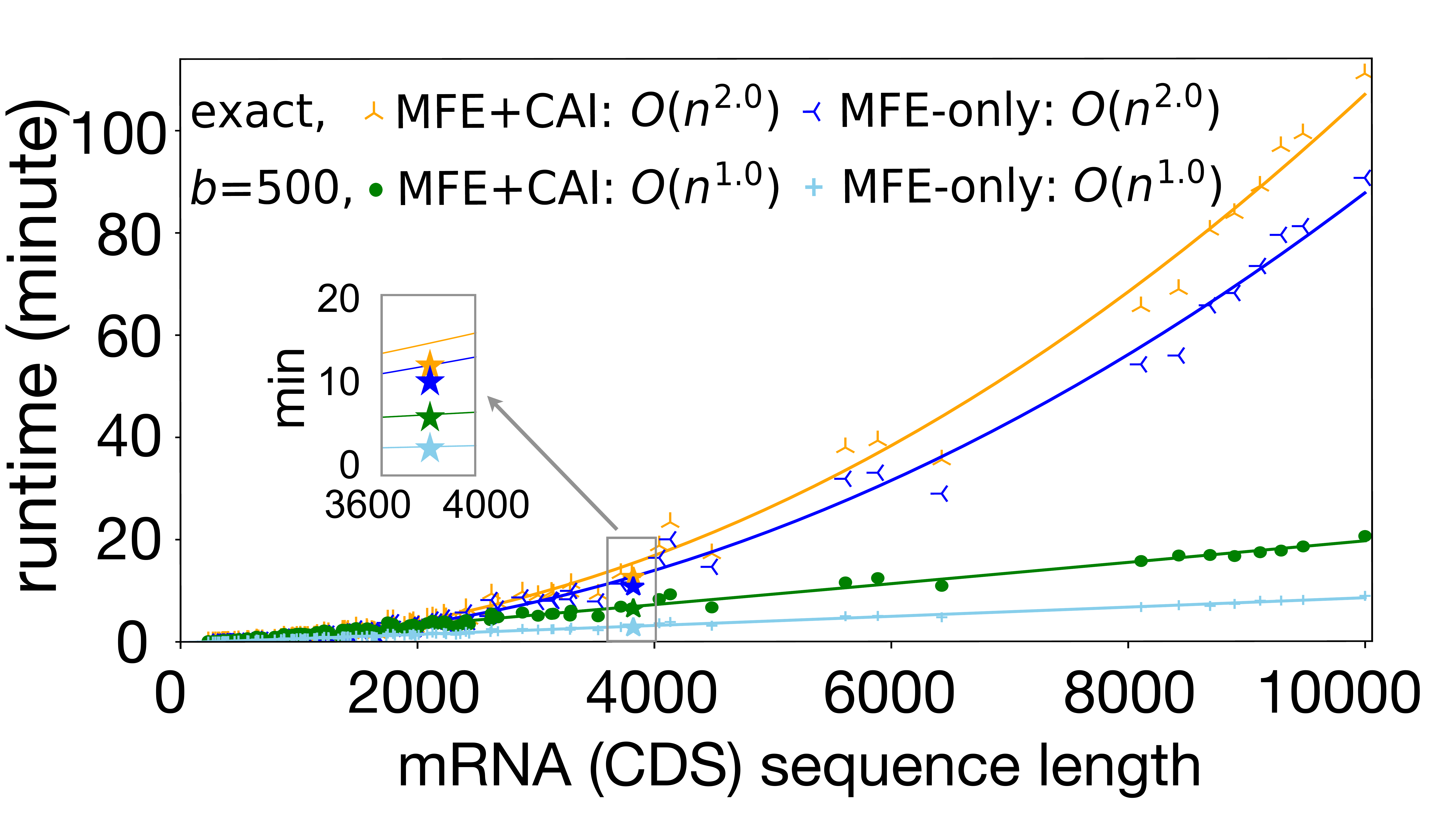}}&\raisebox{-3.2cm}{\hspace{-0.2cm}\includegraphics[width=.52\linewidth]{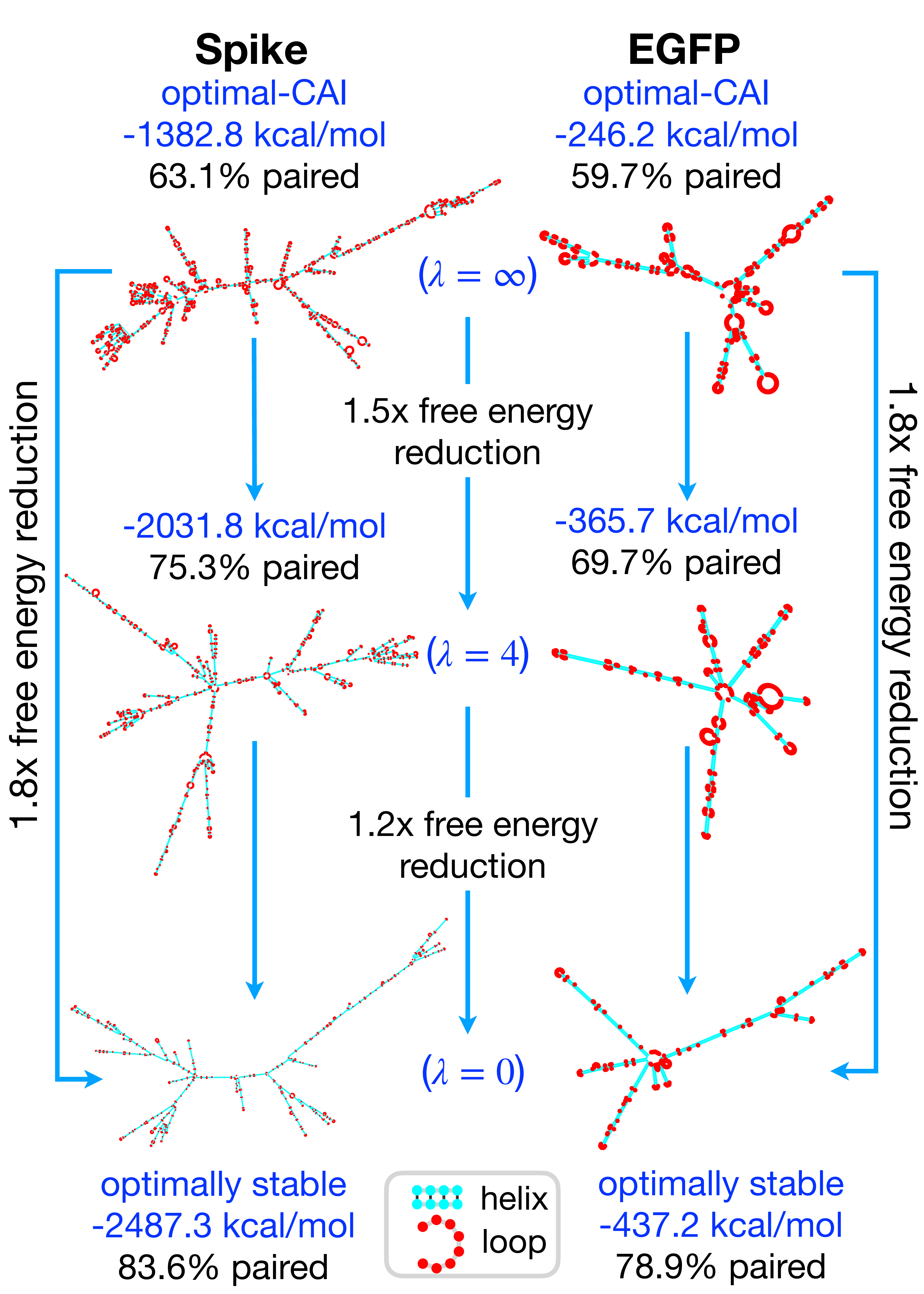}}\\[-7.8cm]
 \raisebox{0.cm}{\hspace{0.cm}\includegraphics[width=.44\linewidth]{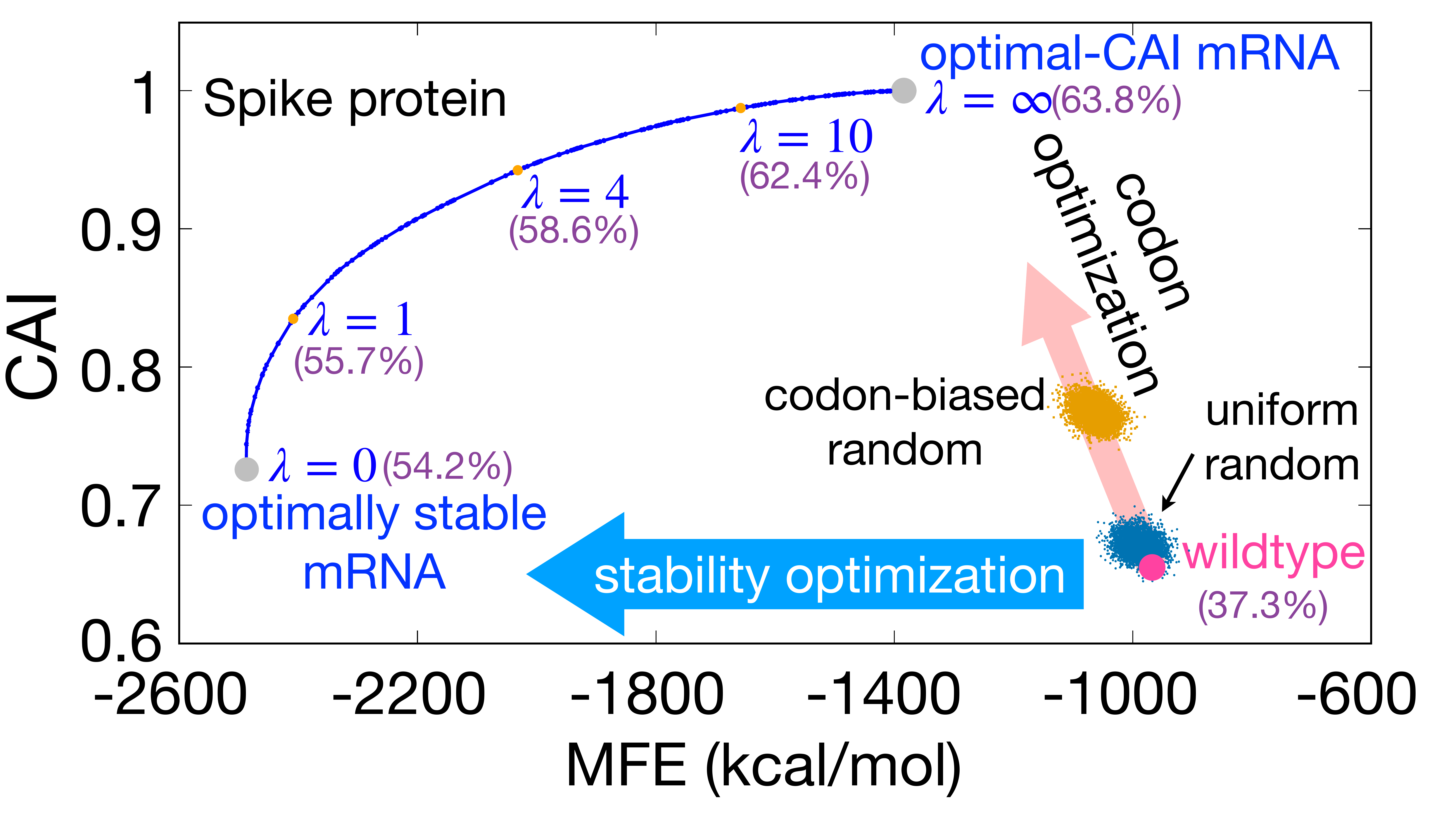}}&\\[-0.2cm]
  \raisebox{0.cm}{\hspace{0.cm}\includegraphics[width=.44\linewidth]{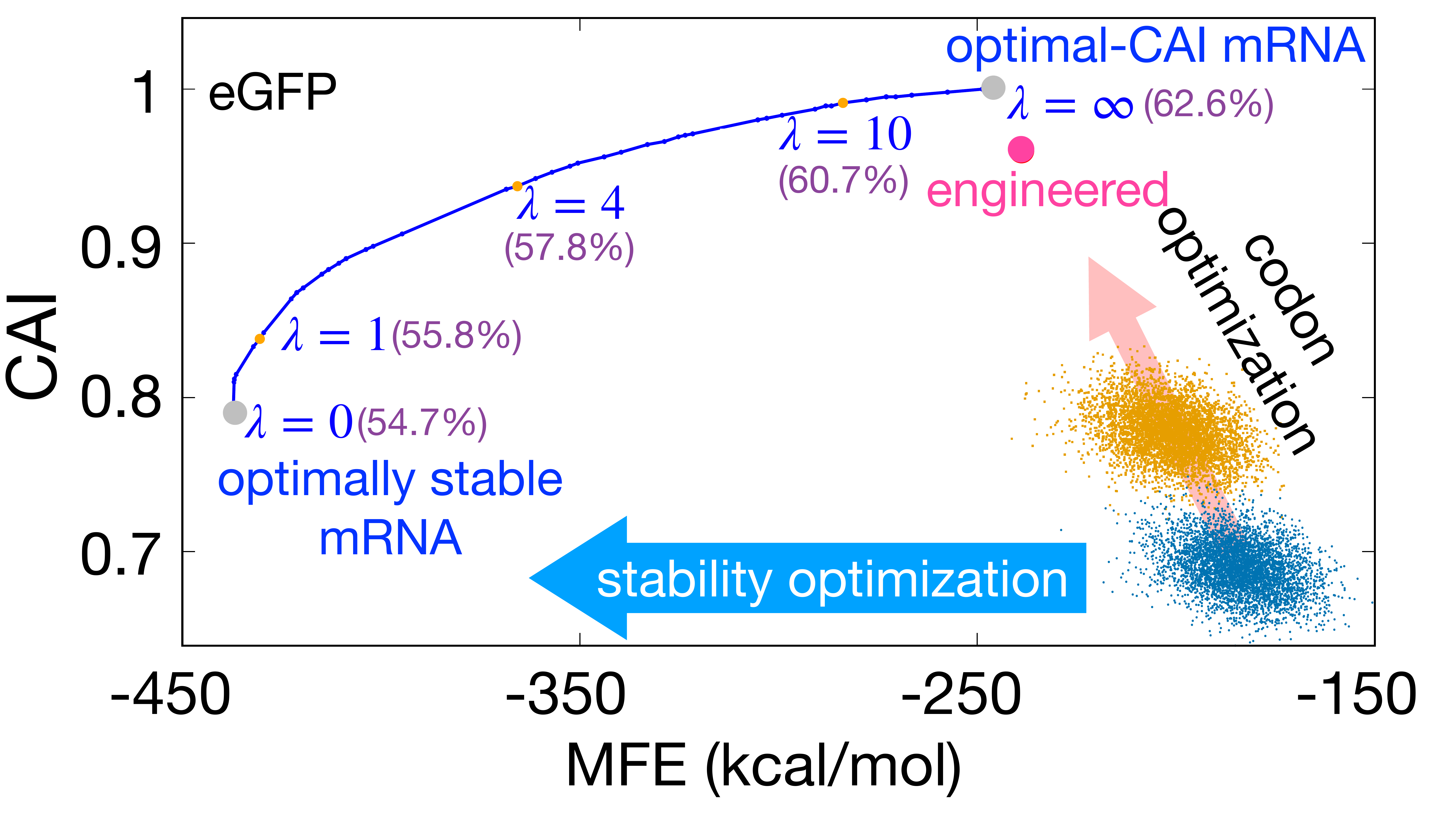}}&\\[-12.6cm]

  \hspace{-6.6cm}\raisebox{-.6cm}{\panel{A}} &\raisebox{-.6cm}{\hspace{-8.3cm}\panel{D}} \\[3.3cm]
  
   \hspace{-6.6cm}\raisebox{-.6cm}{\panel{B}} &\raisebox{-.3cm}{\hspace{-7.3cm}\panel{}} \\[3.4cm]
   
   \hspace{-6.6cm}\raisebox{-.6cm}{\panel{C}} &\raisebox{-.3cm}{\hspace{-7.3cm}\panel{}} \\[3.8cm]

%
%
%
%

\end{tabular}
\caption{\textit{In silico} analysis of \lineardesign.
{\bf A}: \myedit{Runtime visualization of 
\lineardesign on UniProt proteins. 
Overall, our exact search only scales  quadratically with sequence length 
in practice 
(see Fig~\ref{fig:LD_CDSfold_items_generated_table} for analysis),
and our MFE+CAI mode 
(with \textit{$\lambda$} = 4)
is only $\sim$15\% slower than our MFE-only version.
Moreover, beam search ($b=500$) significantly speeds up long sequences, with minor search errors (see also Fig.~\ref{fig:breakdown_gap}).}
\myedit{{\bf B--C}: 2D (\MFE--\CAI) visualizations of Spike protein ({\bf B}) and eGFP ({\bf C}) designs, respectively (both using human codon preference).}
\myedit{The blue curves (\myedit{varying $\lambda$ from 0 to $\infty$}) achieve
the best MFE for a given CAI, and vice versa, thus forming the feasibility limit.}
\myedit{GC\% are shown in parentheses for LinearDesign-generated sequences.
The human genome prefers GC-rich codons, 
therefore codon optimization (the pink arrows in {\bf B--C}) also improves stability, but only marginally,
as the two optimization directions (codon vs.~stability) are largely orthogonal.
By contrast, Fig.~\ref{fig:human_yeast_codon_oval_full}B shows that 
with an AU-rich codon preference, codon optimization decreases stability.
{\bf D}: Secondary structures of example designs. The optimal-CAI designs (top, $\lambda\!=\!\infty$) are largely single-stranded ($\sim\! 60\%$ paired), while the optimally stable designs (bottom, $\lambda\!=\!0$) are mostly double-stranded ($\sim\! 80\%$ paired).
We also show intermediate designs (e.g., $\lambda\!=\!4$) that compromise between stability and CAI.
See also Fig.~\ref{fig:human_yeast_codon_oval_full}A--B for the case of negative $\lambda$'s.}
\label{fig:insilico}
}
\end{figure}

%% file: result_wetlab.tex

We performed experimental assays   
to study the chemical stability, protein expression, and immunogenicity of \lineardesign-generated mRNA molecules 
(Fig.~\ref{fig:wetlab}).
All mRNA sequences in this study encode the \sarscovtwo \spike protein without the 2P mutation \cite{wrapp+:2020}, with unmodified (natural) nucleotides and the same UTRs. 
We used \lineardesign (with beam search plus a \kbest algorithm \cite{huang+chiang:2005} for suboptimal candidates) to generate seven mRNA sequences 
({\bluediamond \sc a--g} in Fig.~\ref{fig:wetlab}A),
which were compared to the benchmark sequence (baseline {\sc h\blueDiamond}) 
designed 
by a commonly-used codon optimization algorithm, \optimumgene.
The seven \lineardesign sequences were selected to be widely distributed in the unexplored high-stability territory 
(the region in Fig.~\ref{fig:wetlab}A where $\MFE<-1,400$~kcal/mol, inaccessible to codon optimization),
and to separate the impacts of \MFE and \CAI, 
we designed sequences that are almost identical in either \MFE ({\sc b--c} /\!/ {\sc d--e--f}) or \CAI ({\sc a--c--f} /\!/ {\sc b--e} /\!/ {\sc d--g--h}).
It is well-known that translation efficiency drops if the 5'-leader region
is particularly structured~\cite{mauger+:2019}, 
so we did not optimize for the first 5 amino acids
and used a heuristic to select the first 15 nucleotides. 
It is also suggested that long stems may induce unwanted innate immune responses~\cite{liu+:2008}, 
so we avoided them in our designs, 
which is why we did not study the lowest-MFE candidates closest to the optimal boundary (the blue curve)
which usually contain long stems. 
See Methods~\ref{methods:design-constraints} for details. 
\myedit{
It is worth mentioning that some UTR structures are crucial for translation~\cite{mignone2002untranslated},
and we observe that our stable designs ({\sc a--f}), having more structured coding regions, form fewer base pairs with, 
and thus interfere less with the structures of, 
commonly used UTRs
than codon-optimized ones (Tab.~\ref{tab:si_utrs}).
This suggests that LinearDesign is likely to be effective independent of the choice of UTRs.}

\paragraph{\Insolution Structure Compactness and Chemical Stability} 

\input{fig_wetlab}

We then studied the structure compactness of mRNA molecules, 
\myedit{which is hypothesized} to be correlated with the folding free energy change. 
We performed Electrophoretic Mobility Shift Assay (EMSA) using Agarose gel electrophoresis at 30\celsius.
An mRNA molecule with a lower MFE contains more secondary structures and thus exhibits more compact shape
and smaller hydrodynamic size, 
which makes it move 
faster in the agarose gel matrix. 
We can see that the gel mobility pattern in Fig.~\ref{fig:wetlab}B correlated with the MFEs of the eight designs almost perfectly:
\myedit{sequence {\sc a}, with the lowest MFE, moved the fastest, followed by sequences {\sc b} and {\sc c}, 
then by sequences {\sc d}, {\sc e}, and {\sc f}, 
then by sequence {\sc g}, 
and finally by the baseline sequence {\sc h}, which has the highest MFE.}


To evaluate the chemical stability, 
we incubated the designed mRNA molecules in RNA storage buffer (Thermo Fisher) at 37\celsius followed by quantification of intact mRNA as a function of time.
Overall, the chemical stability results also correlated well with the MFEs.
The lowest-MFE design ({\sc a}) degraded the slowest, 
and the highest-MFE baseline ({\sc h}) the fastest, followed by the design ({\sc g}) with the second highest MFE.
For example, after day 4, 
there was only 2.8\% intact molecules of {\sc h} and 18.1\% of {\sc g},
while design {\sc a} still had 36.9\%  and four other designs ({\sc b--d} and {\sc f}) 
also had around 30\%.
After day 5, sequence {\sc h} had 0\% remaining, sequence {\sc g} had 8.8\%,
but the lower-MFE designs ({\sc a--f}) still had 14.7\%--19.2\% left (Fig.~\ref{fig:wetlab}C).
These results suggest that low-MFE designs have longer half-lives,
which will contribute to higher protein expression levels (see below).
%

\paragraph{\frankedit{Cellular} Protein Expression}
Sufficient antigen expression is one of the prerequisites for the induction of antibody responses so we evaluated
the designed mRNA molecules 
for protein expression. 
The \sarscovtwo \spike protein is a transmembrane protein that can be directly detected on viable mRNA-transfected cells by flow cytometry. 
Following transfection into HEK-293 cells,
we measured the amount of \spike proteins generated by all designs after 24 hours.
Of note, 5 out of 7 \frankedit{mRNA molecules (designs {\sc a--d} and {\sc g}) showed substantially higher protein expression levels than the benchmark~{\sc h} (Fig.~\ref{fig:wetlab}).}
For example, 
designs {\sc d} and {\sc g} (with CAIs almost identical to {\sc h}, but lower MFEs) were 1.5$\times$ and 1.4$\times$ better, 
respectively, and the lowest MFE design {\sc a} was also 1.5$\times$ better. 
In general, our results are consistent with Mauger et al.'s \cite{mauger+:2019} 
that low MFE and high CAI synergize to improve 
protein expression,
but we were able to test this hypothesis \frankedit{using mRNA molecules 
with much lower MFEs than they could, 
thanks to \lineardesign's ability to explore 
the previously 
unreachable design space.}


\paragraph{\Invivo Immunogenicity}
Ultimately, we tested whether our designs could endow mRNA vaccines with higher immunogenicity,
which is usually represented by the magnitude of elicited antibody and T cell responses.
We employed a 
\frankedit{lipid based delivery system for \lineardesign-generated mRNA vaccines~\cite{rana+:2021}},
and further evaluated and compared the immunogenicity of each design side-by-side.  
C57BL/6 mice were intramuscularly immunized with two doses of vaccines at an interval of 2 weeks.
Neutralizing antibodies, \spike protein-specific Immunoglobulin G (IgG),  as well as antigen-specific interferon (IFN)-$\gamma$-secreting T cells were assessed (see Fig.~\ref{fig:wetlab}).
Interestingly, \frankedit{all vaccine candidates carrying 
\lineardesign-generated mRNA molecules were able to elicit high levels of binding IgG and neutralizing antibodies.
While in contrast, none or low levels of neutralizing antibodies were induced by the benchmark {\sc h}.
Similar results were also observed on} the antigen-specific T cell response, where a robust T helper Type 1 (Th1)-biased T cell response was induced only by vaccines containing \lineardesign-generated mRNAs.
Overall, our designs {\sc a--d}, which are closer to the optimal boundary (shaded in Fig.~\ref{fig:wetlab}A),
\myedit{i.e., strong in both MFE and CAI,}
led to 
a surprising $9\sim20\times$ increase in neutralizing antibody titers
\myedit{and $19\sim23\times$ increase in OD values for Spike-specific binding antibody than the benchmark (codon-optimized baseline {\sc h}).}
These ratios were much higher than the corresponding ratios for cellular protein expression ($1.2\sim1.5\times$),
\frankedit{because 
the latter was \frankedit{only} measured for 24 hours.
We suspect that \lineardesign-generated mRNA molecules have longer functional half-lives \invivo (beyond 24 hours), 
which induce higher levels of antibody response.}


%% file: fig_wetlab.tex

\setlength\tabcolsep{4pt}
\begin{figure}[!t]
\centering
\begin{tabular}{cccc}
  \raisebox{2.1cm}{\hspace{-.4cm}\multirow{4}{*}{\includegraphics[width=.45\linewidth]{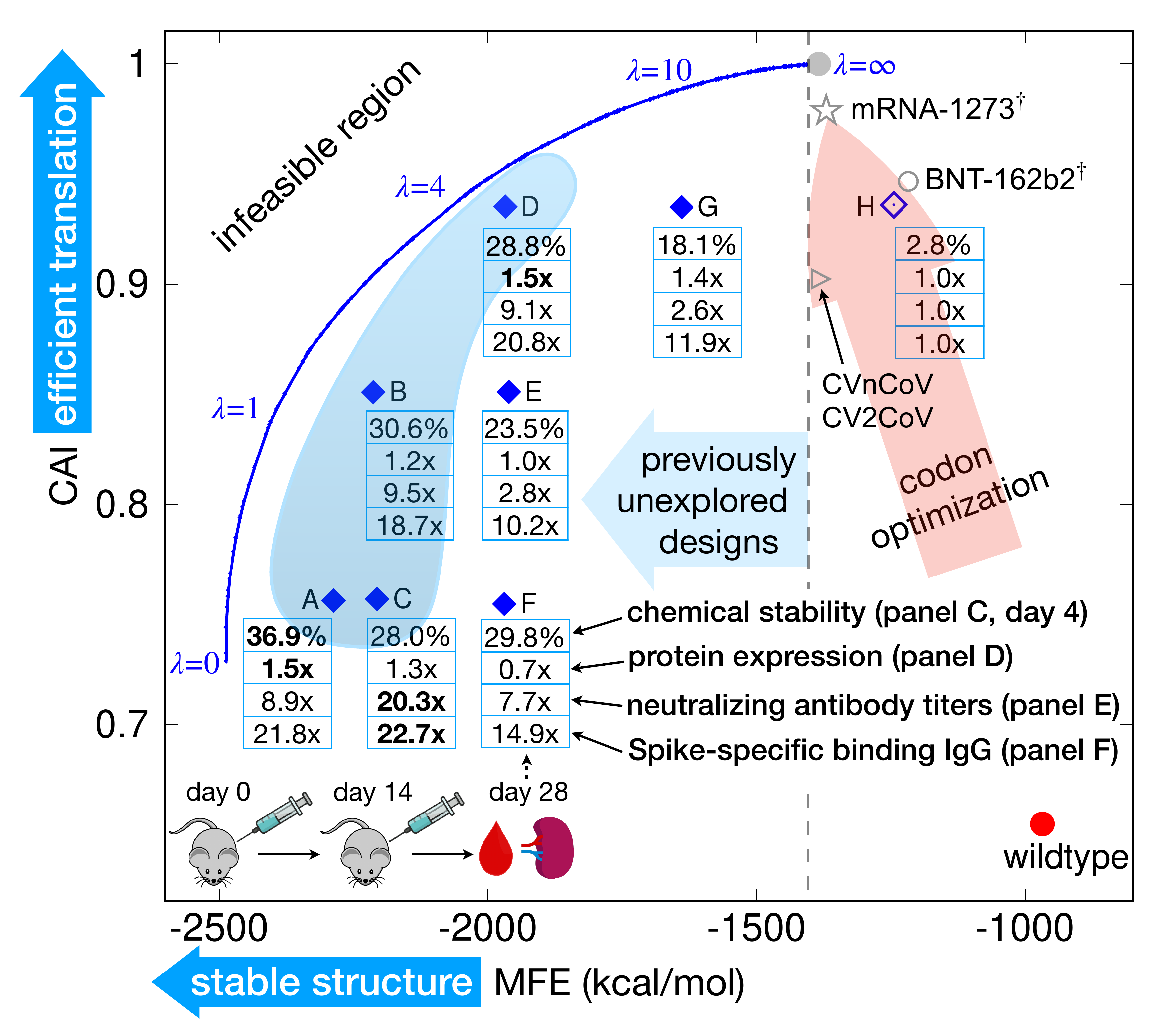}}}
  &
    \hspace{-.5cm}\raisebox{-.1cm}{\includegraphics[width=.17\linewidth]{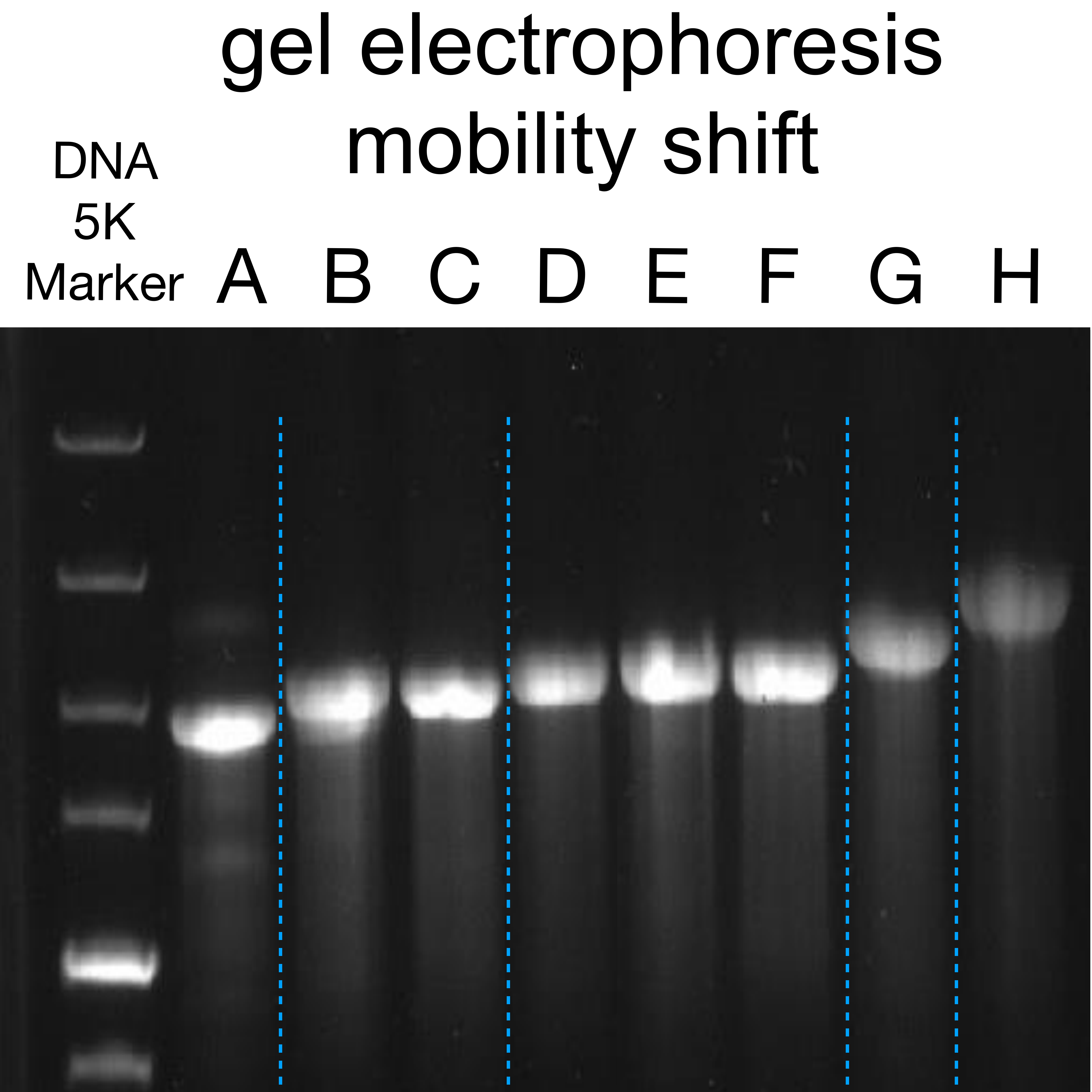}} & \hspace{-1cm}\raisebox{-.5cm}{\includegraphics[width=.2\linewidth]{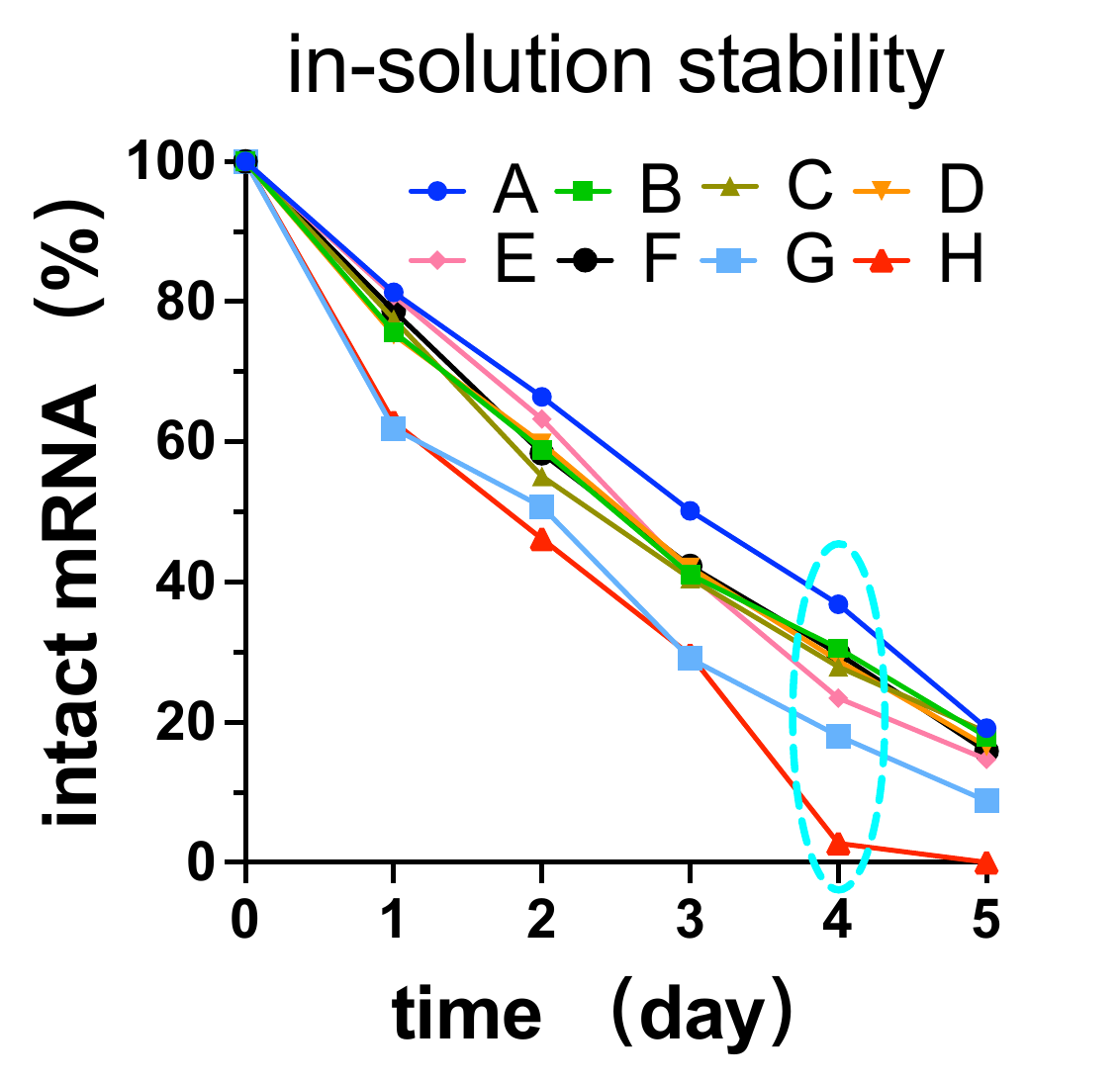}} &
  \hspace{-.9cm}\raisebox{-.11cm}{\includegraphics[width=.165\linewidth]{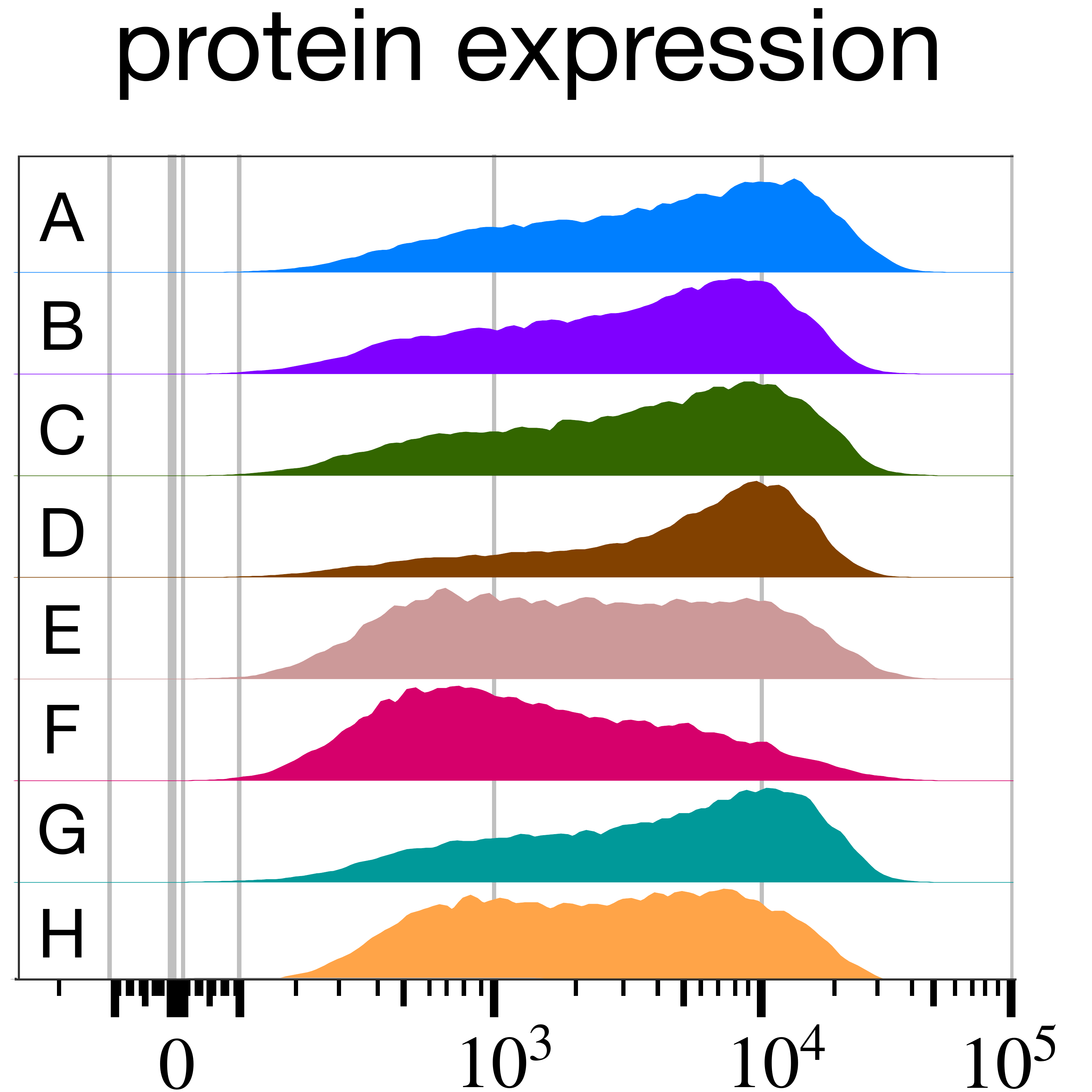}}\\[0.1cm]
  &  
   \hspace{-.5cm}\raisebox{0.1cm}{\includegraphics[width=.23\linewidth]{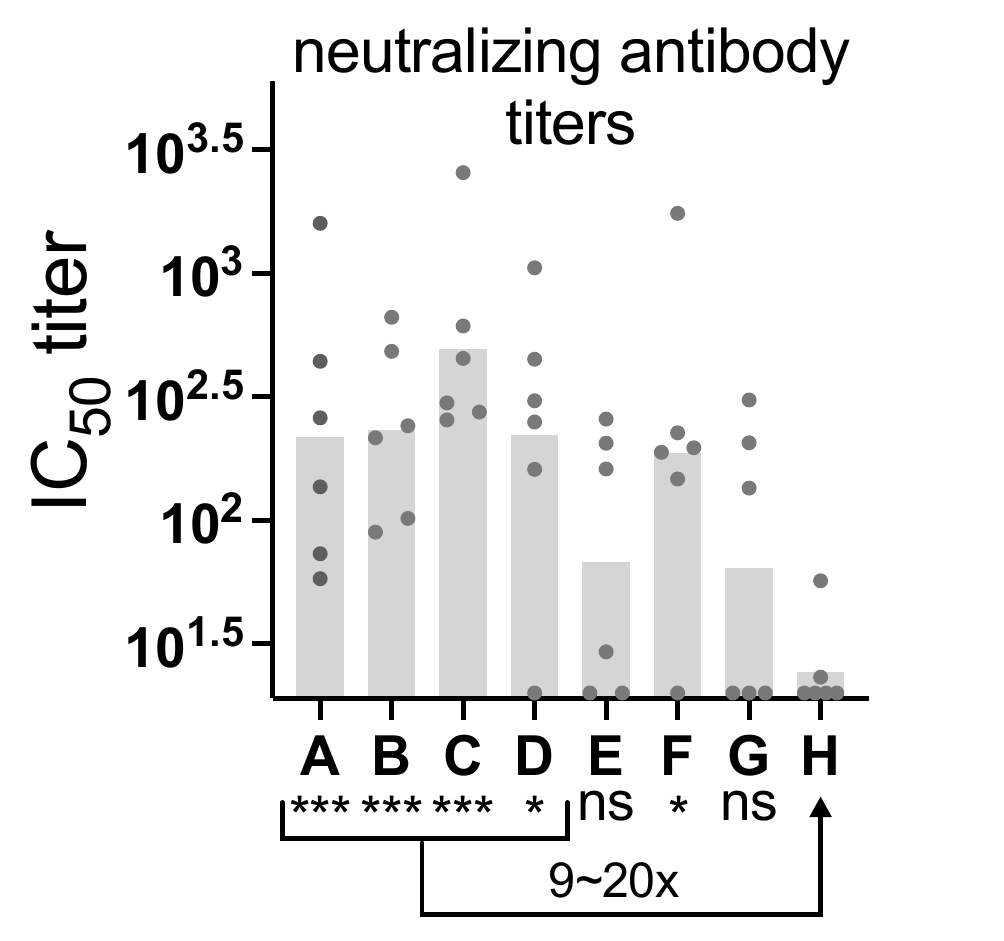}}
  &
   \hspace{-.8cm}\raisebox{0.15cm}{\includegraphics[width=.22\linewidth]{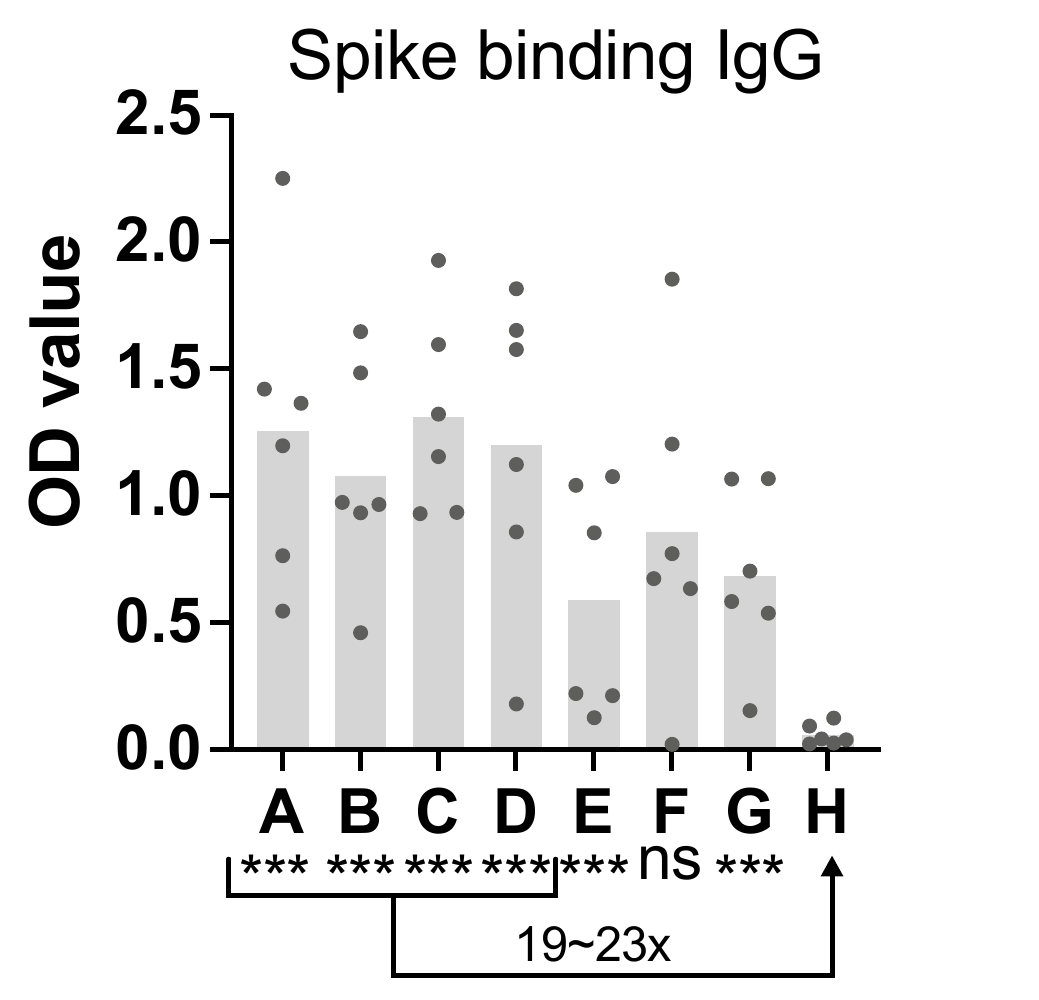}}
  &
    \hspace{-.82cm}\raisebox{0.14cm}{\includegraphics[width=.22\linewidth]{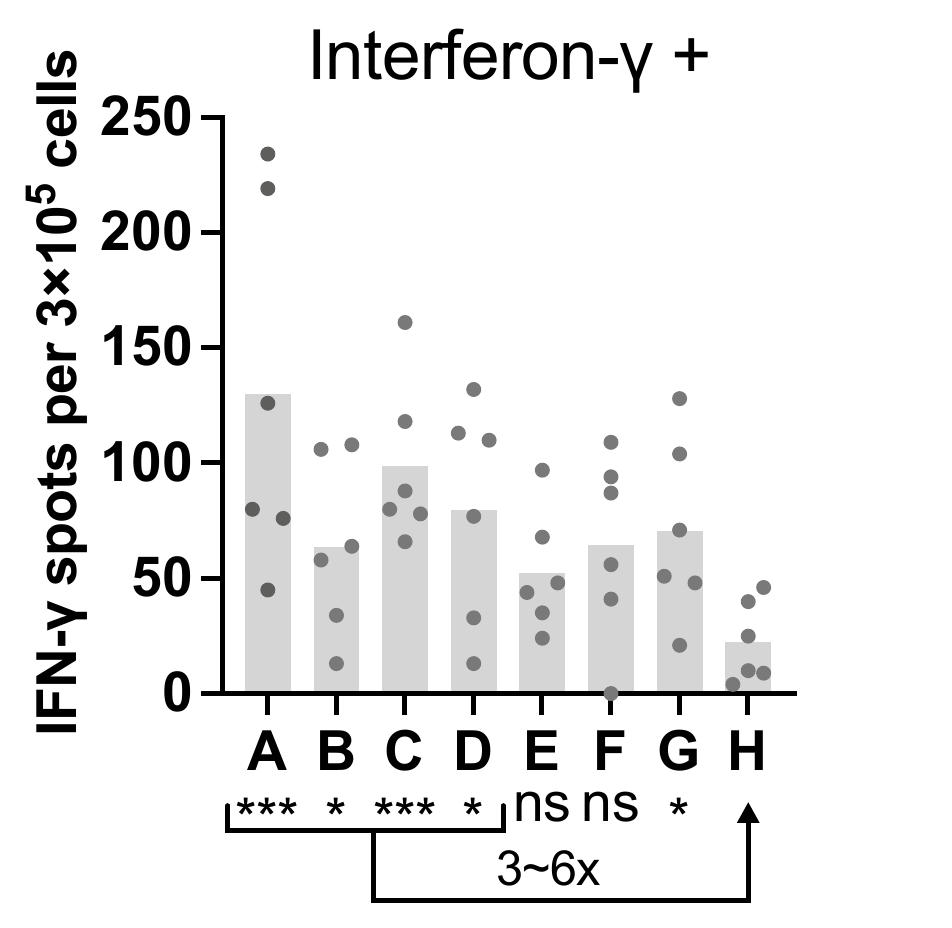}}\\[-6.8cm]
    
    \hspace{-6.9cm} \raisebox{.5cm}{\panel{A}} & \hspace{-3.7cm} \raisebox{.5cm} {\panel{B}}  &\hspace{-3.8cm} \raisebox{.5cm} { \panel{C} } & \hspace{-3.5cm} \raisebox{.5cm} {\panel{D}} \\[-.1cm]
    \hspace{-7.5cm} \raisebox{3.2cm}{\panel{ }} & \hspace{-3.7cm}\raisebox{.8cm} {\panel{\myedit{E}}}  &\hspace{-3.8cm} \raisebox{.8cm}{ \panel{\myedit{F}} } & \hspace{-3.5cm} \raisebox{.8cm}{\panel{\myedit{G}}} \\[2.07cm]

\hspace{-6.9cm}\raisebox{0cm}{\panel{H}}&\\[-.3cm]

\multicolumn{4}{c}{
\hspace{-1cm}
\resizebox{.96\textwidth}{!}{
\begin{tabular}{c|ccccc|ccccccc}
  &           					&       				 &  						&     					&		  				&  (\textit{Panel B})				  &		\multicolumn{2}{c}{(\textit{Panel C})}	&			 (\textit{Panel D}) & (\textit{Panel E})   & (\textit{Panel F}) & (\textit{Panel G})   \\
  &           	 \multicolumn{1}{c}{MFE}				&       				 &  						&     					&Molecular & \multicolumn{1}{c}{Mobi.~shift} &\multicolumn{2}{c}{Intact\%} 						& \multicolumn{1}{c}{Protein} & \multicolumn{1}{c}{NAb}    & \multicolumn{1}{c}{Bind.} & \multicolumn{1}{c}{IFN-$\gamma +$}   \\
  design&   (\textit{kcal/mol})   & \multicolumn{1}{c}{CAI}    & \multicolumn{1}{c}{GC\%}   & \multicolumn{1}{c}{U\%}   & weight& \multicolumn{1}{c}{relat.~dist.}& \multicolumn{1}{c}{(D4)} &  \multicolumn{1}{c}{(D5)} & \multicolumn{1}{c}{expr.~(MFI)}   & \multicolumn{1}{c}{titers} & \multicolumn{1}{c}{IgG}     & \multicolumn{1}{c}{T cells} \\
  \hline \hline
{\sc a} & -2,287.3 & 0.756 & 54.8 & 22.7 &1,229,541&\textBF{3.20}& \textBF{36.9}   &\textBF{19.2}& \textBF{6,837}    & 217.9 & 1.26    & \textBF{130.0}              \\
\hline
{\sc b} & -2,213.2 & 0.851 & 57.0 & 20.8 &1,229,943&3.00& 30.6   & 18.0 & 5,328    & 231.5 & 1.08    & \ 63.8               \\
{\sc c} & -2,206.0 & 0.757 & 55.3 & 22.2 &1,230,380&2.97& 28.0   &18.6 & 5,939    & \textBF{496.4} & \textBF{1.31}    & \ 98.5               \\
\hline
{\sc d} & -1,967.4 & 0.935 & 58.4 & 19.1 & 1,229,927&2.80 & 28.8    &16.6& 6,801    & 221.1 & 1.20    & \ 79.7              \\
{\sc e} & -1,961.3 & 0.851 & 56.8 & 21.1 &1,229,328&2.72& 23.5   & 14.7 & 4,499    & \ 68.2  & 0.59    & \ 52.7               \\
{\sc f} & -1,969.3 & 0.755 & 54.7 & 23.0 &1,230,153&2.75  &29.8    &15.9 & 2,973    & 187.5 & 0.86    & \ 64.5             \\
\hline
{\sc g} & -1,639.3 & 0.935 & 58.6 & 18.9 &1,229,526& 2.50 & 18.1   & \ 8.8 & 6,446    & \ 64.0  & 0.69    & \ 70.5        \\
\hline
{\sc h} & -1,244.4 & 0.936 & 55.0 & 21.2 &1,228,180&2.20& {\ 2.8}   & \ 0.0 & 4,491    & \ 24.4  & 0.06    & \ 22.3              
\end{tabular}
}}\\[-3cm]

\end{tabular}
\vspace{2.8cm}
\caption{Experimental 
results of \lineardesign-generated molecules and vaccines
for the SARS-CoV-2 Spike protein, using unmodified nucleotides.
{\bf A}: Summary of chemical stability (day 4), protein expression, and neutralizing and binding antibody levels
for our designs (\bluediamond{\sc a--g}) compared to the codon-optimized baseline ({\sc h\blueDiamond}).
{\bf B}: Gel mobility shift 
using gel electrophoresis correlates perfectly
with folding free energy changes.
{\bf C}: Chemical stability of mRNAs upon storage in buffer. 
{\bf D}: Flow cytometric analysis of expression level of Spike protein on cell surface following mRNA transfection into immortalized HEK-293 cells. 
{\bf E}: Neutralizing antibody titers. 
{\bf F}: OD values of Spike protein-specific binding IgG.
{\bf G}: Frequency of Interferon (IFN)-$\gamma$-secreting T cells.
We used a two-tailed Mann-Whitney test for significance testing of our designs against the baseline {\small (`ns':\! not significant, `$\star$':\! $0.01 \!\leq\! p \!<\!0.05$, `$\star\!\star$':\! $0.001 \!\leq\! p \!<\! 0.01$, `$\star\!\!\star\!\!\star$':\! $p \!<\!0 .001$)}.
Overall, our best designs ({\sc a--d}, in shades) saw substantial improvements
in half-life,  protein expression, and antibody response (up to $23\times$).
{\bf H}: detailed computational and experimental data for the eight designs (MFI: mean fluorescence intensity);
see Fig.~\ref{fig:seqA_H_structure} for their secondary structures.
{\small $^\dagger$The vaccines of Moderna and BioNTech use modified nucleotides\protect\cite{jeongassemblies}, 
but their MFEs here are calculated with the standard energy model~\protect\cite{turner+:2010}.}
  \label{fig:wetlab}}
\end{figure}

%% file: discussion.tex


An effective mRNA design strategy is of utmost importance, especially for the development of mRNA vaccines that have shown great promise in fighting the current and future pandemics.
However, it is extremely challenging due to the prohibitively large search space. 
We, instead, presented a surprisingly simple solution to this problem
by formulating the design space as DFAs
and reducing the mRNA design problem to lattice parsing 
in computational linguistics. 
\myedit{This unexpected cross-disciplinary analogy provides an efficient algorithm 
that scales $O(n^2)$ for practical applications, 
taking only 10.7 minutes for the \sarscovtwo \spike protein,
and can jointly optimize stability and codon optimality using weighted DFAs.
Our DFA framework can 
also apply to non-standard genetic codes,
modified nucleotides, and coding constraints such as avoiding certain adjacent codon pairs.
Finally, to provide suboptimal candidates for vaccine development and further speed up the design for long sequences,
we provide an $O(n)$-time approximate variant.} 


The mRNA sequences generated by \lineardesign were comprehensively characterized in this study and demonstrated superiority over the commonly-used codon optimization benchmark in three attributes critical for vaccine performance: 
 chemical stability, translation efficiency, and immunogenicity.
In particular, four of our seven designs showed $9\!\sim\!20\times$ increase in neutralizing antibody titers
and $19\!\sim\!23\times$ increase in binding antibody levels over the benchmark.
Given that  
chemical modification is widely believed to be critical to the recent success of mRNA vaccines\myedit{~\cite{kariko2008incorporation, pardi+:2018, baden+:2021, polack+:2020}},\footnote{\myedit{CureVac does not use modification either, but uses codon optimization and engineers the non-coding regions~\cite{gebre+:2021};
our algorithm is orthogonal to, and can be used with, their engineering efforts.}}
it is intriguing that our designed mRNAs 
without chemical modification (\myedit{with reduced cost in manufacturing}) still showed high levels of stability, translation efficiency, and immunogenicity. 
\myedit{On the other hand, our algorithm is {\em orthogonal} to chemical modification 
and can be combined with it once the corresponding energy model is available.}
\myedit{By unleashing the previously inaccessible region of highly stable and efficient sequences,}
\lineardesign 
is a timely and promising tool for mRNA vaccine development 
which is of utmost importance to the current and future pandemics. 
But more importantly, it is also a general and principled method for  molecule design in mRNA medicine, 
and can be used for 
all therapeutic proteins including monoclonal antibodies and anti-cancer drugs.


%% file: methods.tex



\renewcommand{\thesubsection}{\S\arabic{subsection}}

\renewcommand{\theparagraph}{\S\arabic{subsection}.\arabic{paragraph}}
\setcounter{secnumdepth}{4}




\section*{Methods}
\label{sec:methods}

\subsection{Details of \lineardesign Algorithm}
\label{methods:design}

\paragraph{Optimization Objectives}
\label{methods:obj}

There are two objectives in mRNA design: stability and codon optimality.
The optimal-stability mRNA design problem can be formalized as follows. Given a protein sequence $\vecp=p_0 \ldots p_{|\vecp|-1}$ where each $p_i$ is an amino acid residue, 
we find  
the optimal mRNA sequence $\vecr^\star(\vecp)$
that has the {\em lowest
minimum folding free energy change} (\MFE)
among all possible mRNA sequences 
encoding that protein:
\begin{eqnarray}
 \vecr^\star(\vecp) & = \argmin_{\vecr \in \mRNA(\vecp)} \mfe(\vecr)  \label{eq:design}\\
 \mfe(\vecr) & = \min_{\vecs \in \structures(\vecr)} \freeenergy(\vecr, \vecs) \label{eq:fold}
\end{eqnarray}
where  $\mRNA(\vecp) = \{ \vecr \mid \protein(\vecr) = \vecp \}$ is the set of candidate mRNA sequences,
$\structures(\vecr)$ is the set of all possible secondary structures for mRNA sequence \vecr,
and $\freeenergy(\vecr, \vecs)$ is the free energy change of structure~\vecs for mRNA~\vecr according to an energy model. This is clearly a double minimization objective involving the per-sequence minimization over all of its possible structures (i.e., RNA folding; Eq.~\ref{eq:fold}) which has well-known dynamic programming solutions,
and the global minimization over all sequences (i.e., optimal mRNA design; Eq.~\ref{eq:design}) which we will solve using lattice parsing (\ref{methods:CFG-DFA-DP}).

Next, we integrate codon optimality by adding Codon Adaptation Index (\CAI) \mycite{sharp+li:1987}, defined as the geometric mean of the codon optimality of each codon in the mRNA \vecr:
\[
  \CAI(\vecr) = \ ^{\frac{|\vecr|}{3}\!\!\!\!} \sqrt{\textstyle\prod_{0\leq i < \frac{|\vecr|}{3}} w \big(\codon(\vecr, i)\big)} \label{eq:CAI} 
\]
where $\codon(\vecr, i) = r_{3i} r_{3i+1} r_{3i+2}$ 
is the $i$th triplet codon in \vecr,
and $w(c)$ is the relative adaptiveness of codon $c$, defined as the frequency of $c$ divided by
the frequency of its most frequent synonymous codon ($0\leq w(c) \leq 1$). 
Because CAI is always between 0 and 1 but MFE is generally proportional to the mRNA sequence length, 
we 
scale CAI by the number of codons and 
use a hyperparameter $\lambda$ to balance \MFE and \CAI ($\lambda = 0$ being purely \MFE), and define a novel joint objective:
\[
   \MFECAIlambda(\vecr)  = \MFE(\vecr) - \tfrac{|\vecr|}{3} \lambda \log {\CAI(\vecr)} 
\]
which can be simplified by expanding CAI:
\begin{eqnarray}
      \MFECAIlambda(\vecr) & = &\MFE(\vecr) -  \tfrac{|\vecr|}{3} \lambda \log \ ^{\frac{|\vecr|}{3}\!\!\!\!} \sqrt{\displaystyle\prod_{0\leq i < \tfrac{|\vecr|}{3}} w \big(\codon(\vecr, i)\big)} \notag\\
                                        & = &  \MFE(\vecr) - \lambda \displaystyle\sum_{0\leq i < \tfrac{|\vecr|}{3}} {\log}\, w \big(\codon(\vecr, i) \big)
                                        \label{eq:joint}
\end{eqnarray}
This joint objective is basically \MFE plus (a scaled) sum of the negative logarithm of each codon's relative adaptiveness.
Now the joint optimization can be defined as:
\begin{equation*}
\begin{split}
\vecr^\star_\lambda(\vecp ) & = \argmin_{\vecr \in \mRNA(\vecp)} \MFECAIlambda (\vecr)
                    = \argmin_{\vecr \in \mRNA(\vecp)}\Big({\MFE(\vecr)}- \lambda \displaystyle\sum_{0\leq i < \tfrac{|\vecr|}{3}} {\log}\, w \big(\codon(\vecr, i)\big) \Big)
\end{split}
\end{equation*}
See Fig.~\ref{fig:main_CFG_DFA}D for examples of relative adaptiveness calculation.

\paragraph{DFA Representations for Codons and mRNA Candidate Sequences}
\label{methods:DFA}
Informally, a DFA is a directed graph with labeled edges and distinct start and end states. 
For our purpose each edge is labeled by a nucleotide,
so that for each codon DFA, each start-to-end path represents a triplet codon.
Formally, a DFA is a 5-tuple $\langle Q, \Sigma, \delta, q_0, F \rangle$, 
where $Q$ is the set of states,
$\Sigma$ is the alphabet (here $\Sigma=\{\nucA, \nucC, \nucG, \nucU\}$),
$q_0$ is the start state (always $\dfastate{(0,0)}$ in this work), 
$F$ is the set of end states (in this work the end state is unique, i.e., $F=\{\dfastate{(3,0)}\}$),
and $\delta$ is the transition function 
that takes a state $q$ and a symbol $a \in \Sigma$ and returns the next state $q'$,
i.e., $\delta(q, a) = q'$ encodes a labeled edge $\laedge{q}{a}{q'}$.


After building  DFAs for each amino acid,
we can concatenate them into
a single DFA $D(\vecp)$ for a protein sequence \vecp,
which represents all possible mRNA sequences that translate into that protein
\[
D(\vecp) = D(p_0) \circ D(p_1) \circ \cdots \circ D(p_{|\vecp|-1}) \circ D(\textsc{stop})
\]
by stitching the end state of each DFA with the start state of the next.
The new end state of the mRNA DFA is $\dfastate{(3|\vecp|\!+\!3, 0)}$.

We also define $\outedges(q)$ to be the set of outgoing edges from state $q$, and $\inedges(q)$ to be the set of incoming edges
(which will be used in the pseudocode, see Figs.~\ref{fig:algorithm}--\ref{fig:update_backtrace}):
\begin{align*}
\outedges(q) & = \{\laedge{q}{a}{q'} \mid \delta(q, a) = q' \}\\
\inedges(q) & = \{\laedge{q'}{a}{q} \mid \delta(q', a) = q \}
\end{align*} 
For the mRNA DFA 
in Fig.~\ref{fig:CFG_DFA}D,
$\outedges(\dfastate{(3,0)}) = \{ \laedge{(3,0)}{\nucU}{(4,0)}, \allowbreak\, \laedge{(3,0)}{\nucC}{(4,1)}\}$
and 
$\inedges(\dfastate{(9,0)}) = \{ \laedge{(8,0)}{\nucA}{(9,0)}, \allowbreak\, 
                                \laedge{(8,0)}{\nucG}{(9,0)}, \allowbreak\,
                                \laedge{(8,1)}{\nucA}{(9,0)}\}$.

\paragraph{Objective 1 (Stability): Stochastic Context-Free Grammar, Lattice Parsing, and Intersection} 
\label{methods:CFG-DFA-DP}
A stochastic context-free grammar (SCFG) is a context-free grammar in which each rule is augmented with a weight.
More formally, an SCFG is a 4-tuple $\langle N, \Sigma, P, S\rangle$
where $N$ is the set of non-terminals, $\Sigma$ is the set of terminals (identical to the alphabet in the DFA, in this case $\Sigma=\{\nucA, \nucC, \nucG, \nucU\}$), $P$ is the set of weight-associated context-free writing rules,
and $S\in N$ is the start symbol.
Each rule in $P$ has the form $A \goestow{w} (N\cup \Sigma)^*$ where 
$A\in N$ is a non-terminal that can be rewritten according to this rule into a sequence of non-terminals and terminals 
(the star $^*$ means repeating zero or more times) 
and $w\in \mathbb{R}$ is the weight associated with this rule.

SCFGs are commonly used to represent the RNA folding energy model.
The weight of a derivation (parse tree, or a secondary structure in this case) is the sum of weights of the productions used in that derivation~\mycite{Rivas:2013}. 
For example, for a very simple Nussinov-Jacobson-style model~\mycite{nussinov+jacobson:1980},
which simplifies the energy model to individual base pairs, 
we can define this SCFG $G$ as in Fig.~\ref{fig:CFG_DFA}E,
where each \nucG\nucC\/ pair gets a score of $-3$, and each \nucA\nucU\/ pair gets a score of $-2$.
Thus, 
the standard RNA secondary structure prediction problem can be cast as a parsing problem: given the above SCFG $G$ and an input RNA sequence, find the minimum-weight derivation in $G$ that can generate the sequence.
This can be solved by the classical CKY algorithm from computational linguistics \myedit{\mythreecite{kasami:1966}{younger:1967}{Rivas+:2012}}.




The optimal-stability mRNA design problem is now a simple extension of the above single-sequence folding problem to the case of multiple inputs:
instead of finding the minimum free energy structure (minimum weight derivation) for a given sequence, 
we find the minimum free energy structure (and its corresponding sequence) among all possible structures for all possible sequences
(see Fig.~\ref{fig:CFG_DFA}). 
This can be solved by lattice parsing on the DFA, which is a generalization of CKY from a single sequence to a DFA.
Take the bifurcation rule $S \goesto N \, P$ for example.
In CKY, if you have derived non-terminal $N$ for span $[i, j]$, 
notated \labeledspan{N}{i}{j}, 
and if you have also derived \labeledspan{P}{j}{k}, 
you can combine the two spans,
i.e.,   \labeledtwospans{N}{i}{j}{P}{k}, 
and use the above rule to derive
\labeledspan{S}{i}{k}.  
Similarly, in lattice parsing,
if you have derived both \labeledpath{N}{q_i}{q_j} 
(i.e., there is a  $q_i \leadsto q_j$ 
path that can be derived from $N$)
and \labeledpath{P}{q_j}{q_k}, 
you can combine them 
to a longer path \labeledtwopaths{N}{q_i}{q_j}{P}{q_k} 
and derive \labeledpath{S}{q_i}{q_k} 
with the above rule. 
While the runtime for CKY scales $O(|G| n^3)$ where $|G|$ is the grammar constant (the number of rules) and $n$ is the RNA sequence length,
the runtime for lattice parsing similarly scales $O(|G| |D|^3)$ where $|D|$ is the number of states in the DFA.
For mRNA design with the standard genetic code, $n \leq |D| \leq 2n$ because each position $i$ has either one or two states ($(i,0)$ and $(i,1)$), so its time complexity is also actually identical to single-sequence folding, just with a larger constant. 
See Methods~\ref{methods:left-right} for details of this algorithm and Figs.~\ref{fig:algorithm}--\ref{fig:update_backtrace} for the pseudocode.

More formally, in theoretical computer science, lattice parsing with an CFG $G$ on a DFA $D$ 
is also known as the intersection between the languages of $G$ and $D$
(i.e., the sets of sequences allowed by $G$ and $D$), notated $L(G) \cap L(D)$, which was solved by the Bar-Hillel construction in 1961 \mycite{bar-hillel+:1961}.
In order to adapt it to mRNA design, we need to extend this concept to the case of weighted (i.e., stochastic) grammars and weighted DFAs (the latter is needed for CAI integration; see below).
While the language $L(G)$ of CFG $G$ is the set of sequences generated by $G$,
the language of the SCFG for RNA folding free energy model defines a mapping from each RNA sequence to its MFE, i.e.,
$L_w(G) : \Sigma^* \mapsto \mathbb{R}$. This can be written as a relation:
\[
L_w(G) = \{ \vecr \sim \MFE(\vecr)  \mid \vecr \in \Sigma^* \}
\]
And we also extend the language of a DFA to a trivial weighted language (which will facilitate the incorporation of CAI into DFA below):
\[
L_w(D) = \{ \vecr \sim 0  \mid \vecr \in L(D) \}
\]
Next we extend the intersection from two sets to two weighted sets $A$ and $B$:
\[
A \cap_w B = \{ \vecr \sim (w_1 + w_2) \mid \vecr \sim w_1 \in A, \vecr \sim w_2 \in B \}
\]
Now we can show that optimal-stability mRNA design problem can be solved via weighted intersection between $L_w(G)$ and $L_w(D)$,
i.e., we can construct a new ``intersected'' stochastic grammar $G'$ that has the same weights (i.e., energy model) as the original grammar but only generates sequences in the DFA:
\[
L_w (G') = L_w(G) \cap_w L_w(D) = \{ \vecr \sim \MFE(\vecr) \mid \vecr \in L(D) \}
\]

\paragraph{Adding Objective 2 (Codon Optimality): Weighted DFA for CAI Integration}
\label{methods:cai}

As described in the main text and Fig.~\ref{fig:main_CFG_DFA}D, our novel joint optimization objective (Eq.~\ref{eq:joint}) 
factors the CAI of each mRNA candidate onto the relative adaptiveness of each of its codons,
and thus can be easily incorporated into the DFA as edge weights.
To do this we need to extend the definition of DFA to weighted DFA,
where the transition function $\delta$ now returns a state and a weight,
i.e., $\delta(q, a) = (q', w)$, which encodes a weighted label edge $\laedgew{q}{a: w}{q'}$. 
Now the set of outgoing and incoming edges are also updated to:

\vspace{-1cm}
\begin{align*}
\outedges(q) & = \{\laedgew{q}{a:w}{q'} \mid \delta(q, a) = (q', w) \}\\
\inedges(q) & = \{\laedgew{q'}{a:w}{q} \mid \delta(q', a) = (q, w) \}
\end{align*} 
In this case, the weighted DFA defines a mapping from each candidate mRNA sequence to its negative logarithm of CAI scaled by the number of codons, i.e., $L_w(D): L(D) \mapsto \mathbb{R}$. More formally,
\[
L_w(D) = \{ \vecr \sim - \tfrac{|\vecr|}{3} \log \CAI(\vecr) \mid \vecr \in L(D) \}
\]
Now the weighted intersection defined above can be extended to incorporate the hyperparameter $\lambda$ and
derive the joint objective:
\[
L^\lambda_w(G') = L_w(G) \cap^\lambda_w L_w(D) = \{ \vecr \sim \big(\MFE(\vecr) - \lambda \tfrac{|\vecr|}{3} \log \CAI(\vecr)\big) \mid \vecr \in L(D) \}
\]

\paragraph{Bottom-Up Dynamic Programming}
\label{methods:bottom-up}

Next, we describe how to implement the dynamic programming algorithm behind lattice parsing (or equivalently, intersection between the languages of a stochastic context-free grammar and a weighted DFA) to solve the joint optimization problem.
For simplicity reasons, here we use bottom-up dynamic programming on a modified Nussinov-Jacobson energy model. 
Fig.~\ref{fig:algorithm} gives the pseudocode for this simplified version.
We first build up the mRNA DFA for the given protein, 
and initialize two hash tables, 
${\mathit best}$ to store the best score of each state, 
and ${\mathit back}$ to store the best backpointer.
For the base cases ($S \goestow{0} N\ N\ N$),
we set ${\mathit best}[S, q_i, q_{i+3}] \gets 0$ for optimal-stability design,
and  ${\mathit best}[S, q_i, q_{i+3}] \gets \mincost(q_i, q_{i+3}, \lambda)$ for the joint optimization
where 
\begin{equation}
     \mincost(q_i, q_{i+3}, \lambda) \defeq \min_{q_i \stackrel{a: w_1}{\longrightarrow} q' \stackrel{b: w_2}{\longrightarrow} q'' \stackrel{c: w_3}{\longrightarrow}  {q_{i+3}}} \lambda (w_1 + w_2 + w_3)
     \label{eq:mincost}
\end{equation}    
is the minimum ($\lambda$-scaled) cost of any $q_i \leadsto q_{i+3}$ path in the CAI-integrated DFA.
Next, 
for each state $(q_i, q_j)$ it goes through the pairing rule and bifurcation rules,
and updates if a better score is found.
After filling out the hash tables bottom-up,
we can backtrace the best mRNA sequence stored with the backpointers.
See Fig.~\ref{fig:update_backtrace} for details of {\sc Update} and {\sc Backtrace} functions. 

\paragraph{Left-to-Right Dynamic Programming and Beam Search}
\label{methods:left-right}
Inspired by our previous work, \linearfold~\mycite{huang+:2019},
we further developed 
a linear-time approximation algorithm for mRNA design.
We apply beam pruning~\mycite{huang+:2012}, 
a classical pruning technique, 
to significantly narrow down the search space
without sacrificing too much search quality. 

Fig.~\ref{fig:linear} gives the pseudocode of simplified \lineardesign algorithm for the Nussinov model,
based on 
left-to-right dynamic programming and beam pruning. 
\lineardesign replaces bottom-up dynamic programming with a left-to-right parsing.
At each step $j$ (the $j$th position of mRNA sequence),
we only keep the top $b$ states with the lowest cost 
and prune out the less promising states,
since they are unlikely to be the optimal sequence.
Here $b$, the beam size, is a user-adjustable parameter to balance runtime and search quality.
Notice that we use $b=100$ as default in \linearfold~\citesiref{huang+:2019},
but in \lineardesign we usually use 
a larger beam size of $b=500$ because the search space is larger.

Our real system uses a left-to-right dynamic programming with beam pruning
on the Turner nearest neighbor free energy model~\mytwocite{mathews+:1999}{Mathews+:2004}.
We implement the thermodynamic parameters following Vienna RNAfold~\mycite{lorenz+:2011},
except for the dangling ends.
Dangling ends refer to stabilizing interactions for multiloops and external loops\mycite{turner+:2010},
which require knowledge of the nucleotide sequence outside of the state $(q_i, q_j)$.
Though it could be integrated in \lineardesign,
the implementation becomes more involved.

\iftrue 
\paragraph{DFAs for Other Genetic Codes, Coding Constraints, and Modified Nucleotides}
\label{methods:generality_dfa}

The DFA framework 
can also represent less common cases such as alternative genetic codes,
 modified nucleotides, and coding constraints.
{First, DFA can encode non-standard genetic codes, such as 
alternative nuclear code for 
some yeast~\mycite{kawaguchi+:1989} 
and mitochondrial codes~\mycite{bonitz+:1980} (Fig.~\ref{fig:si_alter_aa}A).}
Second, we may want 
to avoid some unwanted or rare codons 
(such as the amber stop codon) 
which is an easy change on the codon DFAs (Fig.~\ref{fig:si_alter_aa}B),
or certain adjacent codon pairs that modulate translation efficiency~\mycite{gamble2016adjacent},
which is beyond the scope of single codon DFAs but 
easy on
the mRNA DFA (Fig.~\ref{fig:si_alter_aa}C).
Similarly, we may want to disallow certain restriction enzyme recognition sites,
which span across multiple codons (\ref{fig:si_alter_aa_enzyme}).
Finally, 
chemically modified nucleotides such as 
pseudouridine ($\Psi$) 
have been widely used in mRNA vaccines~\mycite{kariko2008incorporation},
which can also be incorporated in the DFA (Fig.~\ref{fig:si_alter_aa}D).
\fi

\paragraph{Related Work}
\label{methods:prevwork}
{
Two previous studies also tackled our objective 1 (optimal-stability mRNA design) via dynamic programming,
but their algorithms are ad-hoc and complicated~\mytwocite{cohen+skiena:2003}{Terai+:2016}. 
By contrast, our work solves the harder and more general problem of joint optimization between stability and codon optimality
(which subsumes their objective as a special case),
yet using a much simpler and extendable solution.
First, our work is the first to use automata theory to compactly and conveniently represent
the exponentially large mRNA design space.
Second, our reduction of mRNA design to lattice parsing 
induces a simple yet efficient solution based on classical results in 
computational linguistics and theoretical computer science, 
which scales $O(n^2)$ rather than $O(n^3)$ for practical applications (Figs.~\ref{fig:insilico} \& \ref{fig:LD_CDSfold_items_generated_table}).
Third, we define a novel joint optimization objective that factors the (logarithm of) CAI of an mRNA {\em additively} onto its individual codons,
making it possible to incorporate codon optimality into dynamic programming.
Codon usage is an important factor in mRNA design \mycite{mauger+:2019} that previous work was unable to jointly optimize.\footnote{CDSfold~\mycite{Terai+:2016} uses simulated annealing to improve CAI by fine-tuning from the MFE solution,
but this is a heuristic with no guarantees, and \myedit{their objective formulation, unlike ours, does not factor onto individual codons,
thus cannot be incorporated into dynamic programming}.}
Fourth, our DFA framework is so general that it can also represent arbitrary (non-standard) genetic codes, modified nucleotides, and coding constraints such as adjacent codon pair preference, which previous work could not handle even with major modifications.}
Fifth,
we further develop a faster, linear-time, approximate version which greatly reduces runtime for long sequences with small sacrifices in search quality, 
which we also use to generate multiple suboptimal candidates with varying folding stability and codon optimality as candidates for experimentation.
Last but not least, extensive experiments confirm that compared to the standard codon optimization benchmark,
our designs are substantially better in chemical stability and protein expression \invitro, 
and the corresponding mRNA vaccines elicit up to 23$\times$ higher antibody responses \invivo (see Fig.~\ref{fig:wetlab}).

\paragraph{Benchmark Dataset and Machine}
To estimate the time complexity of \lineardesign, we collected 114 human protein sequences from UniProt~\mycite{UniProt:2005}, 
with length from 78 to 3,333 amino acids (not including the stop codon). 
We 
benchmarked \lineardesign on a Linux machine with 2 Intel Xeon E5-2660 v3 CPUs (2.60 GHz) and 377 GB memory, and used Clang (11.0.0) to compile.


\paragraph{Additional Design Constraints}
\label{methods:design-constraints}
Some studies have shown that protein expression level drops if the 5'-end leader region
has more secondary structure~\myfivecite{Ding+:2013}{Wan+:2014}{Shah+:2013}{Tuller+Zur:2014}{mauger+:2019}. 
To design sequences with less structures at 5'-end leader region,  
we take a simple ``design, enumerate and concatenate'' strategy to loose structure of the leader region:
(1) design the CDS region except for the 5'-end leader region (i.e., the first 15 nucleotides);
(2) enumerate all possible subsequences in the 5'-end leader region;
and (3) concatenate each subsequence with the designed sequence, refold, and choose the one whose 5'-end leader region has the most unpaired nucleotides.

In addition, it has been revealed that long double-stranded region may induce unwanted innate immune responses by 
previous studies~\mythreecite{liu+:2008}{husain+:2012}{hur:2019}.
Considering this, we do not allow long double-stranded regions that include 33 or more base pairs in our design algorithms.

\subsection{Details of \Invitro and \Invivo Experiments}
\label{methods:experiments}

\paragraph{Preparation of mRNA Vaccine} 
mRNA molecules were synthesized from corresponding linearized plasmid DNA template using T7 RNA polymerase, 
which flanked the open-reading frame (ORF) of \spike antigen with the 5’ and 3’ untranslated regions and a 70~$nt$ poly-A tail. 
Cleancap analog was included to obtain capped RNA. 
The transcription reaction was incubated at 37\textdegree{}C for 6--8 hours, 
followed by treatment with DNase. 
RNA was then purified using Carboxylated microspheres. 
For the preparation of mRNA vaccines, lipopolyplex delivery platform was used to encapsulate the mRNA cargo as previously reported~\mycite{persano2017lipopolyplex}.

\paragraph{Electrophoretic Mobility Shift Assay (EMSA) and Integrity assay of mRNA}
To compare the electrophoretic mobility of mRNA molecules, mRNA samples were stored in Ambion\textregistered \ 
RNA storage buffer (Cat. \# AM7001, Thermo Fisher, Mg$^{2+}$ = 0mM).
After denaturing at 70\textdegree{}C for 5 mins followed by cooling on ice, the mRNA samples were loaded on 1\% agrose gel to run at 130V for 1h at 4\textdegree{}C.
Gel image was taken by Gel Doc XR+ Gel Documentation System (Bio-Rad). (Fig.~\ref{fig:wetlab}B)
 
RNA integrity was assayed by Qsep100\textsuperscript{TM} Capillary Electrophoresis System. Intact mRNA was calculated as the percentage of full-length mRNA in solution. (Fig.~\ref{fig:wetlab}C)

\paragraph{Protein Expression Assay} 

\subparagraph{Cell Culture} 
Human embryonic kidney (HEK)293T cells were cultured in Dulbecco’s modified Eagle’s medium (DMEM) (Hyclone) containing 10\% fetal bovine serum (FBS) (GEMINI) and 1\% Penicillin-Streptomycin (Gibco). 
All cells were cultured at 37\textdegree{}C in a 5\% CO$_2$ condition.

\subparagraph{Measurement of Protein Expression from mRNA} 
Cells were transfected with mRNA molecules using Lipofectamine MessengerMAX (Thermo Scientific). 
Briefly, 2 $\mu g$ of mRNA was mixed with 6 $\mu L$ of Lipofectamine reagent first and then incubated with cells for 24 or 48 hours. 
For flow cytometric analysis, cells were collected and stained with live/dead cell dye (Fixable Viability Stain 510, BD) for 5 min. 
After washing, cells were incubated with anti-RBD chimeric mAb (1:100 dilution, Sino Biological) for 30 min, 
followed by washing and incubation with PE-anti-human IgG Fc (1:100 dilution, Biolgend) for 30 min. 
Samples were then acquired on BD Canto II (BD Biosciences). 
Data were analyzed using FlowJo V10.1 (Tree Star). (Fig.~\ref{fig:wetlab}A and D)

\paragraph{Immunization} 
C57BL/6 mice (6-8 weeks) were intramuscularly immunized twice with 10 $\mu g$ of mRNA vaccines at a 2-week interval. 
Sera and spleens were collected 14 days after boost immunization. 
All experiments using mouse model were conducted under the ethical regulations and were approved by local ethical committees.

\subparagraph{Surrogate Virus Neutralization (sVNT) Assay (Neutralizing Antibody Assay)} 
Neutralizing antibody titer was measured using sVNT assay as previously resported~\mycite{tan2020sars} with some modifications. 
Briefly, 96-well plates (Greiner Bio-one) were coated with hACE2 protein (100ng/well, Genscript) overnight at 4\textdegree{}C. 
Plates were washed with PBST and blocked with 2\% BSA for 2 hours at RT. 
HRP-conjugated RBD (100 ng/ml) were incubated with serially diluted serum from immunized mice at an equal volume (60 $\mu L$ each) for 30 min at 37\textdegree{}C. 
Sera collected from mice receiving PBS injection were used as negative control. 
Following this, a 100-$\mu L$ mixture of RBD and serum was added into each well and incubated for 15 min at 37\textdegree{}C. 
After washing, TMB substrate (Invitrogen) was used for development and the absorbance was read at 450 nm using BioTek microplate reader. 
The IC50 value was calculated using 4 parameter logistic non-linear regression. (Fig.~\ref{fig:wetlab}E)

\subparagraph{Enzyme-linked Immunosorbent Assays (Binding Antibody Assay)} 
Recombinant \sarscovtwo \spike ectodomain protein (100 ng/well, Genscript) diluted in coating buffer (Biolegend) were coated into 96-well EIA/RIA plates (Greiner Bio-one) overnight at 4\textdegree{}C. 
The plates were then washed with PBS-T (0.05\% Tween-20) and were blocked with 2\% BSA in PBST for 2 hours at room temperature (RT). 
Serum samples (1:6400 diluted) were added and incubated for 2 hours at RT. 
After washing, total IgG was evaluated using HRP-conjugated goat anti-mouse IgG Ab (1:10,000) for 1 hour. 
TMB substrate (Invitrogen) was used for development and the absorbance was read at 450 nm using BioTek microplate reader. (Fig.~\ref{fig:wetlab}F)

\subparagraph{Enzyme-linked Immunospot (ELISpot) Assay (T Cell Response Assay)} 
Frequency of \spike antigen-specific IFN-$\gamma$-secereting T cells was evaluated using Mouse IFN-$\gamma$ ELISpotplus Kit (Mabtech) according to the manual. 
Briefly, 3$\times$105 murine splenocytes were added to wells pre-coated with anti-mouse IFN-$\gamma$ capturing Abs and were incubated with \spike protein (10 $\mu g/ml$) for 20 hours. 
After washing, plates were incubated with Streptavidin-ALP (1:1000) for 1 hour at RT. 
Spots were developed with BCIP/NBT substrate solution and counted using Immunospot S6 analyzer (CTL). (Fig.~\ref{fig:wetlab}G)


%% file: si.tex
\onecolumn
\newpage
\captionsetup[figure]{font={stretch=1}} 
\renewcommand{\thefigure}{S\arabic{figure}}
\renewcommand{\thetable}{S\arabic{table}}

\renewcommand\thesection{\S\arabic{section}}

\begin{singlespace}
\ifarXiv
\else
\begin{linenumbers}
\fi

\setcounter{figure}{0} 

\noindent

\section*{Supplementary Figures and Tables}

\colorlet{StartNode}{purple!20}
\colorlet{NodeGray}{gray!20}
\colorlet{EndNode}{purple!20}

\input{fig_word_lattice}

\input{fig_dfa_cfg_20220209}


\input{fig_pseudocode}

\input{si_update_backtrace}

\input{fig_lineardesign}

\input{fig_si_alter_codon}


\begin{figure}[!hbt]
\centering
\begin{tabular}{cc}
\hspace{-9cm}\panel{\myedit{A}} & \hspace{-8cm}\panel{\myedit{B}} \\[-.1cm]
\hspace{0cm}\raisebox{0cm}{\includegraphics[width=.60\linewidth]{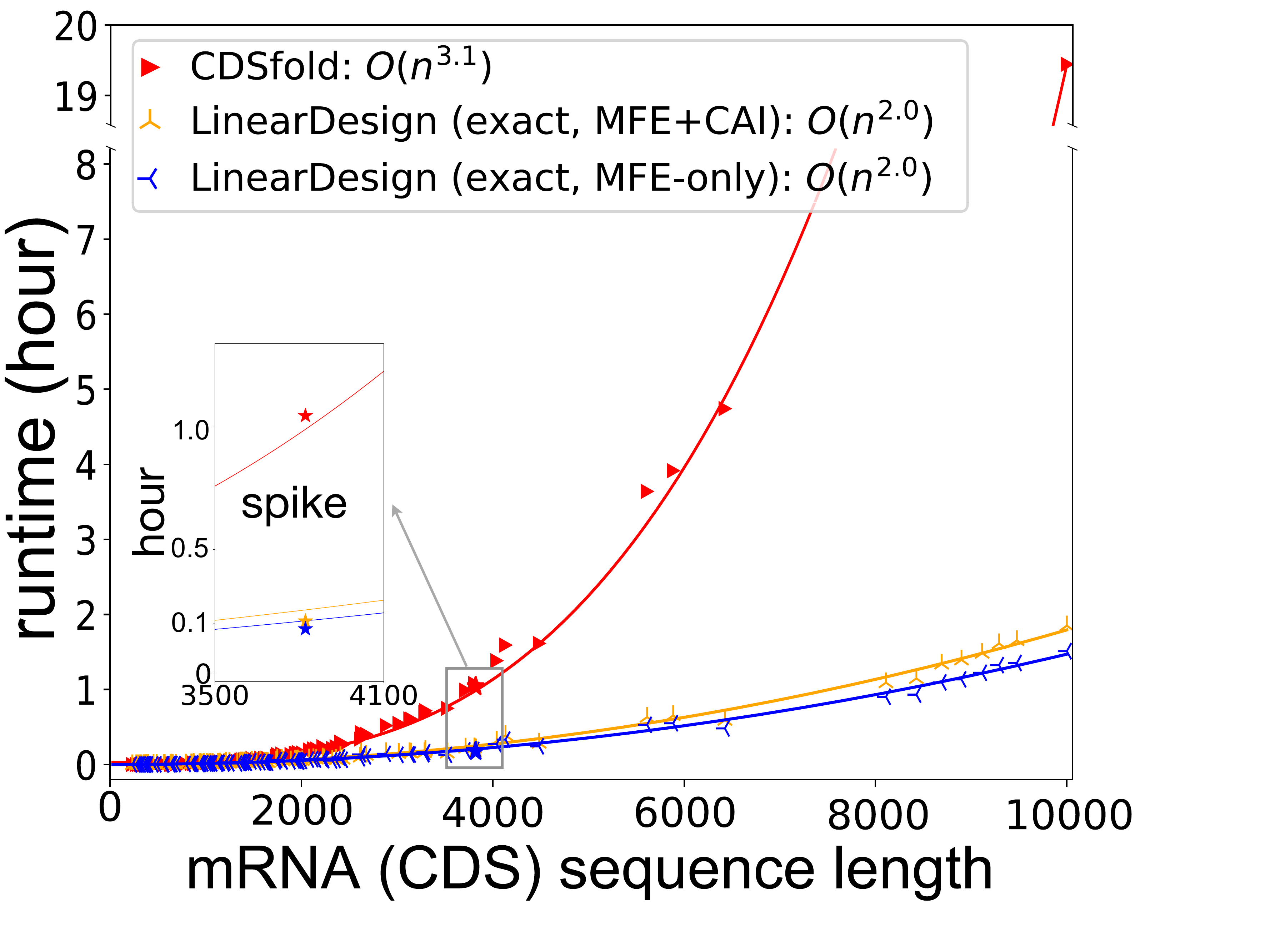}}&

\hspace{-1.5cm}
\raisebox{6.cm}{
\resizebox{!}{6ex}{
\begin{tabular}{c|c|c}
        \!\!{\em $\#$ of items generated}       & \!bifurcation\! & pairing \\
        \!\!{\em in Nussinov model}	               & \footnotesize $S \goesto S\, P$ & \footnotesize \!\!$P \goesto \nucC \, S \, \nucG \mid ...$\!\!\! \\
         \hline
         \!\!single-seq.~folding\!\!       &   $n^3$   &  $n^2$ \\
         \hline
	CDSfold  & $96 n^3$ & $16 n^2$ \\
	LinearDesign & $8 n^3$  & $36 n^2$
\end{tabular}}}
\\[-5.2cm]
\hspace{-9cm}\panel{} & \hspace{-8cm}\panel{C} \\[-.8cm]
&
\hspace{-1.5cm}\raisebox{2.5cm}{\includegraphics[width=.450\linewidth]{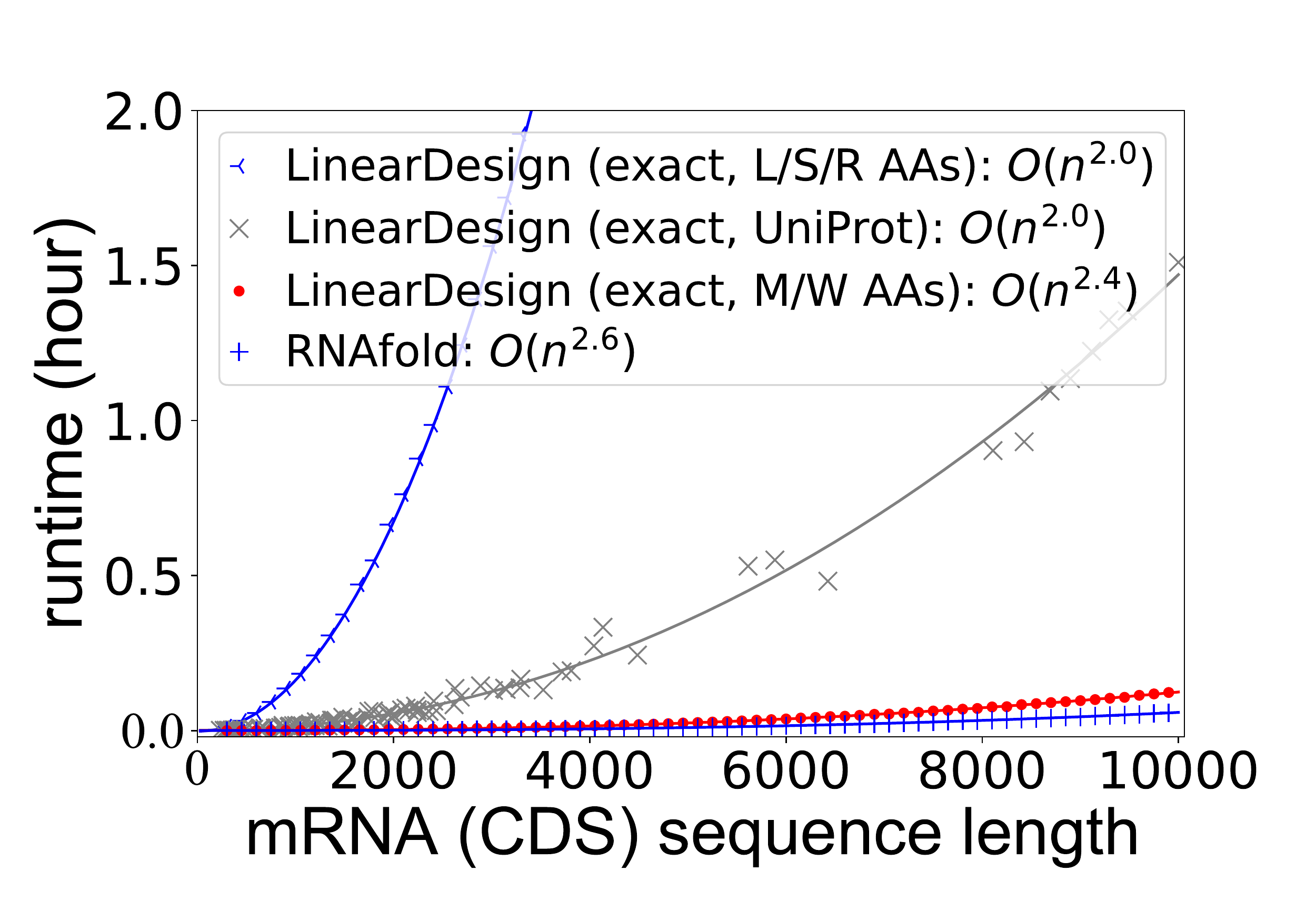}}
\\[-2.8cm]

\end{tabular}
\caption{More \textit{In silico} time complexity analysis. 
{\bf A}: Runtime comparison between \lineardesign and CDSfold on UniProt proteins. 
Overall, \lineardesign is substantially faster than CDSfold, 
and more importantly, \lineardesign scales $O(n^2)$ empirically, where $n$ is mRNA sequence length,
while CDSfold runs in $O(n^{3.1})$; 
this difference can be explained by the analysis in {\bf B}.
On the other hand, our MFE+CAI mode (with $\lambda = 3$)
is only slightly slower than our MFE-only version, 
while CDSfold cannot jointly optimize MFE and CAI.
{\bf B}: \lineardesign's cubic-time bifurcation rule is so efficient that it is dominated by the quadratic-time pairing rule in practice.
{\bf C}: Runtime comparison of \lineardesign on proteins with most ambiguous (6-codon) amino acids,
natural (UniProt) proteins, and proteins with unambiguous (1-codon) amino acids,
the last of which is equivalent to single sequence folding.
We also ran \rnafold on RNA sequences that encode the unambiguous amino acids.
See also Fig.~\ref{fig:insilico} for more \insilico results of \lineardesign.
\label{fig:LD_CDSfold_items_generated_table}
}
\end{figure}

\begin{figure}[!hbt]
\centering
\begin{tabular}{lll}
\hspace{0cm}\panel{A} & \hspace{-.7cm}\panel{B} & \hspace{-.7cm}\panel{C}\\[-.6cm]
\hspace{0cm}\includegraphics[width=.33\linewidth]{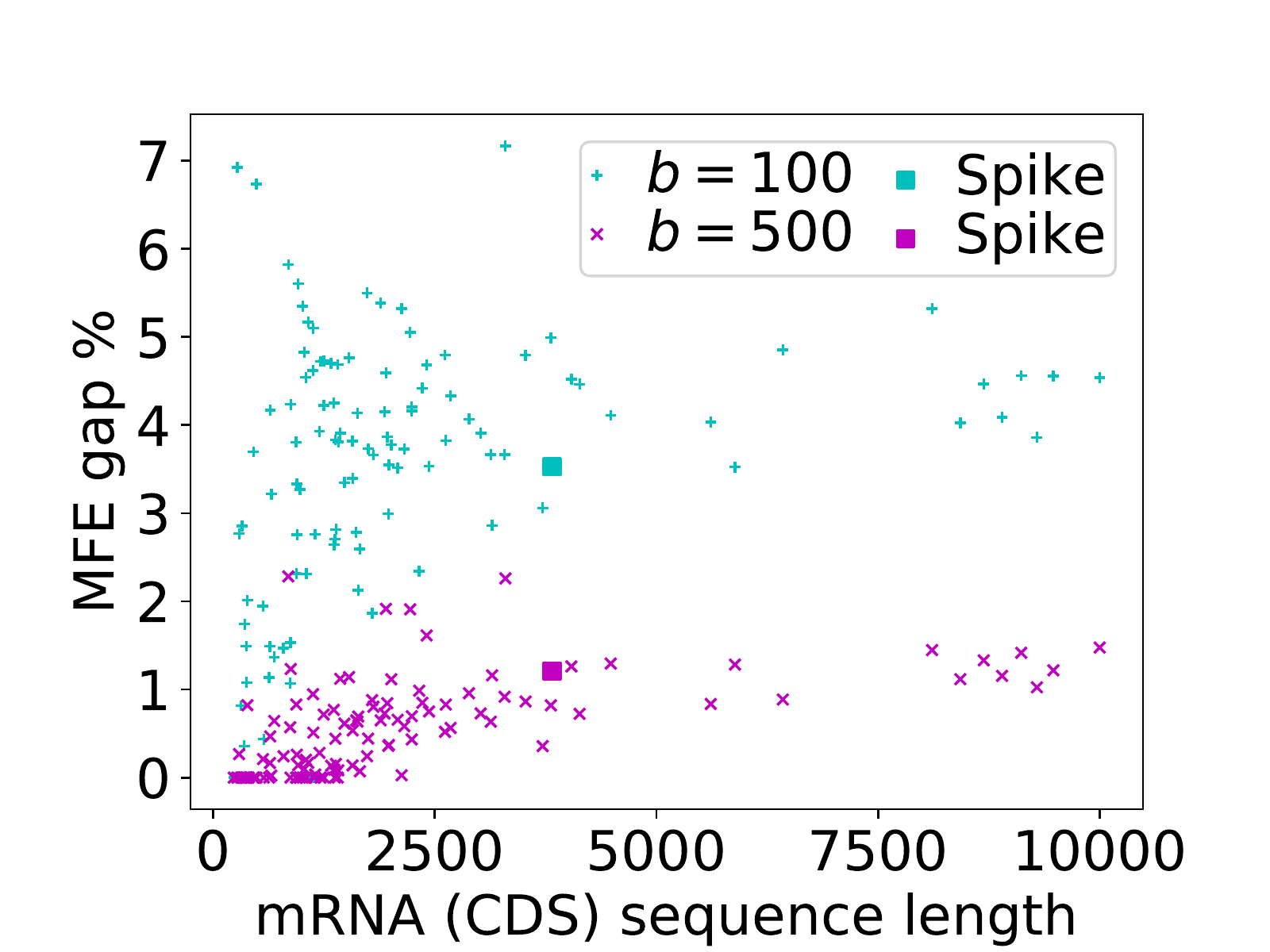} &\hspace{-.7cm}\includegraphics[width=.33\linewidth]{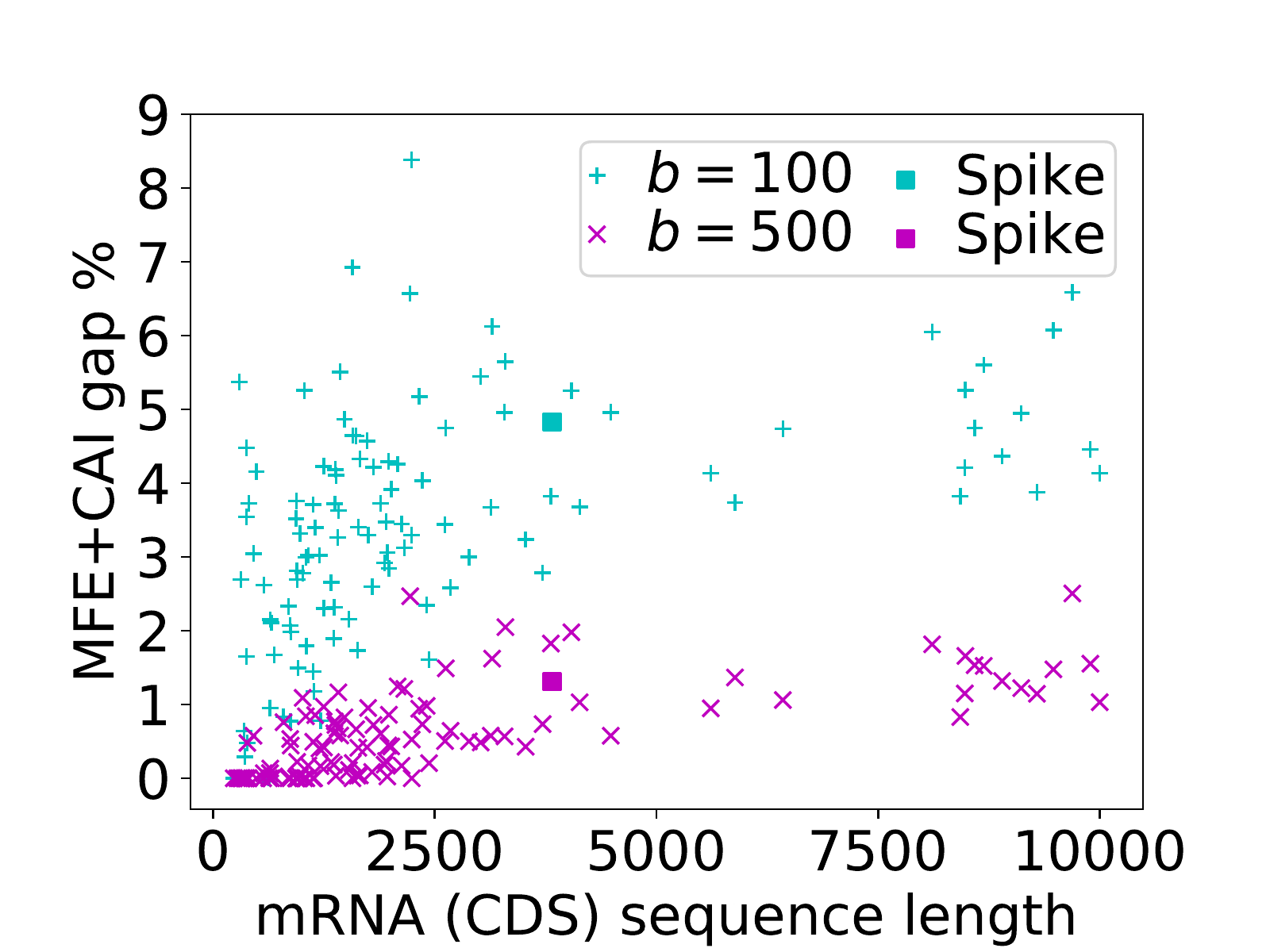} & \hspace{-.7cm}\includegraphics[width=.33\linewidth]{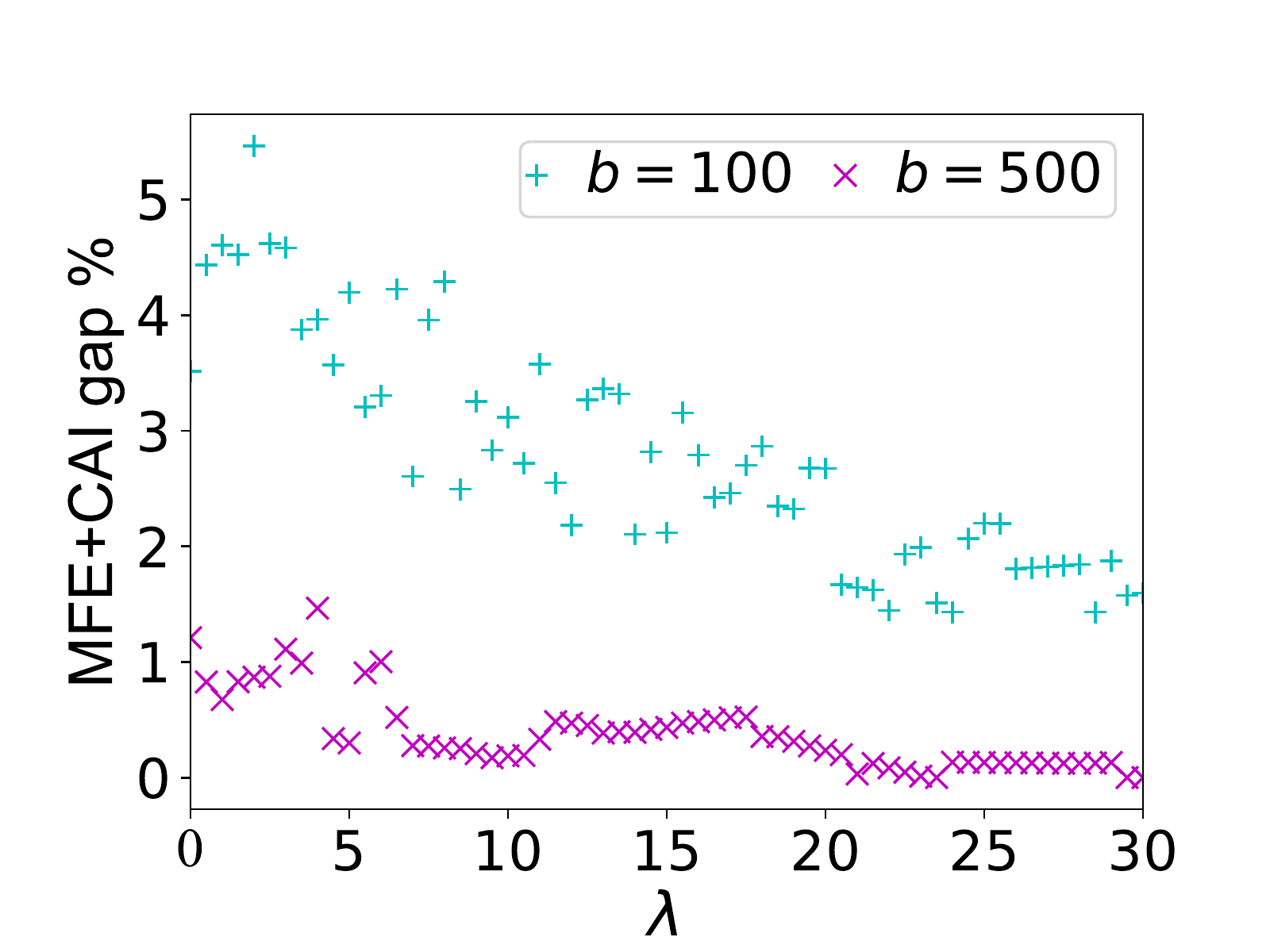} \\[-.2cm]

\end{tabular}
\caption{Search error of LinearDesign's beam search mode against sequence length and $\lambda$ (the weight of CAI in the joint optimization). 
{\bf A--B}: Search error is relatively small, and does not deteriorate with sequence length.
Here we used beam sizes 500 (purple) and 100 (cyan), on UniProt proteins (crosses) and SARS-CoV-2 Spike protein (squares),
and we use $\lambda=3$ for {\bf B}.
{\bf C}: Search error decreases with $\lambda$.
\myedit{Note that the search error in {\bf A} is the free energy gap \% for $\lambda=0$, defined as $1- \MFE(\vecr_\text{approx\_design}) / \MFE(\vecr_\text{exact\_design})$;
the search error in {\bf B--C}  is defined as $1- \MFECAIlambda (\vecr_\text{approx\_design}) / \MFECAIlambda (\vecr_\text{exact\_design})$,
where $\vecr_\text{approx\_design}$ and $\vecr_\text{exact\_design}$ are designed mRNAs from the beam search mode and the exact search mode, respectively.}
See also Fig.~\ref{fig:insilico} for more \insilico results.
  \label{fig:breakdown_gap}}
\end{figure}

\begin{figure}[!hbt]
\centering
\begin{tabular}{cc}

\hspace{-0.cm}\includegraphics[width=.5\linewidth]{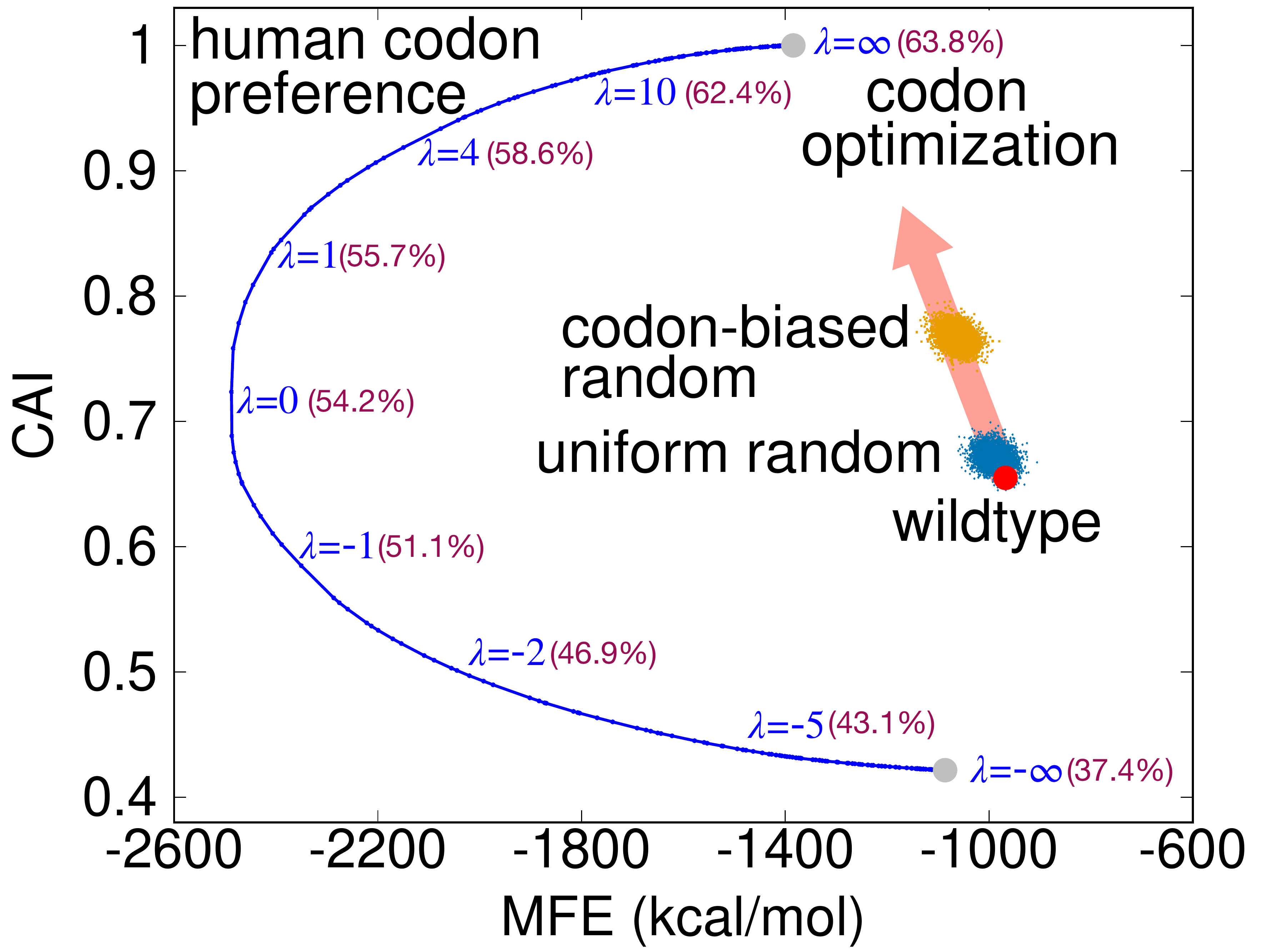} & 
\hspace{-.55cm}\includegraphics[width=.5\linewidth]{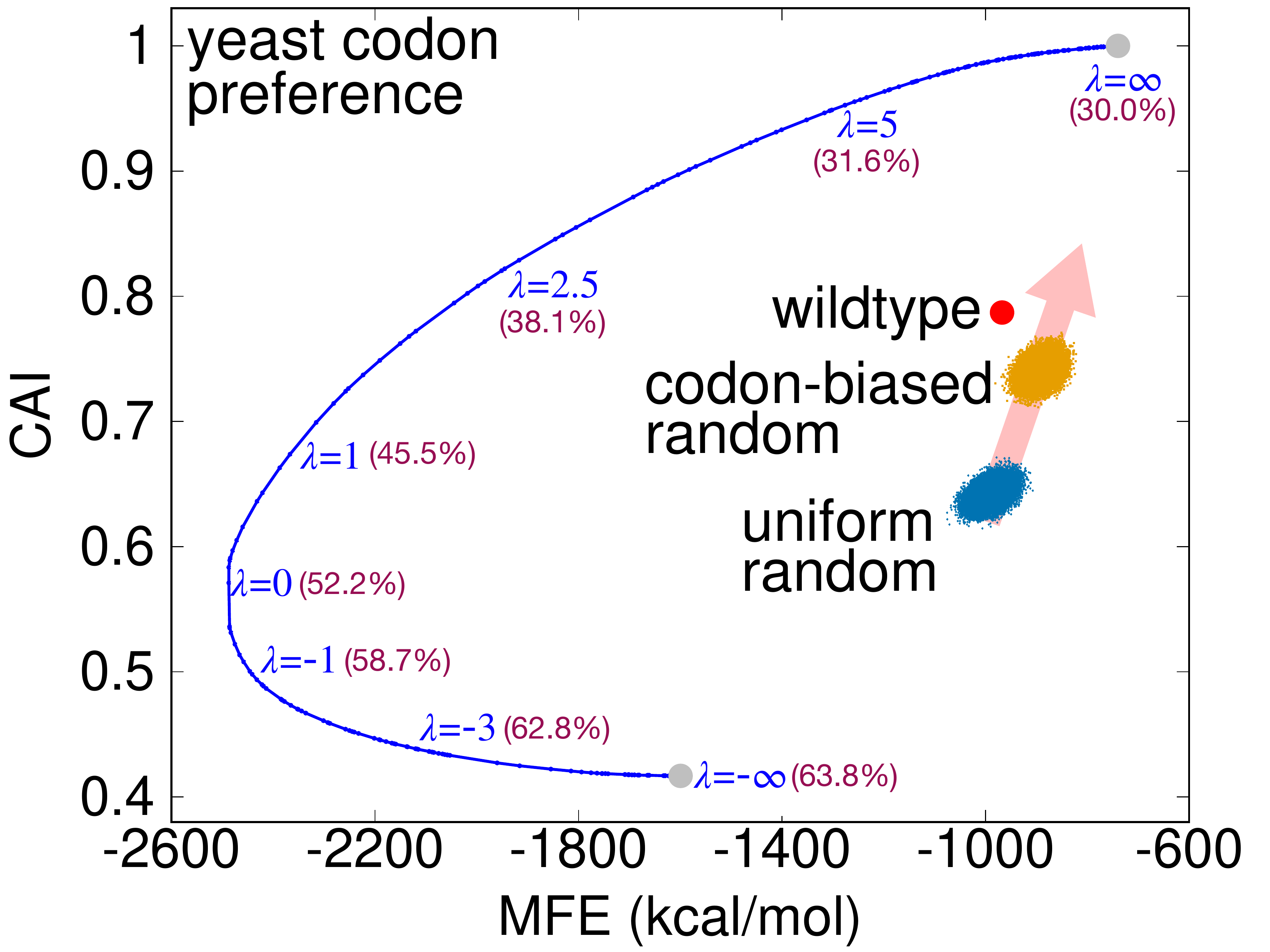}\\[-6.5cm]
\hspace{-7.3cm}\panel{A} & \hspace{-7.6cm}\panel{B} \\[5.5cm]
\end{tabular}
\caption{\MFE-\CAI two dimentional visualizations of Spike designs using human codon preference ({\bf A}) and yeast codon preference ({\bf B}) with positive and negative $\lambda$'s. GC\% are shown in parentheses. 
The human genome prefers GC-rich codons that leads to higher CAI designs are with higher GC\%, while the yeast genome prefers AU-rich codons that exhibits an opposite relationship between CAI and GC\%.
See also Fig.~\ref{fig:insilico} for more \insilico results of \lineardesign.
  \label{fig:human_yeast_codon_oval_full}}
\end{figure}

\begin{figure}[!hbt]
\centering
\hspace{.1cm}\includegraphics[width=1.03\linewidth]{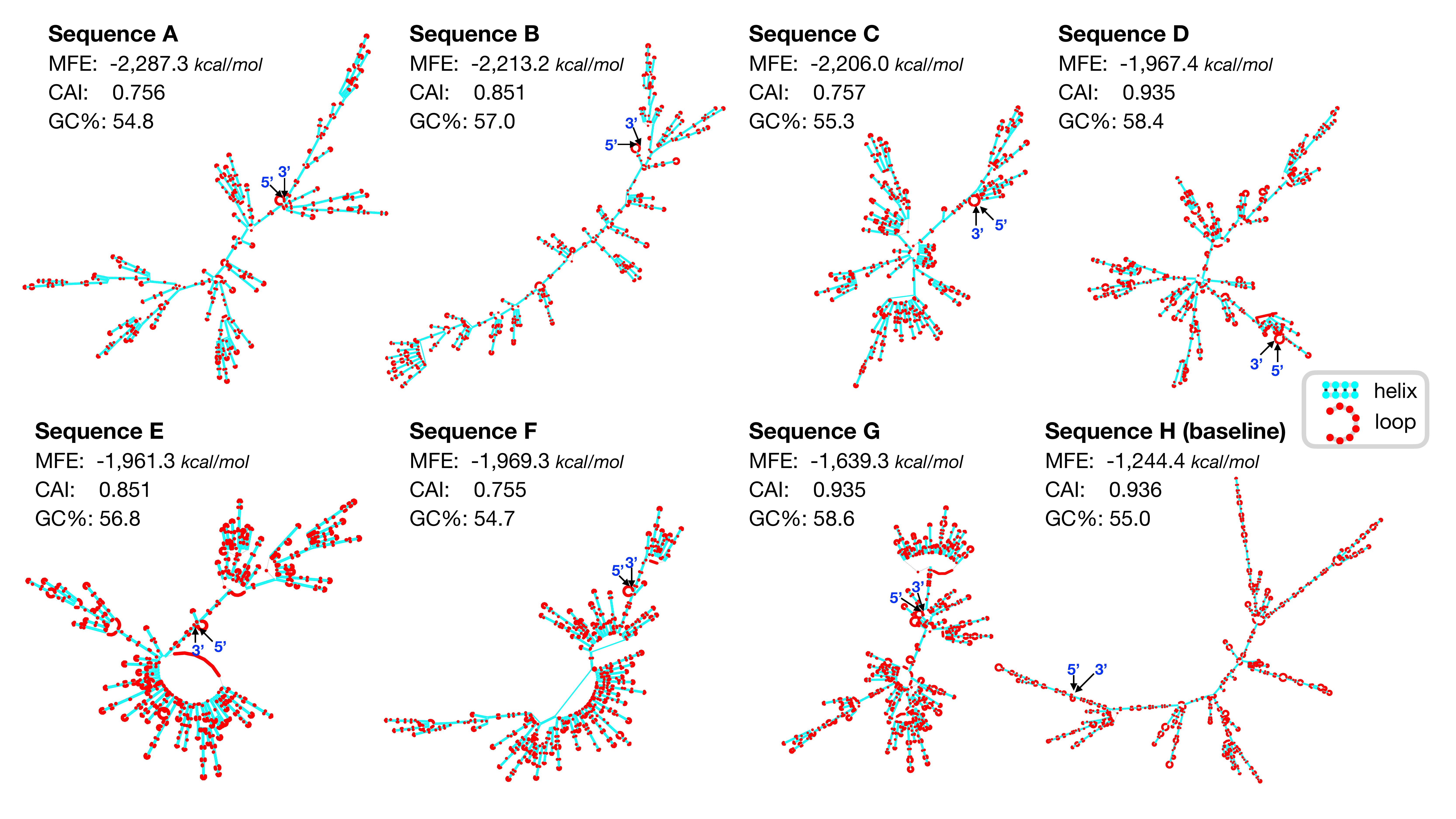}\\[-.2cm]
\caption{The secondary structures of \lineardesign-generated sequences ({\sc a}--{\sc g}) and the baseline sequence ({\sc h}) used in the wet lab experiments. 
The stable helices are in cyan and unstable loops are in red.
Sequences {\sc a--d} clearly have less and smaller loops, 
and they have higher levels of antibody responses compared to the baseline {\sc h} (see Fig.~\ref{fig:wetlab}E--G).
The secondary structures are predicted by \viennarnafold (-d0 mode) and visulized by RNAplot.
  \label{fig:seqA_H_structure}}
\end{figure}

\FloatBarrier
\input{tab_si_1}

\FloatBarrier

\input{tab_si_utrs}

\newpage

\ifarXiv
\else
\end{linenumbers}
\fi
\end{singlespace}

%% file: fig_word_lattice.tex

\tikzset{every picture/.style={/utils/exec={\sffamily}}}

\colorlet{DarkGray}{black!30}
\colorlet{ColorGray}{blue!30}

\begin{figure}[!htb]
\vspace{-.5cm}
\centering
\begin{tabular}{ccc}
\hspace{-5.5cm}{\panel{A}} &  & \hspace{-7.5cm}{\panel{D}}\\[-.8cm]
\hspace{.5cm}\resizebox{.38\textwidth}{!}{
\begin{tikzpicture}[->,>=stealth',shorten >=1pt,auto,semithick]
\node[state, initial, initial text=, inner sep=-10pt, fill=StartNode] (n0) {\fontsize{18.5}{0}\selectfont 0};
\node[state, right=1.3cm of n0, inner sep=-10pt, fill=NodeGray] (n1) {\fontsize{18.5}{0}\selectfont 1};
\node[state, right=1.3cm of n1, inner sep=-10pt, fill=NodeGray] (n2) {\fontsize{18.5}{0}\selectfont 2};
\node[state, right=1.3cm of n2, inner sep=-10pt, fill=NodeGray] (n3) {\fontsize{18.5}{0}\selectfont 3};
\node[state, below=.4cm of n2, inner sep=-10pt, fill=NodeGray] (n5) {\fontsize{18.5}{0}\selectfont 5};
\node[state, below=1.4cm of n1, inner sep=-10pt, fill=NodeGray] (n6) {\fontsize{18.5}{0}\selectfont 6};
\node[state, below=1.4cm of n3, inner sep=-10pt, fill=NodeGray] (n7) {\fontsize{18.5}{0}\selectfont 7};
\node[state, accepting, right=1.3cm of n3, inner sep=-10pt, fill=EndNode] (n4) {\fontsize{18.5}{0}\selectfont 4};
\draw (n0) edge[above] node{\fontsize{16.5}{0}\selectfont I} (n1)
(n1) edge[above] node{\fontsize{16.5}{0}\selectfont like} (n2)
(n2) edge[above] node{\fontsize{16.5}{0}\selectfont this} (n3)
(n3) edge[above] node{\fontsize{16.5}{0}\selectfont meal} (n4)
(n3) edge[above, bend right=40] node{\fontsize{16.5}{0}\selectfont veal} (n4)
(n0) edge[below] node{\fontsize{16.5}{0}\selectfont alike} (n5)
(n5) edge[above] node{\fontsize{16.5}{0}\selectfont his} (n7)
(n0) edge[below] node{\fontsize{16.5}{0}\selectfont alike \ \ } (n6)
(n6) edge[below] node{\fontsize{16.5}{0}\selectfont this} (n7)
(n7) edge[below] node{\fontsize{16.5}{0}\selectfont \ \ veal} (n4);
\end{tikzpicture}}
&
\hspace{-3.cm}
\raisebox{-2.9cm}{
\resizebox{.24\textwidth}{!}{
\centering
\begin{forest}
  for tree={
  	s sep=2.em,    
  	l = -.1em,
    if n children=0{
      tier=terminal,
      s sep=2em, 
    }{},
  }
  [\fontsize{20.5}{0}\selectfont{\it S}
    [\fontsize{20.5}{0}\selectfont{\it NP}
      [\fontsize{20.5}{0}\selectfont{\it PRP}
      	[\fontsize{20.5}{0}\selectfont{I}
      	]
      ]
    ]
    [\fontsize{20.5}{0}\selectfont{\it VP}
		  [\fontsize{20.5}{0}\selectfont{\it VB}
		  	[\fontsize{20.5}{0}\selectfont{like}
		  	]
		  ]
		  [\fontsize{20.5}{0}\selectfont{\it NP}
		  	[\fontsize{20.5}{0}\selectfont{\it DT}
		  		[\fontsize{20.5}{0}\selectfont{this}
		  		]
		  	]
		  	[\fontsize{20.5}{0}\selectfont{\it NN}
		  		[\fontsize{20.5}{0}\selectfont{meal}
		  		]
		  	]
		  ]
    ]
  ]
\end{forest}}}
&
\hspace{-1.5cm}\resizebox{.5\textwidth}{!}{
\raisebox{-6cm}{
\begin{tikzpicture}[->,>=stealth',shorten >=1pt,auto,semithick,
no arrow/.style={-,every loop/.append style={-}}]
\node[state, initial, initial text=, inner sep=-10pt, fill=StartNode] (n0) {\fontsize{18.5}{0}\selectfont 0};
\node[state, right=1.3cm of n0, inner sep=-10pt, fill=NodeGray] (n1) {\fontsize{18.5}{0}\selectfont 1};
\node[state, right=1.3cm of n1, inner sep=-10pt, fill=NodeGray] (n2) {\fontsize{18.5}{0}\selectfont 2};
\node[state, right=1.3cm of n2, inner sep=-10pt, fill=NodeGray] (n3) {\fontsize{18.5}{0}\selectfont 3};
\node[state, below=.4cm of n2, DarkGray, inner sep=-10pt, fill=NodeGray] (n5) {\fontsize{18.5}{0}\selectfont 5};
\node[state, below=1.4cm of n1, DarkGray, inner sep=-10pt, fill=NodeGray] (n6) {\fontsize{18.5}{0}\selectfont 6};
\node[state, below=1.4cm of n3, DarkGray, inner sep=-10pt, fill=NodeGray] (n7) {\fontsize{18.5}{0}\selectfont 7};
\node[state, accepting, right=1.3cm of n3, inner sep=-10pt, fill=EndNode] (n4) {\fontsize{18.5}{0}\selectfont 4};
\draw (n0) edge[above] node{I} (n1)
(n1) edge[above] node{like} (n2)
(n2) edge[above] node{this} (n3)
(n3) edge[above] node{meal} (n4)
(n3) edge[above, bend right=40, DarkGray] node{veal} (n4)
(n0) edge[below, DarkGray] node{alike} (n5)
(n5) edge[above, DarkGray] node{his} (n7)
(n0) edge[below, DarkGray] node{alike \ \ } (n6)
(n6) edge[below, DarkGray] node{this} (n7)
(n7) edge[below, DarkGray] node{\ \ veal} (n4);
\draw (n0) edge[no arrow,dashed,above,bend left=40,blue] node{\fontsize{12.5}{0}\selectfont {\it{PRP}}} (n1)
(n2) edge[no arrow,dashed,above,bend left=90,blue] node{\fontsize{12.5}{0}\selectfont {\it{DT}}} (n3)
(n0) [no arrow,dashed,blue] to[out=90, in=180]++ (1.,1.5) to[out=0,in=90] node[pos=0,above, sloped] {\fontsize{12.5}{0}\selectfont {\it{NP}}} (n1)
(n3) edge[no arrow,dashed,above,bend left=90,blue] node{\fontsize{12.5}{0}\selectfont {\it{NN}}} (n4)
(n1) edge[no arrow,dashed,above,bend left=90,blue] node{\fontsize{12.5}{0}\selectfont {\it{VB}}} (n2)
(n2) edge[no arrow,dashed,above,bend left=90,blue] node{\fontsize{12.5}{0}\selectfont {\it{NP}}} (n4)
(n1) edge[no arrow,dashed,above,bend left=90,blue] node{\fontsize{12.5}{0}\selectfont {\it{VP}}} (n4)
(n0) edge[no arrow,dashed,above,bend left=90,blue] node{\fontsize{12.5}{0}\selectfont {\it{S}}} (n4);
%
\draw (n0) edge[no arrow,dashed,below,bend right=40,ColorGray] node{\fontsize{12.5}{0}\selectfont {\it{JJ}}} (n6)
(n6) edge[no arrow,dashed,below,bend right=25,ColorGray] node{\fontsize{12.5}{0}\selectfont {\it{DT}}} (n7)
(n7) edge[no arrow,dashed,below,bend right=35,ColorGray] node{\fontsize{12.5}{0}\selectfont {\it{NN}}} (n4)
(n6) edge[no arrow,dashed,below,bend right=80,ColorGray] node{\fontsize{12.5}{0}\selectfont {\it{NP}}} (n4);
\end{tikzpicture}}}
\\[-4.cm]
\hspace{-6.9cm}{\panel{B}} & \hspace{-6.5cm}{\panel{C}} \\[-.4cm]
\hspace{-3.cm}
\raisebox{-1.cm}{
\resizebox{.25\textwidth}{!}{
\begin{tikzpicture}[->,>=stealth',shorten >=1pt,auto,semithick]
\node(n00) {\fontsize{20}{0}\selectfont {\it{S}}};
\node[right=.8cm of n00] (n02) {\fontsize{20}{0}\selectfont {\goesto}};
\node[right=.5cm of n02] (n01) {\fontsize{20}{0}\selectfont {\it{NP}} \ {\it{VP}}};
\node[below=.1cm of n00] (n10) {\fontsize{20}{0}\selectfont {\it{NP}}};
\node[below=.31cm of n02] (n12) {\fontsize{20}{0}\selectfont {\goesto}};
\node[right=.5cm of n12] (n11) {\fontsize{20}{0}\selectfont {\it{PRP}}
													\ | \ \fontsize{20}{0}\selectfont {\it{DT}} \ {\it{NN}}};
\node[below=.1cm of n10] (n20) {\fontsize{20}{0}\selectfont {\it{VP}}};
\node[below=.31cm of n12] (n22) {\fontsize{20}{0}\selectfont {\goesto}};
\node[right=.5cm of n22] (n21) {\fontsize{20}{0}\selectfont {\it{VB}} \ {\it{NP}}};
\node[below=.1cm of n20] (n30) {\fontsize{20}{0}\selectfont {\it{PRP}}};
\node[below=.31cm of n22] (n32) {\fontsize{20}{0}\selectfont {\goesto}};
\node[right=.5cm of n32] (n31) {\fontsize{22}{0}\selectfont I};
\node[below=.1cm of n30] (n40) {\fontsize{20}{0}\selectfont {\it{DT}}};
\node[below=.31cm of n32] (n42) {\fontsize{20}{0}\selectfont {\goesto}};
\node[right=.5cm of n42] (n41) {\fontsize{22}{0}\selectfont this};
\node[below=.1cm of n40] (n50) {\fontsize{20}{0}\selectfont {\it{NN}}};
\node[below=.31cm of n42] (n52) {\fontsize{20}{0}\selectfont {\goesto}};
\node[right=.5cm of n52] (n51) {\fontsize{22}{0}\selectfont meal \ | \ veal};
\node[below=.1cm of n50] (n60) {\fontsize{20}{0}\selectfont {\it{VB}}};
\node[below=.31cm of n52] (n62) {\fontsize{20}{0}\selectfont {\goesto}};
\node[right=.5cm of n62] (n61) {\fontsize{22}{0}\selectfont like};
\node[below=.1cm of n60] (n80) {\fontsize{20}{0}\selectfont {\it{JJ}}};
\node[below=.31cm of n62] (n82) {\fontsize{20}{0}\selectfont {\goesto}};
\node[right=.5cm of n82] (n81) {\fontsize{22}{0}\selectfont alike};
\node[below=.5cm of n80] (n70) {};
\node[below=.31cm of n82] (n71) {\fontsize{22}{0}\selectfont ...};
\end{tikzpicture}}}
&
\hspace{-2.8cm}
\raisebox{-.5cm}{
\resizebox{.29\textwidth}{!}{
\begin{tikzpicture}[->,>=stealth',shorten >=1pt,auto,semithick,
no arrow/.style={-,every loop/.append style={-}}]
    \node(00) at (0,0) {\fontsize{8}{0}\selectfont 0};
    \node(11) at (.9,0) {\fontsize{8}{0}\selectfont 1};
    \node(22) at (1.9,0) {\fontsize{8}{0}\selectfont 2};
    \node(33) at (2.9,0) {\fontsize{8}{0}\selectfont 3};
    \node(44) at (4,0) {\fontsize{8}{0}\selectfont 4};
\end{tikzpicture}}}
    \\[-.1cm]
\end{tabular}
\caption{The word lattice parsing in natural languague processing. 
{\bf A}: An example of word lattice (sentence DFA) for speech recognition.
{\bf B}: Simplified language grammar.
{\bf C}: Single sentence parsing with between-word indices, which is a special case of word lattice parsing.
{\bf D}: Illustration of word lattice parsing for speech recognition with given word lattice and language grammar;
the dashed blue arcs above
the DFA depict the best parsing structure for the optimal sentence ``{\sf {I like this meal}}'',
while the dashed light-blue arcs below the DFA represent the best parsing structure for a non-optimal sentence ``{\sf {alike this veal}}''.
See also Fig.~\ref{fig:overview}.
\label{fig:word_lattice}
}
\end{figure}

%% file: fig_dfa_cfg_20220209.tex

\usetikzlibrary{positioning, calc, automata, arrows, shapes, backgrounds, quotes}
\tikzset{every picture/.style={/utils/exec={\sffamily}}}

\colorlet{DarkGray}{gray!30}
\colorlet{DarkerGray}{black!60}
\colorlet{ColorGray}{blue!45}

\colorlet{StartNode}{purple!20}
\colorlet{NodeGray}{gray!20}
\colorlet{EndNode}{purple!20}

\begin{figure}[!t]
\vspace{-0.2cm}
\centering
\begin{tabular}{cccc}

&&\raisebox{-0cm}{\hspace{-3.65cm}\includegraphics[width=.455\textwidth]{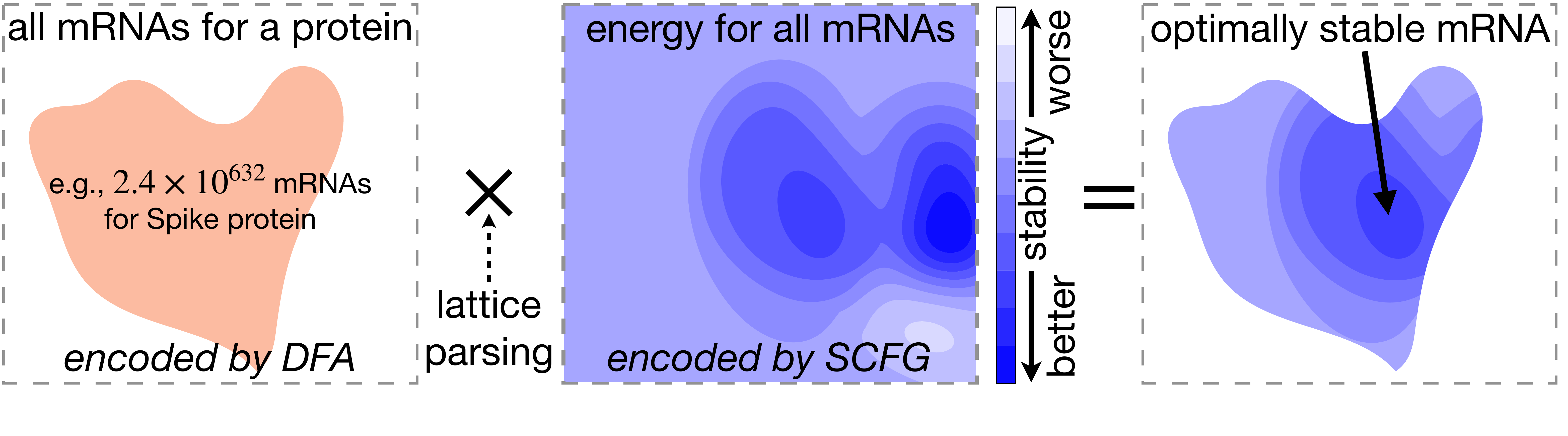}} \raisebox{0cm}{\hspace{.2cm}\includegraphics[width=.455\textwidth]{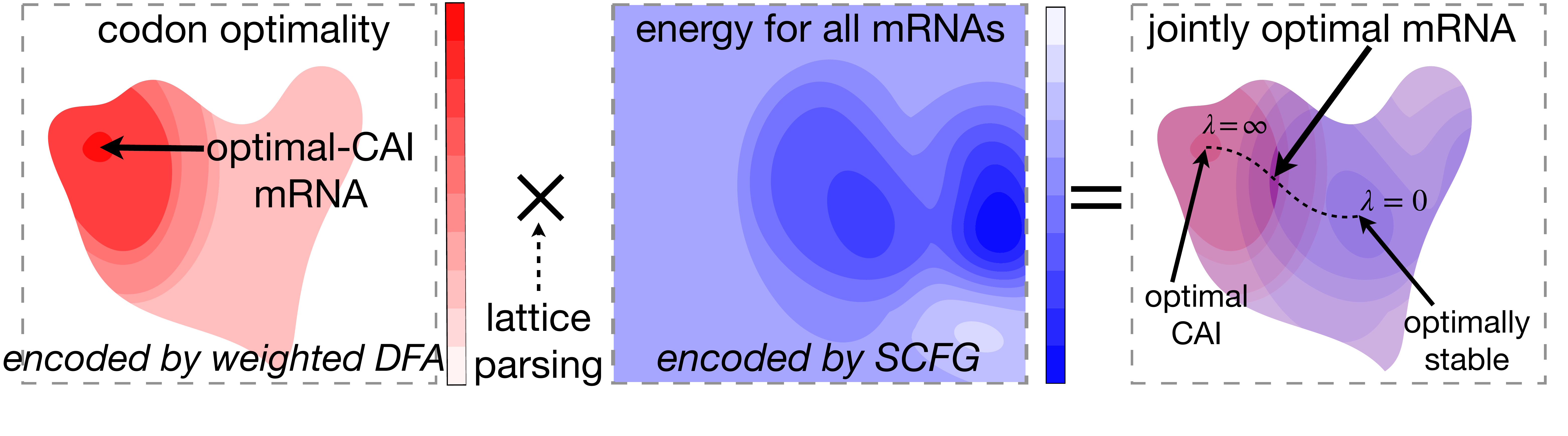}}\\[-2.2cm]

&&\raisebox{0cm}{\hspace{-11.3cm}\panel{A}}  \raisebox{0cm}{\hspace{7.16cm}\panel{B}}\\[1.45cm]

\hspace{0.1cm}\raisebox{2.1cm}{\panel{\myedit{C}} \fontsize{9}{0}\selectfont \textbf{\myedit{\ \ \ codon DFAs}}}
& \hspace{-3.7cm}\raisebox{1.cm}{\resizebox{.25\textwidth}{!}{
\begin{tikzpicture}[->,>=stealth',shorten >=1pt,auto,semithick]
\node[state, initial, initial text=, inner sep=-10pt, fill=StartNode] (n00) {\fontsize{16.5}{0}\selectfont 0,0};
\node[state, right=.8cm of n00, inner sep=-10pt, fill=NodeGray] (n10) {\fontsize{16.5}{0}\selectfont 1,0};
\node[state, right=.8cm of n10, inner sep=-10pt, fill=NodeGray] (n20) {\fontsize{16.5}{0}\selectfont 2,0};
\node[state, accepting, right=.8cm of n20, inner sep=-10pt, fill=EndNode] (n30) {\fontsize{16.5}{0}\selectfont 3,0};
\node[below=-1.6cm of n00, inner sep=-10pt] (n0a) {}; 
\node[right=1.8cm of n0a] (n0b) {$D(\text{valine})$}; 
\draw (n00) edge[above] node[violet]{G} (n10)
(n10) edge[above] node[gray]{U} (n20)
(n20) edge[above, bend left=85] node[red]{A} (n30)
(n20) edge[above, bend left=40] node[blue]{C} (n30)
(n20) edge[below] node[violet, above]{G} (n30)
(n20) edge[below, bend right=40] node[gray, above]{U} (n30);
\end{tikzpicture}}}

\hspace{-4.15cm}\raisebox{-.8cm}{\resizebox{.25\textwidth}{!}{
\begin{tikzpicture}[->,>=stealth',shorten >=1pt,auto,semithick]
\node[state, initial, initial text=, inner sep=-10pt, fill=StartNode] (n00) {\fontsize{16.5}{0}\selectfont 0,0};
\node[state, right=.8cm of n00, inner sep=-10pt, fill=NodeGray] (n10) {\fontsize{16.5}{0}\selectfont 1,0};
\node[state, below=0.2cm of n10, inner sep=-10pt, fill=NodeGray] (n11) {\fontsize{16.5}{0}\selectfont 1,1};
\node[state, right=.8cm of n10, inner sep=-10pt, fill=NodeGray] (n20) {\fontsize{16.5}{0}\selectfont 2,0};
\node[state, below=0.2cm of n20, inner sep=-10pt, fill=NodeGray] (n21) {\fontsize{16.5}{0}\selectfont 2,1};
\node[state, accepting, right=.8cm of n20, inner sep=-10pt, fill=EndNode] (n30) {\fontsize{16.5}{0}\selectfont 3,0};
\node[below=-1.6cm of n00, inner sep=-10pt] (n0a) {}; 
\node[right=1.8cm of n0a] (n0b) {$D(\text{serine})$}; 
\draw (n00) edge[above] node[gray]{U} (n10)
(n00) edge[above] node[red]{A} (n11)
(n10) edge[above] node[blue]{C} (n20)
(n11) edge[above] node[violet]{G} (n21)
(n20) edge[above, bend left=65,inner sep=2pt] node[red]{A} (n30)
(n20) edge[above, bend left=30,inner sep=2pt] node[blue]{C} (n30)
(n20) edge[below,inner sep=2pt] node[violet, above]{G} (n30)
(n20) edge[below, bend right=30,inner sep=2pt] node[gray, above]{U} (n30)
(n21) edge[bend right=30] node[blue, above]{C} (n30)
(n21) edge[below, bend right=60, right=0.3] node[gray, above]{U} (n30);
\end{tikzpicture}}}

& \hspace{-.5cm}\raisebox{2.1cm}{\panel{\myedit{D}} \fontsize{9}{0}\selectfont \textbf{\myedit{\qquad mRNA DFA}}} 
& \hspace{-12cm}\raisebox{-.8cm}{\resizebox{.7\textwidth}{!}{
\begin{tikzpicture}[->,>=stealth',shorten >=1pt,auto,semithick]
\node[state, initial, initial text=, inner sep=-10pt, fill=StartNode] (n00) {\fontsize{15.5}{0}\selectfont 0,0};
\node[below=2.2cm of n00] (n0b) {$\mid$}; 

\node[state, right=.9cm of n00, inner sep=-10pt, fill=NodeGray] (n10) {\fontsize{15.5}{0}\selectfont 1,0};
\node[state, right=.9cm of n10, inner sep=-10pt, fill=NodeGray] (n20) {\fontsize{15.5}{0}\selectfont 2,0};
\node[state, right=.9cm of n20, inner sep=-10pt, fill=NodeGray] (n30) {\fontsize{15.5}{0}\selectfont 3,0};

\node[below=2.2cm of n30] (n3b) {$\mid$}; 
\node[below=2.6cm of n30] (n3c) {$\circ$}; 

\node[state, right=.9cm of n30, inner sep=-10pt, fill=NodeGray] (n40) {\fontsize{15.5}{0}\selectfont 4,0};
\node[state, below=1.2cm of n40, inner sep=-10pt, fill=NodeGray] (n41) {\fontsize{15.5}{0}\selectfont 4,1};
\node[state, right=.9cm of n40, inner sep=-10pt, fill=NodeGray] (n50) {\fontsize{15.5}{0}\selectfont 5,0};
\node[state, below=1.2cm of n50, inner sep=-10pt, fill=NodeGray] (n51) {\fontsize{15.5}{0}\selectfont 5,1};
\node[state, right=.9cm of n50, inner sep=-10pt, fill=NodeGray] (n60) {\fontsize{15.5}{0}\selectfont 6,0};

\node[below=2.2cm of n60] (n6b) {$\mid$}; 
\node[below=2.6cm of n60] (n36) {$\circ$}; 

\node[state, right=.9cm of n60, inner sep=-10pt, fill=NodeGray] (n70) {\fontsize{15.5}{0}\selectfont 7,0};
\node[state, right=.9cm of n70, inner sep=-10pt, fill=NodeGray] (n80) {\fontsize{15.5}{0}\selectfont 8,0};
\node[state, below=1.2cm of n80, inner sep=-10pt, fill=NodeGray] (n81) {\fontsize{15.5}{0}\selectfont 8,1};
\node[state, accepting, right=.9cm of n80, inner sep=-10pt, fill=EndNode] (n90) {\fontsize{15.5}{0}\selectfont 9,0};

\node[below=2.2cm of n90] (n9b) {$\mid$}; 

\draw (n0b) edge[below,<->,dashed] node{$D(\text{methionine})$} (n3b);
\draw (n3b) edge[below,<->,dashed] node{$D(\text{leucine})$} (n6b);
\draw (n6b) edge[below,<->,dashed] node{$D(\textsc{stop})$} (n9b);

\draw (n00) edge[above,line width=1.3pt] node[red]{A} (n10)
(n10) edge[above,line width=1.3pt] node[gray]{U} (n20)
(n20) edge[above,line width=1.3pt] node[violet]{G} (n30);

\draw (n30) edge[above] node[gray]{U} (n40)
(n30) edge[above,line width=1.3pt] node[blue]{C} (n41)
(n40) edge[above] node[gray]{U} (n50)
(n41) edge[above,line width=1.3pt] node[gray]{U} (n51)
(n50) edge[above, bend left=40] node[red]{A} (n60)
(n50) edge[above] node[violet]{G} (n60)
(n51) edge[below, bend left=20,line width=1.3pt] node[red]{A} (n60)
(n51) edge[below, bend right=10,line width=1.3pt] node[blue]{C} (n60)
(n51) edge[below, bend right=40,line width=1.3pt] node[violet]{G} (n60)
(n51) edge[below, bend right=70,line width=1.3pt] node[gray]{U} (n60);
\draw (n60) edge[above,line width=1.3pt] node[gray]{U} (n70)
(n70) edge[above] node[red]{A} (n80)
(n70) edge[above,line width=1.3pt] node[violet]{G} (n81)
(n80) edge[above, bend left=40] node[red]{A} (n90)
(n80) edge[above] node[violet]{G} (n90)
(n81) edge[below,line width=1.3pt] node[red]{A} (n90);
\end{tikzpicture}}}\\[.5cm]
\end{tabular}
\\[.1cm]

%
\begin{tabular}{cccc}
\raisebox{.1cm}{\hspace{-.25cm}{\panel{\myedit{E}} \fontsize{9}{0}\selectfont \textbf{\myedit{stochastic grammar}}}} & \raisebox{.1cm}{\hspace{-.8cm}{\raisebox{0cm}{\panel{\myedit{F}} \fontsize{9}{0}\selectfont \textbf{\myedit{single-sequence folding}}}}}& \raisebox{.1cm}{\hspace{-1cm}{\panel{\myedit{G}} \fontsize{9}{0}\selectfont \textbf{\myedit{from\! single-sequence\! folding\! to\! lattice\! parsing}}}} \\[-.2cm]
\hspace{-.5cm}
\raisebox{.7cm}{
\resizebox{.2\textwidth}{!}{
\begin{tikzpicture}[scale = 1]
\node [draw=none, fill=none] (AA) at (0,0) {$S \stackrel{0}{\goesto} S \  P \mid S \ N \mid P$};
\node [draw=none, fill=none, below=of AA.west, anchor=west] (A)  {$S \stackrel{0}{\goesto}  N \ P \mid N \ N \ P$};
\node [draw=none, fill=none, below=of A.west, anchor=west] (B) {$S \stackrel{0}{\goesto} N \ N\ N$};
\node [draw=none, fill=none, below=of B.west, anchor=west] (C) {$P \stackrel{-3}{\goesto} \textblue \nucC \  S \  \textviolet  \nucG \mid \textviolet \nucG \  S \  \textblue \nucC$};
\node [draw=none, fill=none, below=of C.west, anchor=west] (D) {$P \stackrel{-2}{\goesto} \textred \nucA \  S \  \textGray \nucU \mid \textGray \nucU \  S \  \textred \nucA$};
\node [draw=none, fill=none, below=of D.west, anchor=west] (E) {$P \stackrel{-1}{\goesto} \textviolet \nucG \  S \  \textGray \nucU \mid \textGray \nucU \  S \  \textviolet \nucG$};
\node [draw=none, fill=none, below=of E.west, anchor=west] (F) {$N \stackrel{0}{\goesto} \textred \nucA \  \mid  \textblue \nucC \  \mid  \textviolet \nucG \  \mid \textGray \nucU$};
\node[left=-1.2cm of C] (npair_left) {};
\node[left=-2.2cm of C] (npair_right) {};
\draw (npair_left) edge[above, dotted, bend left = 50,  line width=1pt, ] node{} (npair_right);
\node[left=-2.55cm of C] (npair_left) {};
\node[left=-3.55cm of C] (npair_right) {};
\draw (npair_left) edge[above, dotted, bend left = 50,  line width=1pt, ] node{} (npair_right);
\node[left=-1.2cm of D] (npair_left) {};
\node[left=-2.2cm of D] (npair_right) {};
\draw (npair_left) edge[above, dotted, bend left = 50,  line width=1pt, ] node{} (npair_right);
\node[left=-2.55cm of D] (npair_left) {};
\node[left=-3.55cm of D] (npair_right) {};
\draw (npair_left) edge[above, dotted, bend left = 50,  line width=1pt, ] node{} (npair_right);
\node[left=-1.2cm of E] (npair_left) {};
\node[left=-2.2cm of E] (npair_right) {};
\draw (npair_left) edge[above, dotted, bend left = 50,  line width=1pt, ] node{} (npair_right);
\node[left=-2.55cm of E] (npair_left) {};
\node[left=-3.55cm of E] (npair_right) {};
\draw (npair_left) edge[above, dotted, bend left = 50,  line width=1pt, ] node{} (npair_right);
\end{tikzpicture}
}}
&

\hspace{-1.5cm}
\raisebox{.4cm}{

\hspace{.4cm}
\raisebox{2.6cm}{
\resizebox{.3\textwidth}{!}{
\centering
\footnotesize
\hspace{.5cm}
\newcommand{\initial}[1]{\ensuremath{\alpha_{\textrm{\scriptsize #1}}}} 
\newcommand{\auxiliary}[1]{\ensuremath{\beta_{\textrm{\scriptsize #1}}}} 

\begin{tikzpicture}[level distance=20pt,sibling distance=6pt, level 1/.style={sibling distance=6pt, level distance=15pt}, level 2/.style={sibling distance=6pt, level distance=22pt}, level 3/.style={sibling distance=6pt, level distance=15pt}, level 6/.style={sibling distance=6pt, level distance=16pt}] 

\fill [orange!25] (-.6,-4.05) -- (-.4,-1.5) -- (.9,-.95) -- (3.7,-1.5) -- (3.9,-4.05) -- cycle;
\fill [red!25] (-.4,-4.05) -- (-.195,-2.) -- (.02,-4.05) -- cycle;
\fill [blue!25] (.15,-4.05) -- (.35,-2.3) -- (2,-1.75) -- (3.65,-2.3) -- (3.8,-4.05) -- cycle;

\Tree 
[.\node(a){$S$}; 
  [[[[[.\node (b){$N$};
  	\node(b1){\nucA};  
  ]]]]] 
  [.\node(c){$P$}; 
  	[[[[\node(c1){\nucU}; ]]]]
    [.\node(d){$S$};  
      [[[.\node (e){$N$};
  		\node(e1){\nucG};  
  	  ]]] 
  	  [.\node(f){$P$};  
  	  	[[\node(f1){\nucC};]] 
  	  	[.\node(g){$S$};  
  	  		[.\node (h){$N$};
		  	  \node(h1){\nucU};  
		  	] 
		  	[.\node (i){$N$};
		  	  \node(i1){\nucG};  
		  	] 
		  	[.\node (j){$N$};
		  	  \node(j1){\nucU};  
		  	] 
  	  	]
  	  	[[\node(f2){\nucG}; ]]
  	  ]
    ]  
    [[[[\node(c2){\nucA}; ]]]]
  ] 
]
\end{tikzpicture}
}}

\hspace{-4.86cm}
\raisebox{1.67cm}{
\resizebox{.285\textwidth}{!}{
\begin{tikzpicture}[->,>=stealth',shorten >=1pt,auto,semithick, no arrow/.style={-,every loop/.append style={-}}]
    \node(00) at (-.5,0.35) {\fontsize{8}{0}\selectfont 0};
    \node(11) at (-.1,0.35) {\fontsize{8}{0}\selectfont 1};
    \node(22) at (.4,0.35) {\fontsize{8}{0}\selectfont 2};
    \node(33) at (.95,0.35) {\fontsize{8}{0}\selectfont 3};
    \node(44) at (1.45,0.35) {\fontsize{8}{0}\selectfont 4};
    \node(55) at (2.05,0.35) {\fontsize{8}{0}\selectfont 5};
    \node(66) at (2.58,0.35) {\fontsize{8}{0}\selectfont 6};
    \node(77) at (3.09,0.35) {\fontsize{8}{0}\selectfont 7};
    \node(88) at (3.59,0.35) {\fontsize{8}{0}\selectfont 8};
    \node(99) at (4.,0.35) {\fontsize{8}{0}\selectfont 9};

    \node(00) at (-.32,.1) {\fontsize{8}{0}\selectfont $\bullet$};
    \node(11) at (.2,.1) {\fontsize{8}{0}\selectfont {\textbf (}};
    \node(22) at (.7,.1) {\fontsize{8}{0}\selectfont $\bullet$};
    \node(33) at (1.2,.1) {\fontsize{8}{0}\selectfont {\textbf (}};
    \node(44) at (1.75,.1) {\fontsize{8}{0}\selectfont $\bullet$};
    \node(55) at (2.3,.1) {\fontsize{8}{0}\selectfont $\bullet$};
    \node(66) at (2.85,.1) {\fontsize{8}{0}\selectfont $\bullet$};
    \node(77) at (3.35,.1) {\fontsize{8}{0}\selectfont {\textbf )}};
    \node(88) at (3.87,.1) {\fontsize{8}{0}\selectfont {\textbf )}};

    \node(22up) at (.4,.5) {};
    \node(33up) at (1.03,.5) {};
    \node(33pp) at (.8,.5) {};
    \node(88up) at (3.75,.5) {};
    \draw (22up) edge[above,red] node{} (33up);
    \draw (33pp) edge[above,blue] node{} (88up);
    \node(12down) at (.1,-.01) {};
    \node(34down) at (1.1,-.01) {};
    \node(78down) at (3.45,-.01) {};
    \node(89down) at (3.97,-.01) {};
    \draw (12down) edge[above, dotted, bend right=20, no arrow, line width=1pt] node{} (89down);
    \draw (34down) edge[above, dotted, bend right=20, no arrow, line width=1pt] node{} (78down);
\end{tikzpicture}}
}}

&
\hspace{-2cm}
\raisebox{1.72cm}{
\resizebox{.35\textwidth}{!}{
\begin{tikzpicture}[->,>=stealth',shorten >=1pt,auto,semithick, no arrow/.style={-,every loop/.append style={-}}]
    
    \fill [fill=orange!25] (0,0) -- (1.3,1.99) -- (3.5,0) -- cycle;
    \fill [fill=red!25] (0,0) -- (.6,.6) -- (1,0) -- cycle;
    \fill [fill=blue!25] (1.5,0) -- (2.3,.7) -- (3.5,0) -- cycle;
    
    \coordinate[label=below:\fontsize{10}{0}\selectfont$S$] (S) at (1.3,1.8);
    \coordinate[label=below:\fontsize{10}{0}\selectfont$N$] (N) at (.5,.5);
    \coordinate[label=below:\fontsize{10}{0}\selectfont$P$] (subS) at (2.4,.55);
    
    \draw [no arrow,line width=.7pt] (1.2,1.3) -- (.65,.65);
    \draw [no arrow,line width=.7pt] (1.4,1.3) -- (2.25,.75);

    \node (aa) at (-1.8, -0.05) {\fontsize{9}{0}\selectfont string indices};
    \node (aa) at (-1.8, -.35) {\fontsize{9}{0}\selectfont (b/w nucleotides)};
    \node (a) at (-.25, -.2) {\fontsize{11}{0}\selectfont 2};
    \node (b) at (1.25, -.2) {\fontsize{11}{0}\selectfont 3};
    \node (c) at (3.7, -.2) {\fontsize{11}{0}\selectfont 8};
    \draw (a) edge[above,red] node{} (b)
    (b) edge[above,blue] node{} (c);
    
    \node (aa) at (-1.4, -.8) {\fontsize{9}{0}\selectfont DFA states};
    \node[state, inner sep=-10pt, minimum size=.2cm, fill=NodeGray] (A) at (-.25, -.8) {\fontsize{9}{0}\selectfont 2,0};
    \node[state, inner sep=-10pt, minimum size=.2cm, fill=NodeGray] (B) at (1.25, -.8) {\fontsize{9}{0}\selectfont 3,0};
    \node[state, inner sep=-10pt, minimum size=.2cm, fill=NodeGray] (C) at (3.7, -.8) {\fontsize{9}{0}\selectfont 8,1};


    \fill[orange!25] (A.south) 
    -- (A.south east) 
    -- (C.south west)
    -- (C.south)
    to[out=-135,in=0] (1.4,-1.7)
    to[out=-180,in=-45] 
    cycle;
    \node[right=.04cm of B] (nAB) {};
    \node[below=.44cm of nAB, blue] (nABS) {\fontsize{10}{0}\selectfont $S$};

    \fill[red!25] (A.east) 
    to[out=0, in=-180] (.4,-1.)
    to[out=0,in=-180] (B.west)
    --(B.south west)
    to[out=-135,in=0] (.5,-1.3)
    to[out=-180,in=-45] (A.south east) 
    -- cycle;
    \node[right=.25cm of A] (nAB) {};
    \node[below=.06cm of nAB, blue] (nABS) {\fontsize{10}{0}\selectfont $N$};

    \fill[blue!25] (B.east) 
    to[out=0, in=180]++ (.44,.35)
    to[out=0,in=180]++ (1.,-.4)
    to[out=0,in=-180] (C.west)
    -- (C.south west)
    to[out=-135,in=0] (2.4,-1.4)
    to[out=-180,in=-45] (B.south east) 
    -- cycle;
    \node[right=.8cm of B] (nAB) {};
    \node[below=.13cm of nAB, blue] (nABS) {\fontsize{10}{0}\selectfont $P$};

    \node[state, inner sep=-10pt, minimum size=.2cm, fill=NodeGray] (A) at (-.25, -.8) {\fontsize{9}{0}\selectfont 2,0};
    \node[state, inner sep=-10pt, minimum size=.2cm, fill=NodeGray] (B) at (1.25, -.8) {\fontsize{9}{0}\selectfont 3,0};
    \node[state, inner sep=-10pt, minimum size=.2cm, fill=NodeGray] (C) at (3.7, -.8) {\fontsize{9}{0}\selectfont 8,1};

    \draw[red] (A) to[out=0, in=-180] (.4,-1.) to[out=0,in=-180] node{} (B);
    \draw[blue] (B) to[out=0, in=180]++ (.7,.35) to[out=0, in=180]++ (1.,-.4) to[out=0,in=-180] node{} (C);
\end{tikzpicture}

%

}}
\end{tabular}
\\[-2.2cm]

\begin{tabular}{c}
{\hspace{-12.cm}\raisebox{-2.3cm}{{\panel{\myedit{H}} \fontsize{9}{0}\selectfont \textbf{\myedit{lattice parsing}} } }}\\[-3.4cm]
\hspace{1cm}\raisebox{-2cm}{
\resizebox{.725\textwidth}{!}{
\begin{tikzpicture}[->,>=stealth',shorten >=1pt,auto,semithick,
no arrow/.style={-,every loop/.append style={-}}]
\node[state, initial, initial text=, inner sep=-10pt, fill=StartNode] (n00) {\fontsize{15.5}{0}\selectfont 0,0};
\node[state, right=.9cm of n00, inner sep=-10pt, fill=NodeGray] (n10) {\fontsize{15.5}{0}\selectfont 1,0};
\node[state, right=.9cm of n10, inner sep=-10pt, fill=NodeGray] (n20) {\fontsize{15.5}{0}\selectfont 2,0};
\node[state, right=.9cm of n20, inner sep=-10pt, fill=NodeGray] (n30) {\fontsize{15.5}{0}\selectfont 3,0};

\node[state, right=.9cm of n30, inner sep=-10pt, fill=NodeGray] (n40) {\fontsize{15.5}{0}\selectfont 4,0};
\node[state, below=.9cm of n40, inner sep=-10pt, fill=NodeGray] (n41) {\fontsize{15.5}{0}\selectfont 4,1};
\node[state, right=.9cm of n40, inner sep=-10pt, fill=NodeGray] (n50) {\fontsize{15.5}{0}\selectfont 5,0};
\node[state, below=.9cm of n50, inner sep=-10pt, fill=NodeGray] (n51) {\fontsize{15.5}{0}\selectfont 5,1};
\node[state, right=.9cm of n50, inner sep=-10pt, fill=NodeGray] (n60) {\fontsize{15.5}{0}\selectfont 6,0};
\node[state, right=.9cm of n60, inner sep=-10pt, fill=NodeGray] (n70) {\fontsize{15.5}{0}\selectfont 7,0};
\node[state, right=.9cm of n70, inner sep=-10pt, fill=NodeGray] (n80) {\fontsize{15.5}{0}\selectfont 8,0};
\node[state, below=.9cm of n80, inner sep=-10pt, fill=NodeGray] (n81) {\fontsize{15.5}{0}\selectfont 8,1};
\node[state, accepting, right=.9cm of n80, inner sep=-10pt, fill=EndNode] (n90) {\fontsize{15.5}{0}\selectfont 9,0};
\fill[orange!25] (n20.east) -- (n30.west)
    -- (n30.south east)
    -- (n41.north west)
    -- (n41.east)
    -- (n51.west) 
    -- (n51.north) 
    -- (n81.west)     
    -- (n81.south west)
    to [out=-150,in=0] (10.5,-3.8)
    to [out=-180,in=-80] (n20.south east)
    --  cycle;
\node[below=.94cm of n51, blue] (n51S) {\fontsize{12.5}{0}\selectfont $S$};
\fill[red!25] (n20.east) -- (n30.west) 
    -- (n30.south west) 
    to [out=-135,in=0] (4.83,-.6)
    to [out=-180,in=-35] (n20.south east) 
    -- cycle;
\node[right=.37cm of n20] (n23n) {};
\node[below=.14cm of n23n, blue] (n23N) {\fontsize{12.5}{0}\selectfont $N$};
\fill[blue!25] (n30.south east)
    -- (n41.north west)
    -- (n41.east)
    -- (n51.west) 
    -- (n51.north) 
    -- (n60.west)
    -- (n70.west)
    -- (n70.south east)
    -- (n81.north west)     
    -- (n81.west)
    to [out=-150,in=0] (10.5,-3.4)
    to [out=-180,in=-90] (n30.south)
    -- (n30.south east)
    --  cycle;
    
 \fill[blue!25] (n51.north)
  to [out=80,in=185] (n60);

\node[right=.37cm of n51] (n56s) {};
\node[below=.94cm of n56s, blue] (n56S) {\fontsize{12.5}{0}\selectfont $P$};

\draw (n00) edge[no arrow,dashed,below,bend right=40,blue] node[yshift=8pt, fill=white]{\fontsize{12.5}{0}\selectfont $N$} (n10)
(n41) edge[no arrow,dotted,below,bend right=40,blue] node[yshift=8pt]{\fontsize{12.5}{0}\selectfont $N$} (n51)
(n51) edge[no arrow,dotted,below,bend right=55,blue] node[yshift=8pt]{\fontsize{12.5}{0}\selectfont $N$} (n60)
(n60) edge[no arrow,dotted,below,bend right=40,blue] node[yshift=8pt]{\fontsize{12.5}{0}\selectfont $N$} (n70)
(n41) edge[no arrow,dotted,below,bend right=65,blue] node[yshift=8pt]{\fontsize{12.5}{0}\selectfont $S$} (n70)
(n10) [no arrow,dashed,blue] to[out=-65, in=180]++ (9.,-4.2) to[out=0,in=-90] node[yshift=-8pt, fill=white,pos=0,sloped] {\fontsize{12.5}{0}\selectfont $P$} (n90)
(n00) [no arrow,dashed,blue] to[out=-65, in=180]++ (9.,-4.6) to[out=0,in=-90] node[yshift=-8pt, fill=white,pos=0,sloped] {\fontsize{12.5}{0}\selectfont $S$} (n90);
%

\draw (n00) edge[no arrow,dashed,below,out=40, in=130, ColorGray] node[yshift=8pt, fill=white, pos=.65]{\fontsize{12.5}{0}\selectfont $N$} (n10)
(n30) edge[no arrow,dashed,below,bend left=48, ColorGray] node[yshift=8pt, fill=white]{\fontsize{12.5}{0}\selectfont $N$} (n40)
(n40) edge[no arrow,dashed,below,bend left=55, ColorGray] node[yshift=8pt, fill=white]{\fontsize{12.5}{0}\selectfont $N$} (n50)
(n50) edge[no arrow,dashed,below,bend left=48, ColorGray] node[yshift=8pt, fill=white, pos=.325]{\fontsize{12.}{0}\selectfont $N$} (n60)
(n80) edge[no arrow,dashed,below,out=45, in=140, ColorGray] node[yshift=8pt, fill=white, pos=.35]{\fontsize{12.5}{0}\selectfont $N$} (n90)

(n30) edge[no arrow,dashed,below,bend left=51, ColorGray] node[yshift=8pt, fill=white]{\fontsize{12.5}{0}\selectfont $S$} (n60)

(n20) edge[no arrow,dashed,below,bend left=40, ColorGray] node[yshift=8pt, fill=white, pos=.39]{\fontsize{12.5}{0}\selectfont $P$} (n70)
(n20) edge[no arrow,dashed,below,bend left=52, ColorGray] node[yshift=8pt, fill=white, pos=.4]{\fontsize{12.5}{0}\selectfont $S$} (n70)
(n10) edge[no arrow,dashed,below,bend left=42, ColorGray] node[yshift=8pt, fill=white, pos=.6]{\fontsize{12.5}{0}\selectfont $P$} (n80)
(n00) edge[no arrow,dashed,below,bend left=42, ColorGray] node[yshift=8pt, fill=white]{\fontsize{12.5}{0}\selectfont $S$} (n80)
(n00) edge[no arrow,dashed,below,bend left=42, ColorGray] node[yshift=8pt, fill=white]{\fontsize{12.5}{0}\selectfont $S$} (n90);

\draw (n00) edge[above,line width=1.3pt] node[red]{A} (n10)
(n00) edge[below,line width=.pt] node[black]{$\bullet$} (n10)
(n10) edge[above,line width=1.3pt] node[gray]{U} (n20)
(n10) edge[below,line width=.pt] node[black]{\textbf{(}} (n20)
(n30) edge[above] node[gray]{U} (n40)
(n40) edge[above] node[gray]{U} (n50)
(n50) edge[above, bend left=25] node[red]{\fontsize{9}{0}\selectfont A} (n60)
(n50) edge[above,bend right=5] node[violet]{\fontsize{9}{0}\selectfont G} (n60)
(n70) edge[above] node[red]{A} (n80)
(n80) edge[above, bend left=25] node[red]{\fontsize{9}{0}\selectfont A} (n90)
(n80) edge[above,bend right=5] node[violet]{\fontsize{9}{0}\selectfont G} (n90)
(n20) edge[above,line width=1.3pt, red] node[violet]{G} (n30)
(n20) edge[below,line width=.pt, red] node[black]{$\bullet$} (n30)
(n30) edge[above,line width=1.3pt, blue] node[blue]{C} (n41)
edge[below,line width=.pt, blue] node[black]{\textbf{(}} (n41)
(n41) edge[above,line width=1.3pt, blue] node[gray]{U} (n51)
(n41) edge[below,line width=.pt, blue] node[black]{$\bullet$} (n51)

(n51) edge[above, bend left=30, line width=1.3pt, blue] node[red]{\fontsize{9}{0}\selectfont A} (n60)
(n51) edge[above, bend left=10, line width=1.3pt, blue] node[blue]{\fontsize{9}{0}\selectfont C} (n60)
(n51) edge[above,line width=1.3pt, bend right=10, blue] node[violet]{\fontsize{9}{0}\selectfont G} (n60)
(n51) edge[above, bend right=28, line width=1.3pt, blue] node[gray]{\fontsize{9}{0}\selectfont U} (n60)
(n51) edge[below,bend right=28, line width=.pt, blue] node[black]{$\bullet$} (n60);

\draw (n60) edge[above,line width=1.3pt, blue] node[gray]{U} (n70)
(n60) edge[below,line width=.pt, blue] node[black]{$\bullet$} (n70)
(n70) edge[above,line width=1.3pt, blue] node[violet]{G} (n81)
(n70) edge[below,line width=.pt, blue] node[black]{\textbf{)}} (n81)
(n81) edge[above,line width=1.3pt] node[red]{A} (n90)
(n81) edge[below,line width=.pt] node{\textbf{)}} (n90);

\node[state, right=.9cm of n10, inner sep=-10pt, fill=NodeGray] (n200) {\fontsize{15.5}{0}\selectfont 2,0};
\node[state, right=.9cm of n20, inner sep=-10pt, fill=NodeGray] (n300) {\fontsize{15.5}{0}\selectfont 3,0};
\node[state, below=.9cm of n40, inner sep=-10pt, fill=NodeGray] (n411) {\fontsize{15.5}{0}\selectfont 4,1};
\node[state, below=.9cm of n50, inner sep=-10pt, fill=NodeGray] (n511) {\fontsize{15.5}{0}\selectfont 5,1};
\node[state, right=.9cm of n50, inner sep=-10pt, fill=NodeGray] (n600) {\fontsize{15.5}{0}\selectfont 6,0};
\node[state, right=.9cm of n60, inner sep=-10pt, fill=NodeGray] (n700) {\fontsize{15.5}{0}\selectfont 7,0};
\node[state, below=.9cm of n80, inner sep=-10pt, fill=NodeGray] (n811) {\fontsize{15.5}{0}\selectfont 8,1};
\end{tikzpicture}}}


\end{tabular}\\[0cm]

\caption{
\myedit{{\bf A--B}: Illustrations of mRNA design as optimization problems for stability (objective 1, in {\bf A})
and joint stability and codon optimality (objectives 1 \& 2, in {\bf B})}. 
\myedit{{\bf C--H} show how lattice parsing solves
the first optimization problem (see Fig.~\ref{fig:main_CFG_DFA}D for the second).}
\myedit{{\bf C}: Codon DFAs.} 
{\bf D}: An mRNA DFA 
made of three codon DFAs. 
The thick paths depict the optimal mRNA sequences under the simplified energy model in {\bf E},
\nucA\nucU\nucG\nucC\nucU$\star$\nucU\nucG\nucA\/,
where $\star$
\myedit{could be any nucleotide.} 
{\bf E}: Stochastic context-free grammar (SCFG) 
for a simplified folding free 
energy model. Each rule \myedit{has a cost (i.e., energy term, the lower the better)}, 
and the dotted arcs represent base pairs in RNA secondary structure. 
{\bf F}: Single-sequence folding is equivalent to context-free parsing with 
an SCFG; 
the parse tree represents the best secondary structure for the input mRNA sequence.
{\bf G}: We extend single-sequence parsing (top) to lattice parsing (bottom) by replacing the input string with a DFA, 
where each string index becomes a DFA state, and a span 
becomes a path between two states. 
{\bf H}: Lattice parsing with the grammar in {\bf E} for the DFA in {\bf D}.
The blue arcs below the DFA depict the (shared) best structure for the optimal sequences \nucA\nucU\nucG\nucC\nucU$\star$\nucU\nucG\nucA\/ in the whole DFA,
while the dashed light-blue arcs above the DFA represent the best structure for a suboptimal sequence 
\nucA\nucU\nucG\nucU\nucU\nucA\nucU\nucA\nucA. 
\myedit{Lattice parsing can also 
incorporate codon optimality (objective 2, see {\bf B}), by replacing the DFA with a weighted one (Fig.~\ref{fig:main_CFG_DFA}D).}
\label{fig:CFG_DFA} 
}
\end{figure}

%% file: fig_pseudocode.tex



\begin{figure}[!thb]
\center
\begin{algorithmic}[1]
  \newcommand{\INDSTATE}[1][1]{\State\hspace{#1\algorithmicindent}}
  \setstretch{1.2} 
\Function{BottomUpDesign}{$\vecp$, $\lambda$} \Comment{\vecp: protein sequence; $\lambda$: weight of CAI}
    \State $n \gets 3 \cdot (|\vecp| + 1)$ \Comment{mRNA length; +1 for the stop codon}
    \State $D \gets D(x_1) \circ D(x_2) \circ ... \circ D({\rm stop})$ \Comment{build (CAI-integrated) mRNA DFA}
    \State $\best \gets$ hash() \Comment{hash table: from $[X, q_i, q_j]$ to  score}
    \State $\back \gets$ hash() \Comment{hash table: from $[X, q_i, q_j]$ to $\rm backpointer$}
    \For{$i=0 \ldots (n-3)$} \Comment{base case: $S\goestow{0} N\ N\ N$}
      \ForEach {$q_i \in {\rm nodes}(D, i)$} 
        \ForEach {$q_{i+3} \in {\rm nodes}(D, i+3)$} 
           \State {$\best[S, q_i, q_{i+3}] \gets {\color{blue}\mincost(q_i, q_{i+3}, \lambda)}$} \Comment{best cost of any $q_i \leadsto q_{i+3}$ path (Eq.~\ref{eq:mincost})}  \label{lineno:ckybase}
         \EndFor
       \EndFor
    \EndFor 
    \For{$l=2 \ldots n$} \Comment{$l=(j-i)$ is the span width}
      \For{$i=0 \ldots (n-l)$}
        \State $j \gets i+l$
        \ForEach {$q_i \in {\rm nodes}(D, i)$} 
          \ForEach {$q_j \in {\rm nodes}(D, j)$} 
            \ForEach {$\laedgew{q_{j-1}}{b: w_b}{q_j}  \in \inedges(D, q_j)$} \Comment{$q_i \stackrel{S}{\longleadsto} q_{j-1} \stackrel{b}{\!\leadsto\!} q_j$}
               \State {\textproc{Update}$(S, q_i, q_j, \best[S, q_i, q_{j-1}] {\color{blue}+ \lambda w_b}, (q_{j-1}, b))$} \Comment{$S \goestow{0} S\ N$}
               
               \If {$j-i>4$} \Comment{pairing (no sharp turn)}
                 \ForEach {$\laedgew{q_i}{a: w_a}{q_{i+1}} \in \outedges(D, q_i)$} \Comment{$q_i \stackrel{a}{\!\goesto\!} q_{i+1} \stackrel{S}{\longleadsto} q_{j-1} \stackrel{b}{\!\goesto\!} q_j$}
                  \If {$\Delta G(a, b) < 0$} \label{line:pair} \Comment{$\Delta G(\nucC,\!\nucG)\!=\!-3; \Delta G(\nucA,\!\nucU)\!=\!-2;...$} 
                    \State $\score \gets \best[S, q_{i+1}, q_{j-1}]  {+\color{blue}\lambda(w_a \!+\! w_b)} + \Delta G(a, b) $ \Comment{$P\goestow{-3} \nucC\ S\ \nucG \mid ...$}
                    \State {\textproc{Update}$(P, q_i, q_j, \score, (a, q_{i+1},  q_{j-1}, b))$}
                  \EndIf
                \EndFor
             \EndIf            
            \EndFor
            \For{$k=(i+1) \ldots (j-1)$} \Comment{bifurcation midpoint}
              \ForEach {$q_k \in {\rm nodes}(D, k)$} \Comment{$q_i \stackrel{S}{\longleadsto} q_k \stackrel{P}{\longleadsto} q_j$}
                \State $\score \gets \best[S, q_i, q_k] + \best[P, q_k, q_j]$ \Comment{$S\goestow{0} S\ P$}
                \State {\textproc{Update}$(S, q_i, q_j, \score, q_k)$}
              \EndFor
            \EndFor
          \EndFor
        \EndFor
      \EndFor
    \EndFor
    \State \Return $\best[S, q_0, q_n], \textproc{Backtrace}(S, q_0, q_n)$ 
\EndFunction
\end{algorithmic}
\caption{
The pseudocode of a simplified bottom-up version of our mRNA Design algorithm for the joint optimization between stability and codon optimality. 
The costs in blue are for CAI integration. 
See Methods~\ref{methods:bottom-up} for more algorithm description, and Fig.~\ref{fig:update_backtrace} for \textproc{Update} and \textproc{Backtrace} functions.
\label{fig:algorithm}}
\vspace{-0.3cm}
\end{figure}

%% file: si_update_backtrace.tex

\begin{figure}[!h]
\vspace{-1.5cm}
\centering
\begin{minipage}{1\textwidth}
\begin{algorithmic}[1]
  \newcommand{\INDSTATE}[1][1]{\State\hspace{#1\algorithmicindent}}
  \setstretch{1.2} 
\Function{Update}{$X, q_i, q_j, \score, \backpointer$} 
    \If{key $(X, q_i, q_j)$ not in $\best$ or $\score < \best[X, q_i, q_j]$} \Comment{minimizing weight}
      \State {$\best[X, q_i, q_j] \gets \score$}
      \State {$\back[X, q_i, q_j] \gets \backpointer$}
    \EndIf
\EndFunction
\end{algorithmic}

\begin{algorithmic}[1]
  \newcommand{\INDSTATE}[1][1]{\State\hspace{#1\algorithmicindent}}
  \setstretch{1.2} 
\Function{Backtrace}{$X, q_i, q_j$} \Comment{returns a (sequence, structure) pair}
    \State $\backpointer \gets \back[X, q_i, q_j]$
    \If {$\stateindex{q_j} - \stateindex{q_i} = 3$} \Comment{$\stateindex{q}$: the string index of state $q$}
      \State \Return $\textproc{AnyPath}(q_i, q_j), \text{\tt "\md\md\md"}$ \Comment{$S \goestow{0} N \ N \ N$; any $q_i \longleadsto q_j$ path is fine}
    \EndIf
    \If {{length}$(\backpointer)=4$} \Comment{pairing: $P \goestow{-3} \nucC\ S \ \nucG \mid ...$}
      \State $a, q_{i+1}, q_{j-1}, b \gets \backpointer$ \Comment{$q_i \stackrel{a}{\!\goesto\!} q_{i+1} \stackrel{\seq}{\longleadsto} q_{j-1} \stackrel{b}{\!\goesto\!} q_j$}
      \State $\seq, \struct \gets \textproc{Backtrace}(S, q_{i+1}, q_{j-1})$
      \State \Return {$a + \seq + b, \text{\tt "("} + \struct + \text{\tt ")"}$}
   \ElsIf {length$(\backpointer)=2$} \Comment{skip: $S \goestow{0} S\ N$}
      \State $q_{j-1}, b \gets \backpointer$ \Comment{$q_i \stackrel{\seq}{\longleadsto} q_{j-1} \stackrel{b}{\!\goesto\!} q_j$}
      \State $\seq, \struct \gets \textproc{Backtrace}(S, q_{i}, q_{j-1})$
      \State \Return {$\seq + b, \struct + \text{\tt "."}$}  

    \Else \Comment{bifurcation: $S \goestow{0} S\ P$}
      \State $q_{k} \gets \backpointer$ \Comment{$q_i \stackrel{\seq_1}{\longleadsto} q_k \stackrel{\seq_2}{\longleadsto} q_j$}
      \State $\seq_1, \struct_1 \gets \textproc{Backtrace}(S, q_{i}, q_{k})$
      \State $\seq_2, \struct_2 \gets \textproc{Backtrace}(P, q_{k}, q_{j})$
      \State \Return {$\seq_1 + \seq_2, \struct_1 + \struct_2$}     \EndIf
\EndFunction
\end{algorithmic}

%

\begin{algorithmic}[1]

\Function{beamprune}{$X, j, b$}
    \State $\cands \gets$ hash() \Comment{hash table: from $q_i$ to score $\best[S, q_0, q_i] + \best[X, q_i, q_j]$}
    \ForEach{$q_j \in \nodes(j)$}
	\ForEach{key $(X, q_i, q_j) \in \best$}
        	      \State $\cands[q_i] \gets \best[S, q_0, q_i] + \best[X, q_i, q_j]$ \Comment{$\best[S, q_0, q_i]$ as prefix score}
        \EndFor
    \EndFor
    \State $\cands \gets \textproc{SelectTopB}(\cands, b)$ \Comment{select top-$b$ by score}
    \ForEach{key $(X, q_i, q_j)  \in \best$}
        \If{key $q_i$ not in $\cands$}
            \State delete $(X, q_i, q_j)$ in $\best$ \Comment{prune out low-scoring states}
        \EndIf
    \EndFor
\EndFunction 
\end{algorithmic}
\end{minipage}
\caption{
The pseudocode for \textproc{Update}, \textproc{Backtrace} (used in \textproc{BottomUpDesign}, see Fig.~\ref{fig:algorithm}) and \textproc{BeamPrune} (used in \textproc{LinearDesign}, see Fig.~\ref{fig:linear})
functions.
\label{fig:update_backtrace}}
\vspace{-0.3cm}
\end{figure}

%% file: fig_lineardesign.tex
\renewcommand\thealgorithm{}

\begin{figure}[!htb]
\center
\begin{algorithmic}[1]
  \newcommand{\INDSTATE}[1][1]{\State\hspace{#1\algorithmicindent}}
  \setstretch{1.2} 
\Function{LinearDesign}{$\vecp, \lambda, b$} \Comment{$b$ is beam size}
\setalglineno{10} 
    \For{$j=4 \ldots n$}
      \ForEach {$q_{j-1} \in {\rm nodes}(D, j-1)$}
         \ForEach {$q_i$ such that $(S, q_i, q_{j-1}) \in \best$}
         	\ForEach {$\laedgew{q_{j-1}}{b: w_b}{q_j} \in \outedges(D, q_{j-1})$} \Comment{$q_i \stackrel{S}{\longleadsto} q_{j-1} \stackrel{b}{\!\goesto\!} q_j$}
		     \State {\textproc{Update}$(S, q_i, q_j, \best[q_i, q_{j-1}] {\color{blue} + \lambda w_b}, (q_{j-1}, b))$}\Comment{$S \goestow{0} S\ N$}
		     \ForEach {$\laedgew{q_{i-1}}{a: w_a}{q_i} \in \inedges(D, q_i)$} \Comment{$q_{i-1} \stackrel{a}{\!\goesto\!} q_{i} \stackrel{S}{\longleadsto} q_{j-1} \stackrel{b}{\!\goesto\!} q_j$}
                       \If {$\Delta G(a, b) < 0$} \label{line:pair} \Comment{$\Delta G(\nucC,\!\nucG)\!=\!-3; \Delta G(\nucA,\!\nucU)\!=\!-2;...$} 
                         \State $\score \gets \best[S, q_{i}, q_{j-1}] {\color{blue}+ \lambda (w_a\!+\!w_b)} + \Delta G(a, b) $                     
                        \State {\textproc{Update}$(P, q_{i-1}, q_j, \score, (a, q_{i},  q_{j-1}, b))$} \Comment{$P\goestow{-3} \nucC\ S\ \nucG \mid ...$}

                     \EndIf

		   \EndFor
	        \EndFor
	       \EndFor
	     \EndFor
      \State \textproc{BeamPrune}$(P, j, b)$ \Comment{choose top-$b$ among all $(P, q_i, q_j)$'s}
      \ForEach {$q_j \in \nodes(D, j)$}
         \ForEach {$q_i$ such that $(P, q_i, q_j) \in \best$}
            \ForEach {$q_k$ such that $(S, q_k, q_i) \in \best$}
                \State $\score \gets \best[S, q_k, q_i] + \best[P, q_i, q_j]$  
                \Comment{$q_i \stackrel{S}{\longleadsto} q_k \stackrel{P}{\longleadsto} q_j$}               
                \State {\textproc{Update}$(S, q_k, q_j, \score, q_i)$} \Comment{$S\goestow{0} S\ P$}

           \EndFor          
         \EndFor
      \EndFor
      \State \textproc{BeamPrune}$(S, j, b)$ \Comment{choose top-$b$ among all $(S, q_i, q_j)$'s}

    \EndFor
    \State \Return $\best[S, q_0, q_n], \textproc{Backtrace}(S, q_0, q_n)$ 
\EndFunction
\end{algorithmic}
\caption{
The pseudocode of (simplified) LinearDesign algorithm for the joint optimization between stability and codon optimality. 
The costs in blue are for CAI integration.
The first~\ref{lineno:ckybase} lines are the same as in \textproc{BottomUpDesign} (see Fig.~\ref{fig:algorithm}). 
See Methods~\ref{methods:left-right} for more algorithm description,  and Fig.~\ref{fig:update_backtrace} for \textproc{Update}, \textproc{Backtrace}, and \textproc{BeamPrune} functions.
\label{fig:linear}}
\vspace{-0.3cm}
\end{figure}

%% file: fig_si_alter_codon.tex

\tikzset{every picture/.style={/utils/exec={\sffamily}}}

\begin{figure*}[!htb]
\centering
\begin{tabular}{ccc}
\vspace{.4cm}
\raisebox{-.5cm}{\hspace{-4.2cm}\panel{A}}\\[-.8cm]
\hspace{1.5cm}{\fontsize{9}{0}\selectfont \textbf{alternative genetic code}}\\[-.2cm]

\hspace{-1.cm}\resizebox{.34\textwidth}{!}{\raisebox{1.5cm}{
\begin{tikzpicture}[->,>=stealth',shorten >=1pt,auto,semithick]
\node[state, initial, initial text=, inner sep=-10pt, fill=StartNode] (n00) {\fontsize{16.5}{0}\selectfont 0,0};
\node[state, right=.9cm of n00, inner sep=-10pt, fill=NodeGray] (n10) {\fontsize{16.5}{0}\selectfont 1,0};
\node[state, below=.3cm of n10, inner sep=-10pt, fill=NodeGray] (n11) {\fontsize{16.5}{0}\selectfont 1,1};
\node[state, below=.3cm of n11, inner sep=-10pt, fill=NodeGray] (n12) {\fontsize{16.5}{0}\selectfont 1,2};
\node[state, right=.9cm of n10, inner sep=-10pt, fill=NodeGray] (n20) {\fontsize{16.5}{0}\selectfont 2,0};
\node[state, below=.3cm of n20, inner sep=-10pt, fill=NodeGray] (n21) {\fontsize{16.5}{0}\selectfont 2,1};
\node[state, below=.3cm of n21, inner sep=-10pt, fill=NodeGray] (n22) {\fontsize{16.5}{0}\selectfont 2,2};
\node[state, accepting, right=.9cm of n20, inner sep=-10pt, fill=EndNode] (n30) {\fontsize{16.5}{0}\selectfont 3,0};
\node[below=3.1cm of n00] (n0a) {}; 
\node[right=-.3cm of n0a] (n0b) {\fontsize{16}{0}\selectfont  $D(\text{serine})$\! of\! alternative}; 
\node[below=-.2cm of n0b] (n0c) {\fontsize{16}{0}\selectfont  yeast\! nuclear\! code~\protect\mycite{ohama+:1993}}; 
\draw (n00) edge[above] node[gray]{U} (n10)
(n00) edge[above] node[red]{A} (n11)
(n00) edge[above, line width=1.5pt] node[blue]{C} (n12)
(n10) edge[above] node[blue]{C} (n20)
(n11) edge[above] node[violet]{G} (n21)
(n12) edge[above, line width=1.5pt] node[gray,above]{U} (n22)
(n20) edge[above, bend left=80] node[red,above]{A} (n30)
(n20) edge[above, bend left=40] node[blue,above]{C} (n30)
(n20) edge[below] node[violet,above]{G} (n30)
(n20) edge[below, bend right=30] node[gray,above, inner sep=2pt]{U} (n30)
(n21) edge[below, bend right=22] node[blue,above, inner sep=3pt]{C} (n30)
(n21) edge[below, bend right=50] node[gray,above, inner sep=3pt]{U} (n30)
(n22) edge[below, bend right=40, line width=1.5pt] node[violet,above, inner sep=4pt]{G} (n30);
\end{tikzpicture}}}
&
\hspace{-1.cm}\resizebox{.34\textwidth}{!}{\raisebox{3.3cm}{
\begin{tikzpicture}[->,>=stealth',shorten >=1pt,auto,semithick]
\node[state, initial, initial text=, inner sep=-10pt, fill=StartNode] (n00) {\fontsize{16.5}{0}\selectfont 0,0};
\node[state, right=.9cm of n00, inner sep=-10pt, fill=NodeGray] (n10) {\fontsize{16.5}{0}\selectfont 1,0};
\node[state, right=.9cm of n10, inner sep=-10pt, fill=NodeGray] (n20) {\fontsize{16.5}{0}\selectfont 2,0};
\node[state, accepting, right=.9cm of n20, inner sep=-10pt, fill=EndNode] (n30) {\fontsize{16.5}{0}\selectfont 3,0};
\node[below=0.2cm of n00] (n0a) {}; 
\node[right=1.2cm of n0a] (n0b) {\fontsize{18}{0}\selectfont \qquad   $\Downarrow$}; 
\node[below=-1.6cm of n00] (n0c) {}; 
\node[right=0.3cm of n0c] (n0d) {\fontsize{16}{0}\selectfont {original $D(\text{tryptophan})$}}; 
\draw (n00) edge[above] node[gray]{U} (n10)
(n10) edge[above] node[violet]{G} (n20)
(n20) edge[above] node[violet]{G} (n30);
\raisebox{-2.2cm}{
\node[state, initial, initial text=, inner sep=-10pt, fill=StartNode] (n00) {\fontsize{16.5}{0}\selectfont 0,0};
\node[state, right=.9cm of n00, inner sep=-10pt, fill=NodeGray] (n10) {\fontsize{16.5}{0}\selectfont 1,0};
\node[state, right=.9cm of n10, inner sep=-10pt, fill=NodeGray] (n20) {\fontsize{16.5}{0}\selectfont 2,0};
\node[state, accepting, right=.9cm of n20, inner sep=-10pt, fill=EndNode] (n30) {\fontsize{16.5}{0}\selectfont 3,0};
\node[below=1.cm of n00] (n0a) {}; 
\node[right=-.2cm of n0a] (n0b) {\fontsize{16}{0}\selectfont \myedit{reassignment\! of\! CGG\! to}}; 
\node[below=-0.2cm of n0b] (n0c) {\fontsize{16}{0}\selectfont \myedit{$D(\text{tryptophan})$\! in\! a clade of}};
\node[below=-0.2cm of n0c] (n0d) {\fontsize{16}{0}\selectfont \myedit{Bacilli or in Anaerococcus~\protect\mycite{shulgina+:2021}}}; 
\draw (n00) edge[above] node[gray]{U} (n10)
(n00) edge[below, bend right=40, line width=1.5pt] node[blue]{C} (n10)
(n10) edge[above, line width=1.5pt] node[violet]{G} (n20)
(n20) edge[above, line width=1.5pt] node[violet]{G} (n30);
}
\end{tikzpicture}}}
&
\hspace{-.2cm}\resizebox{.35\textwidth}{!}{\raisebox{4.cm}{
\begin{tikzpicture}[->,>=stealth',shorten >=1pt,auto,semithick]
\node[state, initial, initial text=, inner sep=-10pt, fill=StartNode] (n00) {\fontsize{16.5}{0}\selectfont 0,0};
\node[state, right=1cm of n00, inner sep=-10pt, fill=NodeGray] (n10) {\fontsize{16.5}{0}\selectfont 1,0};
\node[state, right=1cm of n10, inner sep=-10pt, fill=NodeGray] (n20) {\fontsize{16.5}{0}\selectfont 2,0};
\node[state, accepting, right=1cm of n20, inner sep=-10pt, fill=EndNode] (n30) {\fontsize{16.5}{0}\selectfont 3,0};
\node[below=0.5cm of n00] (n0a) {}; 
\node[right=1.2cm of n0a] (n0b) {\fontsize{18}{0}\selectfont \qquad   $\Downarrow$}; 
\node[below=-1.8cm of n00] (n0c) {}; 
\node[right=-0.3cm of n0c] (n0d) {\fontsize{16}{0}\selectfont {original $D(\text{threonine})$}}; 
\draw (n00) edge[above] node[red]{A} (n10)
(n10) edge[above] node[blue]{C} (n20)
(n20) edge[above, bend left=60] node[red]{A} (n30)
(n20) edge[above, bend left=20] node[blue]{C} (n30)
(n20) edge[below] node[violet]{G} (n30)
(n20) edge[below, bend right=40] node[gray]{U} (n30);
\raisebox{-2.8cm}{
\node[state, initial, initial text=, inner sep=-10pt, fill=StartNode] (n00) {\fontsize{16.5}{0}\selectfont 0,0};
\node[state, right=1cm of n00, inner sep=-10pt, fill=NodeGray] (n10) {\fontsize{16.5}{0}\selectfont 1,0};
\node[state, below=.6cm of n10, inner sep=-10pt, fill=NodeGray] (n11) {\fontsize{16.5}{0}\selectfont 1,1};
\node[state, right=1cm of n10, inner sep=-10pt, fill=NodeGray] (n20) {\fontsize{16.5}{0}\selectfont 2,0};
\node[state, below=.6cm of n20, inner sep=-10pt, fill=NodeGray] (n21) {\fontsize{16.5}{0}\selectfont 2,1};
\node[state, accepting, right=1cm of n20, inner sep=-10pt, fill=EndNode] (n30) {\fontsize{16.5}{0}\selectfont 3,0};
\node[below=2cm of n00] (n0a) {}; 
\node[right=-.cm of n0a] (n0b) {\fontsize{16}{0}\selectfont $D(\text{threonine})$ of yeast}; 
\node[below=-.2cm of n0b] (n0c) {\fontsize{16}{0}\selectfont mitochondrial codons}; 
\draw (n00) edge[above] node[red]{A} (n10)
(n00) edge[above, line width=1.5pt] node[blue]{C} (n11)
(n10) edge[above] node[blue]{C} (n20)
(n11) edge[above, line width=1.5pt] node[gray]{U} (n21)
(n20) edge[above, bend left=95] node[red]{A} (n30)
(n20) edge[above, bend left=50] node[blue]{C} (n30)
(n20) edge[above, bend left=20] node[violet]{G} (n30)
(n20) edge[below, bend left=8] node[gray]{U} (n30)
(n21) edge[below, bend left=10, line width=1.5pt] node[red]{A} (n30)
(n21) edge[below, bend right=20, line width=1.5pt] node[blue]{C} (n30)
(n21) edge[below, bend right=50, line width=1.5pt] node[violet]{G} (n30)
(n21) edge[below, bend right=95, line width=1.5pt] node[gray]{U} (n30);
}
\end{tikzpicture}}}
\\[-.8cm]
\end{tabular}

\begin{tikzpicture}
\draw (-5,0) -- (10.4, 0);
\end{tikzpicture}
\\[-.3cm]

\begin{tabular}{cc}
\raisebox{-.5cm}{\hspace{-5.6cm}\panel{B}}\\[-.4cm]
\hspace{-.3cm}{\fontsize{9}{0}\selectfont \textbf{avoiding certain codon}}\\[-.2cm]
\hspace{-.cm}\resizebox{.38\textwidth}{!}{\raisebox{.7cm}{
{\begin{tikzpicture}[->,>=stealth',shorten >=1pt,auto,semithick]
\node[state, initial, initial text=, inner sep=-10pt, fill=StartNode] (n00) {\fontsize{16.5}{0}\selectfont 0,0};
\node[state, right=1cm of n00, inner sep=-10pt, fill=NodeGray] (n40) {\fontsize{16.5}{0}\selectfont 1,0};
\node[state, below=.6cm of n40, inner sep=-10pt, fill=NodeGray] (n41) {\fontsize{16.5}{0}\selectfont 1,1};
\node[state, right=1cm of n40, inner sep=-10pt, fill=NodeGray] (n50) {\fontsize{16.5}{0}\selectfont 2,0};
\node[state, below=.6cm of n50, inner sep=-10pt, fill=NodeGray] (n51) {\fontsize{16.5}{0}\selectfont 2,1};
\node[state, accepting, right=.9cm of n50, inner sep=-10pt, fill=EndNode] (n60) {\fontsize{16.5}{0}\selectfont 3,0};
\node[above=.3cm of n00] (n0a) {}; 
\node[right=.7cm of n0a] (n0b) {\fontsize{16}{0}\selectfont original $D(\text{serine})$}; 
\draw (n00) edge[above] node[red]{A} (n40)
(n00) edge[above] node[gray]{U} (n41)
(n40) edge[above] node[violet]{G} (n50)
(n41) edge[above] node[blue]{C} (n51)
(n50) edge[above, bend left=40] node[blue]{C} (n60)
(n50) edge[above] node[gray]{U} (n60)
(n51) edge[below, bend left=20, red, dashed] node[red]{A} (n60)
(n51) edge[below, bend left=20, red, dashed] node[black, pos=.8, yshift=.23cm]{x} (n60)
(n51) edge[below, bend right=10] node[gray]{U} (n60)
(n51) edge[below, bend right=40] node[blue]{C} (n60)
(n51) edge[below, bend right=70, red, dashed] node[violet]{G} (n60)
(n51) edge[below, bend right=70, red, dashed] node[black, pos=.7, yshift=0.2cm]{x} (n60);

\raisebox{-3.6cm}{
\node[state, initial, initial text=, inner sep=-10pt, fill=StartNode] (n00_mod) {\fontsize{16.5}{0}\selectfont 0,0};
\node[state, right=1cm of n00_mod, inner sep=-10pt, fill=NodeGray] (n40_mod) {\fontsize{16.5}{0}\selectfont 1,0};
\node[state, below=.6cm of n40_mod, inner sep=-10pt, fill=NodeGray] (n41_mod) {\fontsize{16.5}{0}\selectfont 1,1};
\node[state, right=1cm of n40_mod, inner sep=-10pt, fill=NodeGray] (n50_mod) {\fontsize{16.5}{0}\selectfont 2,0};
\node[state, below=.6cm of n50_mod, inner sep=-10pt, fill=NodeGray] (n51_mod) {\fontsize{16.5}{0}\selectfont 2,1};
\node[state, accepting, right=.9cm of n50_mod, inner sep=-10pt, fill=EndNode] (n60_mod) {\fontsize{16.5}{0}\selectfont 3,0};
\node[below=2cm of n00_mod] (n0a_mod) {}; 
\node[right=-.5cm of n0a_mod] (n0b_mod) {\fontsize{15}{0}\selectfont $D(\text{serine})$ after synonymous}; 
\node[below=-.2cm of n0b_mod] (n0c_mod) {\fontsize{15}{0}\selectfont codon compression}; 
\node[below=-1.8cm of n00, inner sep=-10pt] (n0c) {}; 
\node[right=2.7cm of n0c] (n0d) {\fontsize{18}{0}\selectfont $\Downarrow$};
\draw (n00_mod) edge[above] node[red]{A} (n40_mod)
(n00_mod) edge[above] node[gray]{U} (n41_mod)
(n40_mod) edge[above] node[violet]{G} (n50_mod)
(n41_mod) edge[above] node[blue]{C} (n51_mod)
(n50_mod) edge[above, bend left=40] node[blue]{C} (n60_mod)
(n50_mod) edge[above] node[gray]{U} (n60_mod)
(n51_mod) edge[below, bend right=10] node[gray]{U} (n60_mod)
(n51_mod) edge[below, bend right=40] node[blue]{C} (n60_mod);
}
\end{tikzpicture}}}}
&
\hspace{2cm}\resizebox{.38\textwidth}{!}{\raisebox{1.2cm}{
{\begin{tikzpicture}[->,>=stealth',shorten >=1pt,auto,semithick]
\node[state, initial, initial text=, inner sep=-10pt, fill=StartNode] (n00) {\fontsize{16.5}{0}\selectfont 0,0};
\node[state, right=1cm of n00, inner sep=-10pt, fill=NodeGray] (n10) {\fontsize{16.5}{0}\selectfont 1,0};
\node[state, right=1cm of n10, inner sep=-10pt, fill=NodeGray] (n20) {\fontsize{16.5}{0}\selectfont 2,0};
\node[state, below=.2cm of n20, inner sep=-10pt, fill=NodeGray] (n21) {\fontsize{16.5}{0}\selectfont 2,1};
\node[state, accepting, right=1cm of n20, inner sep=-10pt, fill=EndNode] (n30) {\fontsize{16.5}{0}\selectfont 3,0};
\node[below=-1.7cm of n00, inner sep=-10pt] (n0a) {}; 
\node[right=.7cm of n0a] (n0b) {\fontsize{16}{0}\selectfont original $D(\textsc{STOP})$}; 
\node[below=.8cm of n00, inner sep=-10pt] (n0c) {}; 
\node[right=2.0cm of n0c] (n0d) {\fontsize{18}{0}\selectfont$\Downarrow$};

\draw (n00) edge[above] node[gray]{U} (n10)
(n10) edge[above] node[red]{A} (n20)
(n10) edge[above] node[violet]{G} (n21)
(n20) edge[above, bend left=20] node[red]{A} (n30)
(n20) edge[below, red, dashed] node[violet]{G} (n30)
(n20) edge[below, red, dashed] node[black, pos=.5, yshift=.23cm]{x} (n30)
(n21) edge[below] node[red]{A} (n30);

\raisebox{-2.6cm}{
\node[state, initial, initial text=, inner sep=-10pt, fill=StartNode] (n00_mod) {\fontsize{16.5}{0}\selectfont 0,0};
\node[state, right=1cm of n00_mod, inner sep=-10pt, fill=NodeGray] (n10_mod) {\fontsize{16.5}{0}\selectfont 1,0};
\node[state, right=1cm of n10_mod, inner sep=-10pt, fill=NodeGray] (n20_mod) {\fontsize{16.5}{0}\selectfont 2,0};
\node[state, accepting, right=1cm of n20_mod, inner sep=-10pt, fill=EndNode] (n30_mod) {\fontsize{16.5}{0}\selectfont 3,0};
\node[below=.6cm of n00_mod, inner sep=-10pt] (n0a_mod) {}; 
\node[right=.5cm of n0a_mod] (n0b_mod) {\fontsize{15}{0}\selectfont $D(\textsc{STOP})$ without}; 
\node[below=-.2cm of n0b_mod] (n0c_mod) {\fontsize{15}{0}\selectfont amber STOP codon}; 

\draw (n00_mod) edge[above] node[gray]{U} (n10_mod)
(n10_mod) edge[above, bend left=20] node[red]{A} (n20_mod)
(n10_mod) edge[below, bend right=20] node[violet]{G} (n20_mod)
(n20_mod) edge[above] node[red]{A} (n30_mod);
}
\end{tikzpicture}}}}\\[1.8cm]
\end{tabular}

\hspace{8.8cm}
\begin{tikzpicture}
\draw (4.5,0) -- (11, 0);
\end{tikzpicture}
\\[-1.cm]

\begin{tabular}{ccc}
\raisebox{-1.5cm}{\hspace{-7.2cm}\panel{C}} & \raisebox{-1.5cm}{\hspace{-5cm}\panel{D}} \\[-.4cm]
\hspace{.7cm}{\fontsize{9}{0}\selectfont \textbf{avoiding a specific adjacent codon pair}} &  \hspace{1.8cm}{\fontsize{9}{0}\selectfont \textbf{chemically modified nucleotides}}\\[-1.cm]
\hspace{-.0cm}\resizebox{.53\textwidth}{!}{\raisebox{4.cm}{
\begin{tikzpicture}[->,>=stealth',shorten >=1pt,auto,semithick]
\node[state, initial, initial text=, inner sep=-10pt, fill=StartNode] (n00) {\fontsize{16.5}{0}\selectfont 0,0};
\node[state, right=0.7cm of n00, inner sep=-10pt, fill=NodeGray] (n10) {\fontsize{16.5}{0}\selectfont 1,0};
\node[state, below=.6cm of n10, inner sep=-10pt, fill=NodeGray] (n11) {\fontsize{16.5}{0}\selectfont 1,1};
\node[state, right=0.7cm of n10, inner sep=-10pt, fill=NodeGray] (n20) {\fontsize{16.5}{0}\selectfont 2,0};
\node[state, below=.6cm of n20, inner sep=-10pt, fill=NodeGray] (n21) {\fontsize{16.5}{0}\selectfont 2,1};
\node[state, right=0.7cm of n20, inner sep=-10pt, fill=NodeGray] (n30) {\fontsize{16.5}{0}\selectfont 3,0};
\node[state, right=0.7cm of n30, inner sep=-10pt, fill=NodeGray] (n40) {\fontsize{16.5}{0}\selectfont 4,0};
\node[state, right=0.7cm of n40, inner sep=-10pt, fill=NodeGray] (n50) {\fontsize{16.5}{0}\selectfont 5,0};
\node[state, accepting, right=0.7cm of n50, inner sep=-10pt, fill=EndNode] (n60) {\fontsize{16.5}{0}\selectfont 6,0};
\node[below=2cm of n00] (n0a) {}; 
\node[right=4.8cm of n0a] (n0b) {\fontsize{18}{0}\selectfont $\Downarrow$}; 
\node[below=-1cm of n21] (n21a) {}; 
\node[right=3cm of n21a] (n0b) {\fontsize{16}{0}\selectfont original}; 
\node[below=-.5cm of n21] (n21a) {}; 
\node[right=1.5cm of n21a] (n0b) {\fontsize{16}{0}\selectfont $D(\text{leucine})\! \circ \! D(\text{proline})$}; 
\draw (n00) edge[above] node[gray]{\fontsize{10}{0}\selectfont U} (n10)
(n00) edge[above, line width=1.3pt, dashed] node[blue]{\fontsize{10}{0}\selectfont C} (n11)
(n10) edge[above] node[gray]{\fontsize{10}{0}\selectfont U} (n20)
(n11) edge[above, line width=1.3pt, dashed] node[gray]{\fontsize{10}{0}\selectfont U} (n21)
(n20) edge[above, bend left=40] node[red]{\fontsize{10}{0}\selectfont A} (n30)
(n20) edge[above] node[violet]{\fontsize{10}{0}\selectfont G} (n30)
(n21) edge[above, bend left=15] node[red]{\fontsize{10}{0}\selectfont A} (n30)
(n21) edge[above, bend right=10] node[violet]{\fontsize{10}{0}\selectfont G} (n30)
(n21) edge[above, bend right=40] node[gray]{\fontsize{10}{0}\selectfont U} (n30)
(n21) edge[above, bend right=70, line width=1.3pt, dashed, red] node[blue]{\fontsize{10}{0}\selectfont C} (n30)
(n30) edge[above, line width=1.3pt, dashed] node[blue]{\fontsize{10}{0}\selectfont C} (n40)
(n40) edge[above, line width=1.3pt, dashed] node[blue]{\fontsize{10}{0}\selectfont C} (n50)
(n50) edge[above, bend left=60] node[red]{\fontsize{10}{0}\selectfont A} (n60)
(n50) edge[above, bend left=20] node[blue]{\fontsize{10}{0}\selectfont C} (n60)
(n50) edge[below, line width=1.3pt, dashed, red] node[violet]{\fontsize{10}{0}\selectfont G} (n60)
(n50) edge[below, bend right=40] node[gray]{\fontsize{10}{0}\selectfont U} (n60)
(n21) edge[below, red, bend right=70, dashed, line width=.pt] node[black, pos=.3, yshift=.22cm]{x} (n30)
(n50) edge[below, red, dashed, line width=.pt] node[black, pos=.5, yshift=.23cm]{x} (n60);

\raisebox{-3.74cm}{
\node[state, initial, initial text=, inner sep=-10pt, fill=StartNode] (n00) {\fontsize{16.5}{0}\selectfont 0,0};
\node[state, right=0.7cm of n00, inner sep=-10pt, fill=NodeGray] (n10) {\fontsize{16.5}{0}\selectfont 1,0};
\node[state, below=.6cm of n10, inner sep=-10pt, fill=NodeGray] (n11) {\fontsize{16.5}{0}\selectfont 1,1};
\node[state, right=0.7cm of n10, inner sep=-10pt, fill=NodeGray] (n20) {\fontsize{16.5}{0}\selectfont 2,0};
\node[state, below=.6cm of n20, inner sep=-10pt, fill=NodeGray] (n21) {\fontsize{16.5}{0}\selectfont 2,1};
\node[state, right=0.7cm of n20, inner sep=-10pt, fill=NodeGray] (n30) {\fontsize{16.5}{0}\selectfont 3,0};
\node[state, below=.6cm of n30, inner sep=-10pt, fill=NodeGray] (n31) {\fontsize{16.5}{0}\selectfont 3,1};
\node[state, right=0.7cm of n30, inner sep=-10pt, fill=NodeGray] (n40) {\fontsize{16.5}{0}\selectfont 4,0};
\node[state, below=.6cm of n40, inner sep=-10pt, fill=NodeGray] (n41) {\fontsize{16.5}{0}\selectfont 4,1};
\node[state, right=0.7cm of n40, inner sep=-10pt, fill=NodeGray] (n50) {\fontsize{16.5}{0}\selectfont 5,0};
\node[state, below=.6cm of n50, inner sep=-10pt, fill=NodeGray] (n51) {\fontsize{16.5}{0}\selectfont 5,1};
\node[state, accepting, right=0.7cm of n50, inner sep=-10pt, fill=EndNode] (n60) {\fontsize{16.5}{0}\selectfont 6,0};
\node[below=2.3cm of n00, inner sep=-10pt] (n0a) {}; 
\node[right=.9cm of n0a] (n0b) {\fontsize{17}{0}\selectfont modified\! $D(\text{leucine})\circ\! D(\text{proline})$}; 
\node[below=-.2cm of n0b] (n0c) {\fontsize{17}{0}\selectfont modified\! without\! \nucC\nucU\nucC--\nucC\nucC\nucG}; 
\draw (n00) edge[above] node[gray]{\fontsize{10}{0}\selectfont U} (n10)
(n00) edge[above] node[blue]{\fontsize{10}{0}\selectfont C} (n11)
(n10) edge[above] node[gray]{\fontsize{10}{0}\selectfont U} (n20)
(n11) edge[above] node[gray]{\fontsize{10}{0}\selectfont U} (n21)
(n20) edge[above, bend left=38] node[red]{\fontsize{10}{0}\selectfont A} (n30)
(n20) edge[above] node[violet]{\fontsize{10}{0}\selectfont G} (n30)
(n21) edge[above, bend left=15] node[red]{\fontsize{10}{0}\selectfont A} (n30)
(n21) edge[above, bend right=9] node[violet]{\fontsize{10}{0}\selectfont G} (n30)
(n21) edge[above, bend right=34.3] node[gray]{\fontsize{10}{0}\selectfont U} (n30)
(n21) edge[below,  line width=1.5pt] node[blue]{\fontsize{10}{0}\selectfont C} (n31)
(n30) edge[above] node[blue]{\fontsize{10}{0}\selectfont C} (n40)
(n31) edge[above, line width=1.5pt] node[blue]{\fontsize{10}{0}\selectfont C} (n41)
(n40) edge[above] node[blue]{\fontsize{10}{0}\selectfont C} (n50)
(n41) edge[above, line width=1.5pt] node[blue]{\fontsize{10}{0}\selectfont C} (n51)
(n50) edge[above, bend left=70, inner sep = .8pt] node[red]{\fontsize{10}{0}\selectfont A} (n60)
(n50) edge[above, bend left=40, inner sep = .6pt] node[blue]{\fontsize{10}{0}\selectfont C} (n60)
(n50) edge[above, bend left=10, inner sep = .8pt] node[violet]{\fontsize{10}{0}\selectfont G} (n60)
(n50) edge[above, bend right=18, inner sep = .8pt] node[gray]{\fontsize{10}{0}\selectfont U} (n60)
(n51) edge[below, bend left=10, line width=1.5pt] node[red]{\fontsize{10}{0}\selectfont A} (n60)
(n51) edge[below, bend right=20, line width=1.5pt] node[blue]{\fontsize{10}{0}\selectfont C} (n60)
(n51) edge[below, bend right=50, line width=1.5pt] node[gray]{\fontsize{10}{0}\selectfont U} (n60);
}
\end{tikzpicture}}}
&
\hspace{-6.9cm}\raisebox{0cm}{
\begin{tikzpicture}
\draw (0,1) -- (0, -6.);
\end{tikzpicture}}
&
\hspace{-7.cm}\resizebox{.4\textwidth}{!}{\raisebox{2.5cm}{
\begin{tikzpicture}[->,>=stealth',shorten >=1pt,auto,semithick]
\node[state, initial, initial text=, inner sep=-10pt, fill=StartNode] (n00) {\fontsize{16.5}{0}\selectfont 0,0};
\node[state, right=1cm of n00, inner sep=-10pt, fill=NodeGray] (n10) {\fontsize{16.5}{0}\selectfont 1,0};
\node[state, below=1cm of n10, inner sep=-10pt, fill=NodeGray] (n11) {\fontsize{16.5}{0}\selectfont 1,1};
\node[state, right=1cm of n10, inner sep=-10pt, fill=NodeGray] (n20) {\fontsize{16.5}{0}\selectfont 2,0};
\node[state, below=1cm of n20, inner sep=-10pt, fill=NodeGray] (n21) {\fontsize{16.5}{0}\selectfont 2,1};
\node[state, accepting, right=1cm of n20, inner sep=-10pt, fill=EndNode] (n30) {\fontsize{16.5}{0}\selectfont 3,0};
\node[below=2.5cm of n00, inner sep=-10pt] (n0a) {}; 
\node[right=.9cm of n0a] (n0b) {\fontsize{16}{0}\selectfont extended $D(\text{serine})$}; 
\node[below=-.2cm of n0b] (n0c) {\fontsize{16}{0}\selectfont with $\Psi$, m$^6$A, and m$^5$C}; 
\draw (n00) edge[above, bend left=30] node[gray]{U} (n10)
(n00) edge[above, bend right=10, line width=1.5pt] node[gray]{$\Psi$} (n10)
(n00) edge[above] node[red]{A} (n11)
(n00) edge[below, bend right=30, line width=1.5pt, sloped, anchor=center, below] node[red]{m$^6$A} (n11)
(n10) edge[above, bend left=20] node[blue]{C} (n20)
(n10) edge[below, bend right=20, line width=1.5pt] node[blue]{m$^5$C} (n20)
(n11) edge[above] node[violet]{G} (n21)
(n20) edge[above, bend left=105, line width=1.5pt] node[red]{\fontsize{8}{0}\selectfont m$^6$A} (n30)
(n20) edge[above, bend left=60] node[red]{\fontsize{8}{0}\selectfont A} (n30)
(n20) edge[above, bend left=35] node[blue]{\fontsize{8}{0}\selectfont C} (n30)
(n20) edge[above, bend left=8, line width=1.5pt] node[blue]{\fontsize{8}{0}\selectfont m$^5$C} (n30)
(n20) edge[above, bend right=18] node[violet]{\fontsize{8}{0}\selectfont G} (n30)
(n20) edge[above, bend right=40] node[gray]{\fontsize{8}{0}\selectfont U} (n30)
(n20) edge[above, bend right=70, line width=1.5pt] node[gray]{\fontsize{8}{0}\selectfont $\Psi$} (n30)
(n21) edge[above, bend right=25] node[blue]{\fontsize{8}{0}\selectfont C} (n30)
(n21) edge[above, bend right=45] node[gray]{\fontsize{8}{0}\selectfont U} (n30)
(n21) edge[above, bend right=70, line width=1.5pt] node[gray]{\fontsize{8}{0}\selectfont $\Psi$} (n30)
(n21) edge[below, bend right=90, line width=1.5pt, sloped, anchor=center, below] node[blue]{\fontsize{8}{0}\selectfont m$^5$C} (n30);
\end{tikzpicture}}}
\vspace{-.3cm}
\end{tabular}
\caption{\myedit{Examples of the DFA representations for extended codons, modified nucleotides, and coding constraints.
{\bf A}: Alternative genetic codes of serine, tryptophan, and threonine.
{\bf B}: Avoiding certain codon. On the left it shows the original DFA of serine (up), in which the red dashed arrows indicating \nucU\nucC\nucA\  and \nucU\nucC\nucG\ are
chosen to be avoided,  
resulting in a new DFA (down)~\protect\mycite{fredens+:2019}.
On the right it shows removing the rare amber STOP codon (\nucU\nucA\nucG)~\protect\mycite{fredens+:2019}. }
{\bf C}: Avoiding a specific adjacent codon pair.
{\bf D}: Extended serine DFA can include chemically modified nucleotides pseudouridine ($\Psi$), 6-Methyladenosine (\msixA) and 5-methylcytosine (\mfiveC).
\label{fig:si_alter_aa}}
\end{figure*}

\begin{figure*}[!htb]
\centering
\begin{tabular}{c}
\vspace{.4cm}

\raisebox{0cm}{\hspace{-13cm}\panel{\myedit{A}}}\\[-1cm]

\hspace{-.2cm}\raisebox{-1cm}{\resizebox{.85\textwidth}{!}{
\begin{tikzpicture}[->,>=stealth',shorten >=1pt,auto,semithick]
\node[state, initial, initial text=, inner sep=-10pt, fill=StartNode] (n00) {\fontsize{16.5}{0}\selectfont 0,0};
\node[state, right=.8cm of n00, inner sep=-10pt, fill=NodeGray] (n10) {\fontsize{16.5}{0}\selectfont 1,0};
\node[state, right=.8cm of n10, inner sep=-10pt, fill=NodeGray] (n20) {\fontsize{16.5}{0}\selectfont 2,0};
\node[state, right=.8cm of n20, inner sep=-10pt, fill=NodeGray] (n30) {\fontsize{16.5}{0}\selectfont 3,0};
\node[state, right=.8cm of n30, inner sep=-10pt, fill=NodeGray] (n40) {\fontsize{16.5}{0}\selectfont 4,0};
\node[state, right=.8cm of n40, inner sep=-10pt, fill=NodeGray] (n50) {\fontsize{16.5}{0}\selectfont 5,0};
\node[state, right=.8cm of n50, inner sep=-10pt, fill=NodeGray] (n60) {\fontsize{16.5}{0}\selectfont 6,0};
\node[state, right=.8cm of n60, inner sep=-10pt, fill=NodeGray] (n70) {\fontsize{16.5}{0}\selectfont 7,0};
\node[state, right=.8cm of n70, inner sep=-10pt, fill=NodeGray] (n80) {\fontsize{16.5}{0}\selectfont 8,0};
\node[state, accepting, right=.8cm of n80, inner sep=-10pt, fill=EndNode] (n90) {\fontsize{16.5}{0}\selectfont 9,0};
\node[below=2.5cm of n00] (n0a) {}; 
\draw (n00) edge[above] node[violet]{\fontsize{10}{0}\selectfont G} (n10)
(n10) edge[above] node[red]{\fontsize{10}{0}\selectfont A} (n20)
(n20) edge[above, bend left=40] node[red]{\fontsize{10}{0}\selectfont A} (n30)
(n20) edge[above, line width=1.8pt, dashed] node[violet]{\fontsize{10}{0}\selectfont G} (n30)

(n30) edge[above, line width=1.8pt, dashed] node[violet]{\fontsize{10}{0}\selectfont G} (n40)
(n40) edge[above, line width=1.8pt, dashed] node[gray]{\fontsize{10}{0}\selectfont U} (n50)
(n50) edge[above, bend left=60, line width=1.8pt, dashed] node[red]{\fontsize{10}{0}\selectfont A} (n60)
(n50) edge[above, bend left=20] node[blue]{\fontsize{10}{0}\selectfont C} (n60)
(n50) edge[below] node[violet]{\fontsize{10}{0}\selectfont G} (n60)
(n50) edge[below, bend right=40] node[gray]{\fontsize{10}{0}\selectfont U} (n60)

(n60) edge[above, line width=1.8pt, dashed] node[blue]{\fontsize{10}{0}\selectfont C} (n70)
(n70) edge[above, line width=1.8pt, dashed] node[blue]{\fontsize{10}{0}\selectfont C} (n80)
(n80) edge[above, bend left=60] node[red]{\fontsize{10}{0}\selectfont A} (n90)
(n80) edge[above, bend left=20] node[blue]{\fontsize{10}{0}\selectfont C} (n90)
(n80) edge[below] node[violet]{\fontsize{10}{0}\selectfont G} (n90)
(n80) edge[below, bend right=40] node[gray]{\fontsize{10}{0}\selectfont U} (n90);

\node[below=.6cm of n00] (n00b) {};
\node[below=.6cm of n30] (n30b) {};
\node[below=.6cm of n60] (n60b) {};
\node[below=.6cm of n90] (n90b) {};
\draw (n00b) edge[below,<->,dashed] node{$D(\text{glutamic acid})$} (n30b);
\draw (n30b) edge[below,<->,dashed] node{$D(\text{valine})$} (n60b);
\draw (n60b) edge[below,<->,dashed] node{$D(\text{proline})$} (n90b);
\node[below=.35cm of n00] (n00mid) {$\mid$};
\node[below=.35cm of n30] (n30mid) {$\mid$};
\node[below=.75cm of n30] (n30circ) {$\circ$};
\node[below=.35cm of n60] (n60mid) {$\mid$};
\node[below=.75cm of n60] (n60circ) {$\circ$};
\node[below=.35cm of n90] (n90mid) {$\mid$};

\end{tikzpicture}}}\\[-1cm]
\hspace{.2cm}{\fontsize{18}{0}\selectfont $\Downarrow$} \\[-.5cm]

\hspace{-.2cm}\raisebox{0cm}{\resizebox{.85\textwidth}{!}{
\begin{tikzpicture}[->,>=stealth',shorten >=1pt,auto,semithick]
\node[state, initial, initial text=, inner sep=-10pt, fill=StartNode] (n00) {\fontsize{16.5}{0}\selectfont 0,0};
\node[state, right=.8cm of n00, inner sep=-10pt, fill=NodeGray] (n10) {\fontsize{16.5}{0}\selectfont 1,0};
\node[state, right=.8cm of n10, inner sep=-10pt, fill=NodeGray] (n20) {\fontsize{16.5}{0}\selectfont 2,0};
\node[state, right=.8cm of n20, inner sep=-10pt, fill=NodeGray] (n30) {\fontsize{16.5}{0}\selectfont 3,0};
\node[state, below=.7cm of n30, inner sep=-10pt, fill=NodeGray] (n31) {\fontsize{16.5}{0}\selectfont 3,1};
\node[state, right=.8cm of n30, inner sep=-10pt, fill=NodeGray] (n40) {\fontsize{16.5}{0}\selectfont 4,0};
\node[state, below=.7cm of n40, inner sep=-10pt, fill=NodeGray] (n41) {\fontsize{16.5}{0}\selectfont 4,1};
\node[state, right=.8cm of n40, inner sep=-10pt, fill=NodeGray] (n50) {\fontsize{16.5}{0}\selectfont 5,0};
\node[state, below=.7cm of n50, inner sep=-10pt, fill=NodeGray] (n51) {\fontsize{16.5}{0}\selectfont 5,1};
\node[state, right=.8cm of n50, inner sep=-10pt, fill=NodeGray] (n60) {\fontsize{16.5}{0}\selectfont 6,0};
\node[state, right=.8cm of n60, inner sep=-10pt, fill=NodeGray] (n70) {\fontsize{16.5}{0}\selectfont 7,0};
\node[state, right=.8cm of n70, inner sep=-10pt, fill=NodeGray] (n80) {\fontsize{16.5}{0}\selectfont 8,0};
\node[state, accepting, right=.8cm of n80, inner sep=-10pt, fill=EndNode] (n90) {\fontsize{16.5}{0}\selectfont 9,0};
\node[below=2.5cm of n00] (n0a) {}; 

\node[below=2.4cm of n00, inner sep=-10pt] (n0a) {}; 
\node[right=3.1cm of n0a] (n0b) {modified $D(\text{glutamic acid})\circ D(\text{valine})\circ D(\text{proline})$ without $\nucG\!\mid\!\nucG\nucU\nucA\!\mid\!\nucC\nucC$};

\draw (n00) edge[above] node[violet]{\fontsize{10}{0}\selectfont G} (n10)
(n10) edge[above] node[red]{\fontsize{10}{0}\selectfont A} (n20)
(n20) edge[above] node[red]{\fontsize{10}{0}\selectfont A} (n30)
(n20) edge[above, line width=1.5pt] node[violet]{\fontsize{10}{0}\selectfont G} (n31)
(n30) edge[above] node[violet]{\fontsize{10}{0}\selectfont G} (n40)
(n40) edge[above] node[gray]{\fontsize{10}{0}\selectfont U} (n50)
(n31) edge[above, line width=1.5pt] node[violet]{\fontsize{10}{0}\selectfont G} (n41)
(n41) edge[above, line width=1.5pt] node[gray]{\fontsize{10}{0}\selectfont U} (n51)
(n50) edge[above, bend left=60] node[red]{\fontsize{10}{0}\selectfont A} (n60)
(n50) edge[above, bend left=20] node[blue]{\fontsize{10}{0}\selectfont C} (n60)
(n50) edge[below] node[violet]{\fontsize{10}{0}\selectfont G} (n60)
(n50) edge[below, bend right=40] node[gray]{\fontsize{10}{0}\selectfont U} (n60)

(n51) edge[above, bend right=28, line width=1.5pt] node[blue]{\fontsize{10}{0}\selectfont C} (n60)
(n51) edge[above, bend right=52, line width=1.5pt] node[violet]{\fontsize{10}{0}\selectfont G} (n60)
(n51) edge[above, bend right=83, line width=1.5pt] node[gray]{\fontsize{10}{0}\selectfont U} (n60)

(n60) edge[above] node[blue]{\fontsize{10}{0}\selectfont C} (n70)
(n70) edge[above] node[blue]{\fontsize{10}{0}\selectfont C} (n80)
(n80) edge[above, bend left=60] node[red]{\fontsize{10}{0}\selectfont A} (n90)
(n80) edge[above, bend left=20] node[blue]{\fontsize{10}{0}\selectfont C} (n90)
(n80) edge[below] node[violet]{\fontsize{10}{0}\selectfont G} (n90)
(n80) edge[below, bend right=40] node[gray]{\fontsize{10}{0}\selectfont U} (n90);

\end{tikzpicture}}}\\[0cm]

\raisebox{0cm}{\hspace{-13cm}\panel{B}}\\[-.6cm]

\hspace{-.3cm}\raisebox{-1cm}{\resizebox{.85\textwidth}{!}{
\begin{tikzpicture}[->,>=stealth',shorten >=1pt,auto,semithick]
\node[state, initial, initial text=, inner sep=-10pt, fill=StartNode] (n00) {\fontsize{16.5}{0}\selectfont 0,0};
\node[state, right=.8cm of n00, inner sep=-10pt, fill=NodeGray] (n10) {\fontsize{16.5}{0}\selectfont 1,0};
\node[state, right=.8cm of n10, inner sep=-10pt, fill=NodeGray] (n20) {\fontsize{16.5}{0}\selectfont 2,0};
\node[state, right=.8cm of n20, inner sep=-10pt, fill=NodeGray] (n30) {\fontsize{16.5}{0}\selectfont 3,0};
\node[state, right=.8cm of n30, inner sep=-10pt, fill=NodeGray] (n40) {\fontsize{16.5}{0}\selectfont 4,0};
\node[state, right=.8cm of n40, inner sep=-10pt, fill=NodeGray] (n50) {\fontsize{16.5}{0}\selectfont 5,0};
\node[state, right=.8cm of n50, inner sep=-10pt, fill=NodeGray] (n60) {\fontsize{16.5}{0}\selectfont 6,0};
\node[state, right=.8cm of n60, inner sep=-10pt, fill=NodeGray] (n70) {\fontsize{16.5}{0}\selectfont 7,0};
\node[state, right=.8cm of n70, inner sep=-10pt, fill=NodeGray] (n80) {\fontsize{16.5}{0}\selectfont 8,0};
\node[state, accepting, right=.8cm of n80, inner sep=-10pt, fill=EndNode] (n90) {\fontsize{16.5}{0}\selectfont 9,0};
\node[below=2.5cm of n00] (n0a) {}; 
\draw (n00) edge[above] node[violet]{\fontsize{10}{0}\selectfont G} (n10)
(n10) edge[above, line width=1.8pt, dashed] node[violet]{\fontsize{10}{0}\selectfont G} (n20)
(n20) edge[above, bend left=60] node[red]{\fontsize{10}{0}\selectfont A} (n30)
(n20) edge[above, bend left=20] node[blue]{\fontsize{10}{0}\selectfont C} (n30)
(n20) edge[below] node[gray]{\fontsize{10}{0}\selectfont U} (n30)
(n20) edge[below, bend right=40, line width=1.8pt, dashed] node[violet]{\fontsize{10}{0}\selectfont G} (n30)

(n30) edge[above, line width=1.8pt, dashed] node[gray]{\fontsize{10}{0}\selectfont U} (n40)
(n40) edge[above, line width=1.8pt, dashed] node[red]{\fontsize{10}{0}\selectfont A} (n50)
(n50) edge[above, bend left=40, line width=1.8pt, dashed] node[blue]{\fontsize{10}{0}\selectfont C} (n60)
(n50) edge[above] node[gray]{\fontsize{10}{0}\selectfont U} (n60)

(n60) edge[above, line width=1.8pt, dashed] node[blue]{\fontsize{10}{0}\selectfont C} (n70)
(n70) edge[above] node[red]{\fontsize{10}{0}\selectfont A} (n80)
(n80) edge[above, bend left=40] node[blue]{\fontsize{10}{0}\selectfont C} (n90)
(n80) edge[above] node[gray]{\fontsize{10}{0}\selectfont U} (n90);

\node[below=.6cm of n00] (n00b) {};
\node[below=.6cm of n30] (n30b) {};
\node[below=.6cm of n60] (n60b) {};
\node[below=.6cm of n90] (n90b) {};
\draw (n00b) edge[below,<->,dashed] node{$D(\text{glycine})$} (n30b);
\draw (n30b) edge[below,<->,dashed] node{$D(\text{tyrosine})$} (n60b);
\draw (n60b) edge[below,<->,dashed] node{$D(\text{histidine})$} (n90b);
\node[below=.35cm of n00] (n00mid) {$\mid$};
\node[below=.35cm of n30] (n30mid) {$\mid$};
\node[below=.75cm of n30] (n30circ) {$\circ$};
\node[below=.35cm of n60] (n60mid) {$\mid$};
\node[below=.75cm of n60] (n60circ) {$\circ$};
\node[below=.35cm of n90] (n90mid) {$\mid$};

\node[below=1.8cm of n40] (narrow) {};
\node[right=0.4cm of narrow] (arrow) {\fontsize{20}{0}\selectfont$\Downarrow$};

\node[state, below=2.5cm of n00, inner sep=-10pt, fill=StartNode] (n00) {\fontsize{16.5}{0}\selectfont 0,0};
\node[state, right=.8cm of n00, inner sep=-10pt, fill=NodeGray] (n10) {\fontsize{16.5}{0}\selectfont 1,0};
\node[state, right=.8cm of n10, inner sep=-10pt, fill=NodeGray] (n20) {\fontsize{16.5}{0}\selectfont 2,0};
\node[state, right=.8cm of n20, inner sep=-10pt, fill=NodeGray] (n30) {\fontsize{16.5}{0}\selectfont 3,0};
\node[state, below=.7cm of n30, inner sep=-10pt, fill=NodeGray] (n31) {\fontsize{16.5}{0}\selectfont 3,1};
\node[state, right=.8cm of n30, inner sep=-10pt, fill=NodeGray] (n40) {\fontsize{16.5}{0}\selectfont 4,0};
\node[state, below=.7cm of n40, inner sep=-10pt, fill=NodeGray] (n41) {\fontsize{16.5}{0}\selectfont 4,1};
\node[state, right=.8cm of n40, inner sep=-10pt, fill=NodeGray] (n50) {\fontsize{16.5}{0}\selectfont 5,0};
\node[state, below=.7cm of n50, inner sep=-10pt, fill=NodeGray] (n51) {\fontsize{16.5}{0}\selectfont 5,1};
\node[state, right=.8cm of n50, inner sep=-10pt, fill=NodeGray] (n60) {\fontsize{16.5}{0}\selectfont 6,0};
\node[state, right=.8cm of n60, inner sep=-10pt, fill=NodeGray] (n70) {\fontsize{16.5}{0}\selectfont 7,0};
\node[state, right=.8cm of n70, inner sep=-10pt, fill=NodeGray] (n80) {\fontsize{16.5}{0}\selectfont 8,0};
\node[state, accepting, right=.8cm of n80, inner sep=-10pt, fill=EndNode] (n90) {\fontsize{16.5}{0}\selectfont 9,0};
\node[below=2.5cm of n00] (n0a) {}; 

\node[below=2.1cm of n00, inner sep=-10pt] (n0a) {}; 
\node[right=3.1cm of n0a] (n0b) {modified $D(\text{glycine})\circ D(\text{tyrosine})\circ D(\text{histidine})$ without $\nucG\nucG\!\mid\!\nucU\nucA\nucC\!\mid\!\nucC$};
\node[left=.6cm of n00] (n00left) {};

\draw (n00) edge[above] node[violet]{\fontsize{10}{0}\selectfont G} (n10)
(n10) edge[above] node[violet]{\fontsize{10}{0}\selectfont G} (n20)
(n20) edge[above, bend left=60] node[red]{\fontsize{10}{0}\selectfont A} (n30)
(n20) edge[above, bend left=20] node[blue]{\fontsize{10}{0}\selectfont C} (n30)
(n20) edge[below] node[gray]{\fontsize{10}{0}\selectfont U} (n30)
(n20) edge[below, line width=1.5pt] node[violet]{\fontsize{10}{0}\selectfont G} (n31)

(n30) edge[above] node[gray]{\fontsize{10}{0}\selectfont U} (n40)
(n40) edge[above] node[red]{\fontsize{10}{0}\selectfont A} (n50)
(n50) edge[above, bend left=40] node[blue]{\fontsize{10}{0}\selectfont C} (n60)
(n50) edge[above] node[gray]{\fontsize{10}{0}\selectfont U} (n60)

(n31) edge[above, line width=1.5pt] node[gray]{\fontsize{10}{0}\selectfont U} (n41)
(n41) edge[above, line width=1.5pt] node[red]{\fontsize{10}{0}\selectfont A} (n51)
(n51) edge[above, line width=1.5pt] node[gray]{\fontsize{10}{0}\selectfont U} (n60)

(n60) edge[above] node[blue]{\fontsize{10}{0}\selectfont C} (n70)
(n70) edge[above] node[red]{\fontsize{10}{0}\selectfont A} (n80)
(n80) edge[above, bend left=40] node[blue]{\fontsize{10}{0}\selectfont C} (n90)
(n80) edge[above] node[gray]{\fontsize{10}{0}\selectfont U} (n90)
(n00left) edge[above] node[violet]{} (n00);

\end{tikzpicture}}}\\[.8cm]

 \raisebox{0cm}{\hspace{-13cm}\panel{C}}  \\[-.4cm]
 
\hspace{0cm}\raisebox{-1cm}{\resizebox{.6\textwidth}{!}{
\begin{tikzpicture}[->,>=stealth',shorten >=1pt,auto,semithick]
\node[state, initial, initial text=, inner sep=-10pt, fill=StartNode] (n00) {\fontsize{16.5}{0}\selectfont 0,0};
\node[state, right=.8cm of n00, inner sep=-10pt, fill=NodeGray] (n10) {\fontsize{16.5}{0}\selectfont 1,0};
\node[state, right=.8cm of n10, inner sep=-10pt, fill=NodeGray] (n20) {\fontsize{16.5}{0}\selectfont 2,0};
\node[state, right=.8cm of n20, inner sep=-10pt, fill=NodeGray] (n30) {\fontsize{16.5}{0}\selectfont 3,0};
\node[state, right=.8cm of n30, inner sep=-10pt, fill=NodeGray] (n40) {\fontsize{16.5}{0}\selectfont 4,0};
\node[state, right=.8cm of n40, inner sep=-10pt, fill=NodeGray] (n50) {\fontsize{16.5}{0}\selectfont 5,0};
\node[state,accepting, right=.8cm of n50, inner sep=-10pt, fill=EndNode] (n60) {\fontsize{16.5}{0}\selectfont 6,0};
\node[below=2.5cm of n00] (n0a) {}; 
\draw (n00) edge[above, line width=1.8pt, dashed] node[violet]{\fontsize{10}{0}\selectfont G} (n10)
(n10) edge[above, line width=1.8pt, dashed] node[violet]{\fontsize{10}{0}\selectfont G} (n20)
(n20) edge[above, bend left=60] node[red]{\fontsize{10}{0}\selectfont A} (n30)
(n20) edge[above, bend left=20] node[blue]{\fontsize{10}{0}\selectfont C} (n30)
(n20) edge[below] node[violet]{\fontsize{10}{0}\selectfont G} (n30)
(n20) edge[below, bend right=40, line width=1.8pt, dashed] node[gray]{\fontsize{10}{0}\selectfont U} (n30)

(n30) edge[above, line width=1.8pt, dashed] node[red]{\fontsize{10}{0}\selectfont A} (n40)
(n40) edge[above, line width=1.8pt, dashed] node[blue]{\fontsize{10}{0}\selectfont C} (n50)
(n50) edge[above, bend left=60] node[red]{\fontsize{10}{0}\selectfont A} (n60)
(n50) edge[above, bend left=20, line width=1.8pt, dashed] node[blue]{\fontsize{10}{0}\selectfont C} (n60)
(n50) edge[below] node[violet]{\fontsize{10}{0}\selectfont G} (n60)
(n50) edge[below, bend right=40] node[gray]{\fontsize{10}{0}\selectfont U} (n60);

\node[below=.6cm of n00] (n00b) {};
\node[below=.6cm of n30] (n30b) {};
\node[below=.6cm of n60] (n60b) {};
\draw (n00b) edge[below,<->,dashed] node{$D(\text{glycine})$} (n30b);
\draw (n30b) edge[below,<->,dashed] node{$D(\text{threonine})$} (n60b);
\node[below=.35cm of n00] (n00mid) {$\mid$};
\node[below=.35cm of n30] (n30mid) {$\mid$};
\node[below=.75cm of n30] (n30circ) {$\circ$};
\node[below=.35cm of n60] (n60mid) {$\mid$};

\node[below=1.5cm of n30] (narrow) {\fontsize{20}{0}\selectfont$\Downarrow$};

\node[state, below=3cm of n00, inner sep=-10pt, fill=StartNode] (n00) {\fontsize{16.5}{0}\selectfont 0,0};
\node[left=.6cm of n00] (n00left) {};
\node[state, right=.8cm of n00, inner sep=-10pt, fill=NodeGray] (n10) {\fontsize{16.5}{0}\selectfont 1,0};
\node[state, right=.8cm of n10, inner sep=-10pt, fill=NodeGray] (n20) {\fontsize{16.5}{0}\selectfont 2,0};
\node[state, right=.8cm of n20, inner sep=-10pt, fill=NodeGray] (n30) {\fontsize{16.5}{0}\selectfont 3,0};
\node[state, below=.7cm of n30, inner sep=-10pt, fill=NodeGray] (n31) {\fontsize{16.5}{0}\selectfont 3,1};
\node[state, right=.8cm of n30, inner sep=-10pt, fill=NodeGray] (n40) {\fontsize{16.5}{0}\selectfont 4,0};
\node[state, below=.7cm of n40, inner sep=-10pt, fill=NodeGray] (n41) {\fontsize{16.5}{0}\selectfont 4,1};
\node[state, right=.8cm of n40, inner sep=-10pt, fill=NodeGray] (n50) {\fontsize{16.5}{0}\selectfont 5,0};
\node[state, below=.7cm of n50, inner sep=-10pt, fill=NodeGray] (n51) {\fontsize{16.5}{0}\selectfont 5,1};
\node[state, accepting, right=.8cm of n50, inner sep=-10pt, fill=EndNode] (n60) {\fontsize{16.5}{0}\selectfont 6,0};
\node[below=2.5cm of n00] (n0a) {}; 

\node[below=2.1cm of n00, inner sep=-10pt] (n0a) {}; 
\node[right=1cm of n0a] (n0b) {modified $D(\text{glycine})\circ D(\text{threonine})$ without $\nucG\nucG\nucU\!\mid\!\nucA\nucC\nucC$};

\draw (n00) edge[above] node[violet]{\fontsize{10}{0}\selectfont G} (n10)
(n10) edge[above] node[violet]{\fontsize{10}{0}\selectfont G} (n20)
(n20) edge[above, bend left=60] node[red]{\fontsize{10}{0}\selectfont A} (n30)
(n20) edge[above, bend left=20] node[blue]{\fontsize{10}{0}\selectfont C} (n30)
(n20) edge[below] node[violet]{\fontsize{10}{0}\selectfont G} (n30)
(n20) edge[below, line width=1.5pt] node[gray]{\fontsize{10}{0}\selectfont U} (n31)

(n30) edge[above] node[gray]{\fontsize{10}{0}\selectfont A} (n40)
(n40) edge[above] node[red]{\fontsize{10}{0}\selectfont C} (n50)
(n50) edge[above, bend left=70] node[red]{\fontsize{10}{0}\selectfont A} (n60)
(n50) edge[above, bend left=30] node[blue]{\fontsize{10}{0}\selectfont C} (n60)
(n50) edge[below, bend left = 10] node[violet]{\fontsize{10}{0}\selectfont G} (n60)
(n50) edge[below, bend right=30] node[gray]{\fontsize{10}{0}\selectfont U} (n60)

(n31) edge[above, line width=1.5pt] node[gray]{\fontsize{10}{0}\selectfont A} (n41)
(n41) edge[above, line width=1.5pt] node[red]{\fontsize{10}{0}\selectfont C} (n51)
(n51) edge[below, bend right=10, line width=1.5pt] node[red]{\fontsize{10}{0}\selectfont A} (n60)
(n51) edge[below, bend right = 40, line width=1.5pt] node[violet]{\fontsize{10}{0}\selectfont G} (n60)
(n51) edge[below, bend right=70, line width=1.5pt] node[gray]{\fontsize{10}{0}\selectfont U} (n60)

(n00left) edge[above] node[violet]{} (n00);

\end{tikzpicture}}}\\
 
\end{tabular}
\caption{\myedit{Avoiding a restriction enzyme specific recognition sequence (KpnI restriction enzyme recognization site: \nucG\nucG\nucU\nucA\nucC\nucC).
The enzyme recognization sequence is beyond one codon.
{\bf A--C} show three different partitions that split the sequence into different codons.}
See Fig.~\ref{fig:si_alter_aa}C for an example that avoids a specific adjacent codon pair.
\label{fig:si_alter_aa_enzyme}}
\end{figure*}

%% file: tab_si_1.tex
\begin{table}[htb!]
\centering
\renewcommand{\arraystretch}{1.3}
\resizebox{.7\textwidth}{!}{
\begin{tabular}{c|cccccc}
 sequence &   MFE        &        &  && \multicolumn{1}{c}{Molecular}     &  \\
 of CDS &   \multicolumn{1}{c}{(\it{kcal/mol})}    & \multicolumn{1}{c}{CAI}    & \multicolumn{1}{c}{GC\%}   & \multicolumn{1}{c}{U\%}   &  \multicolumn{1}{c}{weight} &  \\
  \hline \hline
{\sc a} & -2,287.3 & 0.756 & 54.8 & 22.7 & 1,229,541        &  \\
\hline
{\sc b} & -2,213.2 & 0.851 & 57.0 & 20.8 &  1,229,943        &   \\
{\sc c} & -2,206.0 & 0.757 & 55.3 & 22.2 & 1,230,380        &            \\
\hline
{\sc d} & -1,967.4 & 0.935 & 58.4 & 19.1 &1,229,927        &       \\
{\sc e} & -1,961.3 & 0.851 & 56.8 & 21.1 & 1,229,328        &            \\
{\sc f} & -1,969.3 & 0.755 & 54.7 & 23.0 & 1,230,153        &         \\
\hline
{\sc g} & -1,639.3 & 0.935 & 58.6 & 18.9 & 1,229,526        &             \\
\hline
{\sc h} & -1,244.4 & 0.936 & 55.0 & 21.2 &  1,228,180        &           \\
\hline \hline
MFE-Optimal &-2,486.7&0.732&54.3&22.9&1,229,980&\\
CAI-Optimal &-1,384.1& 1.000 &63.8&14.5&1,229,426&\\
Wildtype &\ \ -966.7& 0.655 &37.3&33.3&1,221,872&\\
\hline \hline
CV2CoV&-1,384.4& 0.903 &63.9&15.7&1,226,373&\\
mRNA-1273 &-1,369.2&  0.978  &62.3&15.5&1,229,366&\\
BNT-162b2 &-1,217.2& 0.946 &57.0&19.1&1,227,844&\\
\end{tabular}
}
\caption{Details of \lineardesign-generated sequences ({\sc a}--{\sc g} and MFE-Optimal), the baseline sequence ({\sc h}), the CAI-optimal sequence, the wildtype sequence, and three vaccine sequences from CureVac, Moderna, and BioNTech. See Fig.~\ref{fig:wetlab} for more details of \invitro and \invivo experiment results.
CDS sequences have no stop codon added.
  \label{tab:seqs_info}}
\end{table}

%% file: tab_si_utrs.tex
\newcolumntype{N}{>{\centering\arraybackslash}m{1.8cm}}
\newcolumntype{R}[1]{>{\raggedleft\arraybackslash}m{#1}}
	
\definecolor{LightCyan}{rgb}{0.88,1,1}

\ifarXiv
\begin{table}[htb!]
\scalebox{0.71}{
\hspace{-.4cm}
\begin{tabular}{N|R{1.3cm}|R{1.3cm}R{.4cm}R{.39cm}R{.39cm}|R{1.3cm}R{.4cm}R{.39cm}R{.39cm}|R{1.3cm}R{.4cm}R{.39cm}R{.39cm}|R{1.3cm}R{.4cm}R{.39cm}R{.39cm}|R{1.3cm}R{.4cm}R{.49cm}R{.49cm}}
sequence	 & MFE & \multicolumn{4}{c}{Stemirna UTRs} & \multicolumn{4}{c}{BioNTech UTRs} & \multicolumn{4}{c}{Moderna UTRs} & \multicolumn{4}{c}{CureVac UTRs} & \multicolumn{4}{c}{human $\beta$-globin UTRs} \\
of CDS & of CDS & MFE & tot. & 5' & 3' & MFE & tot. & 5' & 3'& MFE & tot. & 5' & 3'& MFE & tot. & 5' & 3'& MFE & tot. & 5' & 3' \\
                & \it{\scriptsize kcal/mol} & &  &  &  &\it{\scriptsize kcal/mol} & & & &\it{\scriptsize kcal/mol}  &  & &  & \it{\scriptsize kcal/mol} &  & && \it{\scriptsize kcal/mol}\\\hline \hline
\rowcolor{LightCyan}
{\sc a} & -2287.3 & -2325.1&14&12&2&-2378.0 & 8  & 5  & 3  & -2333.3 & 12 & 10 & 2  & -2313.4 & 5  & 5  & 0  & -2327.2 & 9  & 0  & 9  \\
\rowcolor{LightCyan}
{\sc b} & -2213.2 &-2252.1 &15&13&2&-2302.0 & 5  & 5  & 0  & -2258.1 & 14 & 12 & 2  & -2235.7 & 5  & 5  & 0  & -2252.0 & 29 & 0  & 29 \\
\rowcolor{LightCyan}
{\sc c} & -2206.0 &-2240.8 &10&8&2&-2294.1 & 6  & 6  & 0  & -2248.1 & 10 & 8  & 2  & -2229.7 & 17 & 17 & 0  & -2242.3 & 11 & 7  & 4  \\
\rowcolor{LightCyan}
{\sc d} & -1967.4 & -2002.5&13&9&4&-2056.4 & 10 & 5  & 5  & -2010.7 & 12 & 10 & 2  & -1991.8 & 3  & 0  & 3  & -2005.8 & 14 & 0  & 14 \\
\rowcolor{LightCyan}
{\sc e} & -1961.3 &-1999.1 &14&12&2&-2057.3 & 15 & 5  & 10 & -2008.9 & 16 & 14 & 2  & -1989.6 & 19 & 12 & 7  & -2002.2 & 18 & 0  & 18 \\
\rowcolor{LightCyan}
{\sc f} & -1969.3 &-2006.5 &11&9&2&-2060.9 & 11 & 5  & 6  & -2014.2 & 6  & 0  & 6  & -1993.8 & 11 & 0  & 11 & -2007.4 & 9  & 0  & 9  \\
{\sc g} & -1639.3 &-1681.5 &23&5&18&-1743.0 & 33 & 4  & 29 & -1686.8 & 61 & 0  & 61 & -1673.4 & 22 & 22 & 0  & -1687.3 & 46 & 11 & 35 \\\hline
{\sc h} & -1244.4 &-1284.9 &18 &8 &10 &-1345.3 & 72 & 8  & 64 & -1292.9 & 25 & 17 & 8  & -1285.5 & 27 & 19 & 8  & -1291.4 & 21 & 0  & 21 \\\hline \hline
CureVac      & -1384.4 & -1424.4&21&5&16&-1479.0 & 33 & 5  & 28 & -1430.5 & 82 & 26 & 56 & -1419.4 & 76 & 61 & 15 & -1425.5 & 20 & 0  & 20 \\
Moderna   & -1369.2 & -1411.6&29&4&25&-1464.3 & 24 & 4  & 20 & -1419.1 & 59 & 12 & 47 & -1406.6 & 63 & 45 & 18 & -1418.3 & 26 & 0  & 26 \\
BioNTech   & -1217.2 & -1259.1&34&5&29&-1316.3 & 98 & 6  & 92 & -1266.3 & 47 & 15 & 32 & -1253.9 & 58 & 54 & 4  & -1265.6 & 42 & 6  & 36 \\\hline \hline
\rowcolor{LightCyan}
MFE-opt. & -2486.7 &-2522.3&3&3&0& -2575.9 & 1  & 1  & 0  & -2532.2 & 3  & 3  & 0  & -2512.5 & 2  & 2  & 0  & -2522.6 & 13 & 0  & 13 \\
CAI-opt. & -1384.1 &-1424.1&29&4&25& -1478.0 & 58 & 0  & 58 & -1430.9 & 35 & 7  & 28 & -1420.4 & 53 & 53 & 0  & -1431.5 & 33 & 5  & 28 \\
Wildtype    & -966.7  & -1007.0&21&19&2&-1060.6 & 29 & 21 & 8  & -1019.1 & 27 & 25 & 2  & -1000.3 & 18 & 18 & 0  & -1011.7 & 56 & 9  & 47
\end{tabular}}
\caption{The numbers of base pairs formed between UTRs and the mRNA coding region, i.e., one base of the pairs is in 5' or 3'-UTR, and the other is in the coding region.
Here we used 5 different UTRs:
Stemirna UTRs used in wet lab experiments,
BNT-162b2 (BioNTech) UTRs, 
mRNA-1273 (Moderna) UTRs, 
CV2CoV (CureVac) UTRs, 
and a widely-used human $\beta$-globin mRNA UTRs.
We tested 14 different sequences of the coding region: 
sequences {\sc a}--{\sc h} in for wet lab experiments,
sequences from three main mRNA vaccine companies,
MFE-opt. and CAI-opt. sequences (i.e., sequences with the lowest folding free energy and with CAI=1, respectively),
and the wildtype sequence.
Most of the \lineardesign-generated mRNA sequences (sequences {\sc a}--{\sc f} and MFE-opt., shaded in light cyan) form fewer base pairs with UTRs. 
The folding free energies and structures are predicted by \viennarnafold (-d0 mode); MFEs of CDS are calculated without stop codon.
	\label{tab:si_utrs}
}
\end{table}

\else
\begin{table}[htb!]
\scalebox{0.64}{
\hspace{-.4cm}
\begin{tabular}{N|R{1.3cm}|R{1.3cm}R{.4cm}R{.39cm}R{.39cm}|R{1.3cm}R{.4cm}R{.39cm}R{.39cm}|R{1.3cm}R{.4cm}R{.39cm}R{.39cm}|R{1.3cm}R{.4cm}R{.39cm}R{.39cm}|R{1.3cm}R{.4cm}R{.49cm}R{.49cm}}
sequence	 & MFE & \multicolumn{4}{c}{Stemirna UTRs} & \multicolumn{4}{c}{BioNTech UTRs} & \multicolumn{4}{c}{Moderna UTRs} & \multicolumn{4}{c}{CureVac UTRs} & \multicolumn{4}{c}{human $\beta$-globin UTRs} \\
of CDS & of CDS & MFE & tot. & 5' & 3' & MFE & tot. & 5' & 3'& MFE & tot. & 5' & 3'& MFE & tot. & 5' & 3'& MFE & tot. & 5' & 3' \\
                & \it{\scriptsize kcal/mol} & &  &  &  &\it{\scriptsize kcal/mol} & & & &\it{\scriptsize kcal/mol}  &  & &  & \it{\scriptsize kcal/mol} &  & && \it{\scriptsize kcal/mol}\\\hline \hline
\rowcolor{LightCyan}
{\sc a} & -2287.3 & -2325.1&14&12&2&-2378.0 & 8  & 5  & 3  & -2333.3 & 12 & 10 & 2  & -2313.4 & 5  & 5  & 0  & -2327.2 & 9  & 0  & 9  \\
\rowcolor{LightCyan}
{\sc b} & -2213.2 &-2252.1 &15&13&2&-2302.0 & 5  & 5  & 0  & -2258.1 & 14 & 12 & 2  & -2235.7 & 5  & 5  & 0  & -2252.0 & 29 & 0  & 29 \\
\rowcolor{LightCyan}
{\sc c} & -2206.0 &-2240.8 &10&8&2&-2294.1 & 6  & 6  & 0  & -2248.1 & 10 & 8  & 2  & -2229.7 & 17 & 17 & 0  & -2242.3 & 11 & 7  & 4  \\
\rowcolor{LightCyan}
{\sc d} & -1967.4 & -2002.5&13&9&4&-2056.4 & 10 & 5  & 5  & -2010.7 & 12 & 10 & 2  & -1991.8 & 3  & 0  & 3  & -2005.8 & 14 & 0  & 14 \\
\rowcolor{LightCyan}
{\sc e} & -1961.3 &-1999.1 &14&12&2&-2057.3 & 15 & 5  & 10 & -2008.9 & 16 & 14 & 2  & -1989.6 & 19 & 12 & 7  & -2002.2 & 18 & 0  & 18 \\
\rowcolor{LightCyan}
{\sc f} & -1969.3 &-2006.5 &11&9&2&-2060.9 & 11 & 5  & 6  & -2014.2 & 6  & 0  & 6  & -1993.8 & 11 & 0  & 11 & -2007.4 & 9  & 0  & 9  \\
{\sc g} & -1639.3 &-1681.5 &23&5&18&-1743.0 & 33 & 4  & 29 & -1686.8 & 61 & 0  & 61 & -1673.4 & 22 & 22 & 0  & -1687.3 & 46 & 11 & 35 \\\hline
{\sc h} & -1244.4 &-1284.9 &18 &8 &10 &-1345.3 & 72 & 8  & 64 & -1292.9 & 25 & 17 & 8  & -1285.5 & 27 & 19 & 8  & -1291.4 & 21 & 0  & 21 \\\hline \hline
CureVac      & -1384.4 & -1424.4&21&5&16&-1479.0 & 33 & 5  & 28 & -1430.5 & 82 & 26 & 56 & -1419.4 & 76 & 61 & 15 & -1425.5 & 20 & 0  & 20 \\
Moderna   & -1369.2 & -1411.6&29&4&25&-1464.3 & 24 & 4  & 20 & -1419.1 & 59 & 12 & 47 & -1406.6 & 63 & 45 & 18 & -1418.3 & 26 & 0  & 26 \\
BioNTech   & -1217.2 & -1259.1&34&5&29&-1316.3 & 98 & 6  & 92 & -1266.3 & 47 & 15 & 32 & -1253.9 & 58 & 54 & 4  & -1265.6 & 42 & 6  & 36 \\\hline \hline
\rowcolor{LightCyan}
MFE-opt. & -2486.7 &-2522.3&3&3&0& -2575.9 & 1  & 1  & 0  & -2532.2 & 3  & 3  & 0  & -2512.5 & 2  & 2  & 0  & -2522.6 & 13 & 0  & 13 \\
CAI-opt. & -1384.1 &-1424.1&29&4&25& -1478.0 & 58 & 0  & 58 & -1430.9 & 35 & 7  & 28 & -1420.4 & 53 & 53 & 0  & -1431.5 & 33 & 5  & 28 \\
Wildtype    & -966.7  & -1007.0&21&19&2&-1060.6 & 29 & 21 & 8  & -1019.1 & 27 & 25 & 2  & -1000.3 & 18 & 18 & 0  & -1011.7 & 56 & 9  & 47
\end{tabular}}
\caption{The numbers of base pairs formed between UTRs and the mRNA coding region, i.e., one base of the pairs is in 5' or 3'-UTR, and the other is in the coding region.
Here we used 5 different UTRs:
Stemirna UTRs used in wet lab experiments,
BNT-162b2 (BioNTech) UTRs, 
mRNA-1273 (Moderna) UTRs, 
CV2CoV (CureVac) UTRs, 
and a widely-used human $\beta$-globin mRNA UTRs.
We tested 14 different sequences of the coding region: 
sequences {\sc a}--{\sc h} in for wet lab experiments,
sequences from three main mRNA vaccine companies,
MFE-opt. and CAI-opt. sequences (i.e., sequences with the lowest folding free energy and with CAI=1, respectively),
and the wildtype sequence.
Most of the \lineardesign-generated mRNA sequences (sequences {\sc a}--{\sc f} and MFE-opt., shaded in light cyan) form fewer base pairs with UTRs. 
The folding free energies and structures are predicted by \viennarnafold (-d0 mode); MFEs of CDS are calculated without stop codon.
	\label{tab:si_utrs}
}
\end{table}
\fi